\PassOptionsToPackage{colorlinks=true,allcolors=blue}{hyperref}
\documentclass[10pt,twocolumn,journal]{IEEEtran}

\usepackage[T1]{fontenc}
\usepackage{microtype}
\usepackage{xcolor}
\usepackage{amsmath,amssymb,amsthm,mathtools,bm}
\usepackage{graphicx}
\usepackage{stfloats,cite}
\usepackage{array,booktabs,tabularx,multirow,makecell}
\usepackage[inline,shortlabels]{enumitem}
\usepackage[ruled,vlined,linesnumbered]{algorithm2e}
\usepackage{braket}
\usepackage[caption=false,font=footnotesize]{subfig}
\IfFileExists{orcidlink.sty}
  {\usepackage{orcidlink}}
  {\newcommand{\orcidlink}[1]{}}
\usepackage[colorlinks=true,allcolors=blue]{hyperref}
\usepackage[capitalize,nameinlink]{cleveref}

\graphicspath{{figures/}}

\allowdisplaybreaks
\renewcommand{\arraystretch}{1.08}

\newtheorem{theorem}{Theorem}
\newtheorem{proposition}{Proposition}
\newtheorem{lemma}{Lemma}
\newtheorem{corollary}{Corollary}
\newtheorem{assumption}{Assumption}

\theoremstyle{definition}

\theoremstyle{remark}

\crefname{algocf}{Algorithm}{Algorithms}
\Crefname{algocf}{Algorithm}{Algorithms}
\crefname{equation}{Eq.}{Eqs.}
\Crefname{equation}{Equation}{Equations}

\DeclareMathOperator{\Tr}{Tr}
\DeclareMathOperator{\Ent}{Ent}
\DeclareMathOperator{\supp}{supp}

\DeclareMathOperator*{\argmin}{arg\,min}

\newcommand{\EX}{\mathbb{E}}

\newcommand{\TV}{D_{\mathrm{TV}}}
\newcommand{\KL}{D_{\mathrm{KL}}}
\newcommand{\best}[1]{\textbf{#1}}
\newcommand{\second}[1]{\underline{#1}}

\newif\ifrev
\revfalse

\ifrev
  
  \newenvironment{revblock}
    {\par\begingroup\color{blue}}
    {\par\endgroup}
\else
  
  \newenvironment{revblock}
    {\par\begingroup}
    {\par\endgroup}
\fi

\newcommand{\blue}[1]{#1}

\title{Robust Belief-State Policy Learning for Quantum Network Routing Under Decoherence and Time-Varying Conditions}

\author{
Amirhossein Taherpour~\orcidlink{0000-0003-4647-102X},
\thanks{Amirhossein Taherpour is with the Department of Electrical Engineering, Columbia University, New York, NY, USA (e-mail: at3532@columbia.edu).}
Abbas Taherpour~\orcidlink{0000-0003-0706-5774},~\IEEEmembership{Member,~IEEE},
\thanks{Abbas Taherpour and Tamer Khattab are with the Intelligent Information Processing Lab (IIPL), Electrical Engineering, Qatar University, Doha, Qatar (e-mail: taherpour@eng.ikiu.ac.ir, tkhattab@ieee.org).}
Tamer Khattab~\orcidlink{0000-0003-2347-9555},~\IEEEmembership{Senior Member,~IEEE},
Mazen Hasna~\orcidlink{0000-0002-3088-6899},~\IEEEmembership{Senior Member,~IEEE}
\thanks{Mazen Hasna is with the Department of Electrical Engineering, Qatar University, Doha, Qatar (e-mail: hasna@qu.edu.qa).}
}

\begin{document}
\maketitle

\begin{abstract}
\blue{
Quantum network routing requires online decisions under probabilistic entanglement generation, finite quantum memories, decoherence, imperfect operations, and classical feedback, while the controller has incomplete knowledge of the physical state. This paper develops a robust belief-state routing framework based on a quantum partially observable Markov decision process (q-POMDP) and a feasibility-masked graph neural network (GNN). The model uses atomic micro-epochs in which each selected operation completes before the next decision boundary. This enables explicit accounting of memory reservations, pair-instance inventories, purification consumption, swapping outcomes, release decisions, queue service, and completion-time delivery fidelity. The controller maintains a classical belief over hidden physical states, including latent environmental conditions, and uses this belief to evaluate feasible actions and update posterior pair states. To make planning scalable, we introduce feasibility-stratified prototypes, identifier-free signatures, and role-aware action matching, which preserve hard resource constraints while enabling value transfer across structurally similar information states. A cached q-POMDP planner is then fused with a role-aware GNN policy through an adaptive trust rule, with a safe fallback for previously unseen feasibility signatures. We provide theoretical guarantees on feasibility, value approximation, policy performance, robustness, regret, and learning variance. Simulations over finite-memory quantum-network topologies show that the proposed hybrid controller improves high-fidelity goodput, reduces below-threshold deliveries, and maintains lower online decision cost than planner-only control, while outperforming heuristic, purification-aware, and learning-based baselines.}
\end{abstract}
\begin{IEEEkeywords}
\blue{
Quantum networks, quantum partially observable Markov decision process (q-POMDP), graph neural network (GNN), entanglement routing, belief-state planning, decoherence-aware control, finite-memory routing, robust policy learning.
}
\end{IEEEkeywords}

\section{Introduction}
\label{sec:introduction}

\IEEEPARstart{Q}{uantum} networks seek to distribute entanglement among remote nodes as a primitive for secure communication, distributed quantum computation, and networked sensing. The quantum-internet vision has evolved from architectural proposals~\cite{Kimble2008QuantumInternet,Wehner2018QuantumInternet} toward experimental demonstrations of multi-node operation, non-neighbor teleportation, network-stack-based entanglement delivery, and metropolitan-scale heralded entanglement~\cite{Pompili2021Multinode,Hermans2022Teleportation,Pompili2022EntanglementDelivery,Stolk2024Metropolitan}. As these systems move from isolated links to controlled networks, routing becomes a dynamic decision problem: a controller must decide when to generate, store, purify, swap, deliver, or release entangled pairs under lossy heralding, finite memories, imperfect local operations, classical feedback, decoherence, and incomplete physical-state information.

A first way to address this problem is to develop the physical and protocol mechanisms that make remote entanglement possible. Quantum repeaters show how long-distance entanglement can be built from shorter lossy segments~\cite{Briegel1998Repeaters}, while entanglement purification protocols explain how noisy pairs can be converted into fewer higher-fidelity pairs~\cite{Bennett1996Purification,Deutsch1996Purification}. Link-layer and network-stack designs then expose heralded physical attempts as controllable services for higher-layer protocols~\cite{Dahlberg2019LinkLayer,Pompili2022EntanglementDelivery}. These works establish the operational ingredients that a routing layer must respect: pair creation is probabilistic, stored entanglement decoheres, purification consumes resources, swapping has branch-dependent outcomes, and delivery quality depends on the state of the pair at service time.

A second way to respond to the routing challenge is to formulate graph-level path selection, resource allocation, and entanglement scheduling problems. Early path-selection and optimal-routing formulations highlight the role of probabilistic links and repeater paths~\cite{VanMeter2013PathSelection,Caleffi2017OptimalRouting,Pant2019RoutingEntanglement}. Flow-based, fidelity-guaranteed, and purification-aware formulations further incorporate multi-demand traffic, finite resources, and service-quality constraints~\cite{Chakraborty2020Multicommodity,Dai2020OptimalRemoteEntanglement,Li2021EffectiveRouting,Li2022FidelityGuaranteed,Zhao2022EFiARP}. More recent work studies online routing, congestion mitigation, concurrent routing, swapping order, asynchronous operation, and reliable large-scale distribution~\cite{Yang2022OnlineEntanglement,Li2023SwappingCongestion,Zeng2024EntanglementRoutingDesign,ShiQian2020ConcurrentRouting,Shi2024ConcurrentRoutingTNET,ChangXue2022OrderMatters,Yang2024AsynchronousRouting,Chen2024REDP}. Together, these approaches show that routing performance is governed not only by graph distance, but also by fidelity evolution, memory pressure, operation ordering, and traffic dynamics.

A third way to make the control problem scalable is to rely on simulation and learning. Discrete-event quantum-network simulators provide reproducible environments for studying link-layer and network-layer behavior~\cite{Coopmans2021NetSquid,DiAdamo2021QuNetSim,Wu2021SeQUeNCe}. Reinforcement learning and graph neural networks (GNNs) are natural tools for such settings because feasible actions and stored-pair graphs change over time, while graph encoders can share structure across network instances~\cite{Wu2021GNN,Zhang2022DeepGraphs}. Recent learning-based routing methods demonstrate the promise of data-driven policies for entanglement distribution in dynamic quantum networks~\cite{Taherpour2026DistributedLearning,LeNguyen2022DQRA,Le2022DeepRL,Shaban2024SPARQ,Roik2024RoutingRL,Meuser2026RELiQ}.

Despite this progress, existing approaches do not jointly provide an operational quantum-network model, partial-observation belief updates, finite-memory resource accounting, scalable graph policies, planner--learner fusion, and theory whose assumptions match the implemented controller. In particular, decoherence alone does not make the control problem non-Markovian: a Markovian quantum partially observable Markov decision process (q-POMDP) can be obtained by augmenting the hidden state with the relevant finite-memory latent environment, while external calibration drift should be treated separately. What is still missing is a routing framework that makes this distinction explicit, updates a classical belief over hidden physical states, preserves pair-instance and memory semantics, and combines model-based planning with graph-based generalization under hard feasibility constraints.

This paper develops such a framework: a robust belief-state q-POMDP and feasibility-masked GNN controller for decoherence-aware quantum-network routing under time-varying physical conditions. The main contributions are as follows.
\begin{itemize}
\item \textbf{Operational q-POMDP model.} We formulate an atomic micro-epoch quantum-network model with on-demand heralded generation, optical-loss success probabilities, finite memory reservations, Deutsch--Ekert--Jozsa--Macchiavello--Popescu--Sanpera (DEJMPS) purification costs, swapping consumption, completion-time delivery fidelity, bounded rewards, and a classical belief over hidden quantum-network states.

\item \textbf{Feasibility-preserving belief aggregation.} We introduce identifier-free feasibility signatures, same-signature prototype assignment, and role-aware action matching. This yields a projected Bellman value bound on represented signatures while preserving hard action feasibility and concrete pair-instance semantics.

\item \textbf{Hybrid planner--GNN routing.} We combine a cached q-POMDP planner with a role-aware, feasibility-masked GNN policy. A planner-to-GNN Kullback--Leibler (KL) trust rule adaptively fuses the two policies, falls back safely on unseen signatures, and supports semi-gradient actor--critic learning.

\item \textbf{Guarantees and empirical validation.} We prove feasibility, projected value approximation, hybrid-policy performance, robustness to model error, nonstationary frozen-model value-regret, and conditional variance bounds. We evaluate the framework in a finite-memory discrete-event simulator against routing, purification-aware, and learning baselines.
\end{itemize}

The rest of the paper is organized as follows. \cref{sec:related_work} reviews related work. \cref{sec:system_model} defines the operational quantum-network model. \cref{sec:problem} formulates the q-POMDP routing objective. \cref{sec:feature_selection_guarantees} develops belief aggregation, signatures, and action matching. \cref{sec:hybrid_pomdp_gnn} presents the hybrid planner--GNN controller. \cref{sec:hybrid_analysis} gives the main analytical guarantees. \cref{sec:simulation} reports numerical results. \cref{sec:conclusion} concludes the paper.


\section{Related Work}
\label{sec:related_work}

\subsection{Quantum Repeater Networks and Entanglement Operations}

Quantum repeaters, entanglement swapping, and purification provide the physical substrate for long-distance quantum networking~\cite{Briegel1998Repeaters,Bennett1996Purification,Deutsch1996Purification}. The routing layer cannot treat these operations as free graph transformations: generation is heralded and probabilistic, purification consumes multiple noisy pairs per successful output, swapping consumes input pairs and has branch-dependent outcomes, and stored pairs decohere while waiting. Link-layer designs and network-stack demonstrations make these operation-level constraints visible to higher layers~\cite{Dahlberg2019LinkLayer,Pompili2022EntanglementDelivery}. Our model takes these constraints as first-class elements of the q-POMDP state, action, and transition kernel.

\subsection{Entanglement Routing and Resource Allocation}

Path selection and entanglement-routing formulations for repeater networks have studied routing metrics, end-to-end success probabilities, and optimal path choices under probabilistic links~\cite{VanMeter2013PathSelection,Caleffi2017OptimalRouting,Pant2019RoutingEntanglement}. Flow-based and optimization approaches further address multiple commodities, congestion, fidelity guarantees, and purification-aware throughput maximization~\cite{Chakraborty2020Multicommodity,Dai2020OptimalRemoteEntanglement,Li2021EffectiveRouting,Li2022FidelityGuaranteed,Zhao2022EFiARP,Zeng2024EntanglementRoutingDesign,Li2023SwappingCongestion}. Online and concurrent routing approaches study scheduling and ordering effects under dynamic traffic and swapping operations~\cite{ShiQian2020ConcurrentRouting,Yang2022OnlineEntanglement,ChangXue2022OrderMatters,Yang2024AsynchronousRouting,Chen2024REDP}. These works motivate the baselines used in \cref{sec:simulation}. However, most of them either assume a more directly observable state or omit belief filtering. More importantly, they do not provide a joint planner--GNN architecture with explicit finite-memory branch ledgers.

\subsection{Learning-Based Quantum Network Control}

Learning-based quantum-network routing has recently used deep reinforcement learning and graph-structured policies to handle large action spaces and dynamic topologies~\cite{LeNguyen2022DQRA,Le2022DeepRL,Shaban2024SPARQ,Roik2024RoutingRL,Meuser2026RELiQ}. Such methods can generalize across graphs and respond quickly to observed network features, but they often lack explicit belief-state semantics or formal guarantees under partial observability. General GNN methodology supports permutation-invariant graph encoders and scalable relational policies~\cite{Wu2021GNN,Zhang2022DeepGraphs}. Our work combines these advantages with a q-POMDP planner and a trust rule that preserves feasibility under both represented and unseen signatures.

\subsection{Quantum Network Simulation}

Discrete-event simulators such as NetSquid, QuNetSim, and SeQUeNCe provide platforms for modeling quantum network protocols with physical timing, channels, and classical control~\cite{Coopmans2021NetSquid,DiAdamo2021QuNetSim,Wu2021SeQUeNCe}. Our experiments follow this discrete-event perspective but impose the atomic micro-epoch abstraction introduced in the theory: exactly one elementary routing action is selected per epoch, completed before the next boundary, and charged to the memory and pair-inventory ledger. In contrast to the approaches above, our framework jointly addresses partial observability, feasibility preservation, and hybrid planner--GNN control under hard resource constraints.


\section{System Model}
\label{sec:system_model}

For convenience, \cref{tab:notation} summarizes the principal notation used throughout the paper. Additional symbols are defined when first introduced.

\begin{table*}[t]
\caption{Principal notation used throughout the paper.}
\label{tab:notation}
\centering
\small
\setlength{\tabcolsep}{3.2pt}
\renewcommand{\arraystretch}{1.16}
\begin{tabularx}{\textwidth}{@{}p{0.15\textwidth}X|p{0.15\textwidth}X@{}}
\toprule
\textbf{Symbol} & \textbf{Meaning} &
\textbf{Symbol} & \textbf{Meaning}\\
\midrule
\multicolumn{4}{@{}l}{\textit{Network, state, and memory}}\\
\midrule
\(\mathcal G^{\mathrm{phy}}\) & Physical graph \((\mathcal V,\mathcal E^{\mathrm{phy}},\boldsymbol\ell)\). &
\(\mathcal G_t\) & Stored-pair instance multigraph \((\mathcal V,\mathcal P_t,\omega_t)\).\\
\(\mathcal P_t\) & Stored Bell-pair instances at epoch \(t\). &
\(\partial p\) & Endpoint set of pair instance \(p\).\\
\(\rho_{b,t}^{p}\) & Belief-induced reduced state of pair \(p\). &
\(\mathbf e_{p,t}\) & Identifier-free pair record: age, depth, eligibility, and error summary.\\
\(m_v^{\max}\) & Quantum-memory capacity at node \(v\). &
\(\mu_v(t)\) & Occupied memory cells at node \(v\).\\
\(r_v(a,c)\) & New reservation demand of action \(a\) at node \(v\). &
\(\Delta\mu_{v,t}^{\mathrm{conv}},\Delta\mu_{v,t}^{\mathrm{rel}}\) & Reservations converted to occupancy or released.\\
\(N_{v,t}^{+},N_{v,t}^{-}\) & Other occupied-cell creations and removals. &
\(q_k(t)\) & Backlog of demand class \(k\).\\
\midrule
\multicolumn{4}{@{}l}{\textit{Physical dynamics and belief state}}\\
\midrule
\(X_t\) & Hidden physical state. &
\(C_t\) & Observable classical state.\\
\(Z_t\) & Latent environmental state. &
\(\boldsymbol\vartheta_t,\widehat{\boldsymbol\vartheta}_t\) & True physical parameter and online estimate.\\
\(\Gamma_v^\phi\) & Physical dephasing rate, \(1/T_{2,v}\). &
\(\gamma\) & Discount factor.\\
\(b_t\) & Classical belief over hidden states. &
\(\varsigma_t\) & Information state \((b_t,c_t)\).\\
\(\mathcal B,\mathcal C,\mathcal S\) & Belief, classical-state, and information-state spaces. &
\(\phi(b,c)\) & Global information-state feature map.\\
\midrule
\multicolumn{4}{@{}l}{\textit{q-POMDP kernels, actions, and rewards}}\\
\midrule
\(\mathsf Q_{\boldsymbol\vartheta}\) & Controlled transition--observation kernel. &
\(\mathsf T_{\boldsymbol\vartheta}\) & Induced information-state transition kernel.\\
\(\mathsf O_{\boldsymbol\vartheta},\mathsf U_{\boldsymbol\vartheta}\) & Outcome kernel and branch-update kernel. &
\(\mathcal A_{\mathrm f}(\varsigma)\) & Feasible atomic action set at state \(\varsigma\).\\
\(\mathcal Y\) & Action-tagged outcome space. &
\(\mathcal K_{a,y}^{z,\boldsymbol\vartheta}\) & Selective CP branch map for action \(a\), outcome \(y\).\\
\(\mathfrak I_{a,z,c,\boldsymbol\vartheta}^{\mathrm{red}}\) & Reduced flagged instrument. &
\(\mathcal E_{a,z,c,\boldsymbol\vartheta}^{\mathrm{ns}}\) & Nonselective local channel with passive evolution.\\
\(R_{\mathrm{rob}}\) & Reward for completed robust actions. &
\(R_{\max},V_{\max}\) & Reward bound and value bound \(R_{\max}/(1-\gamma)\).\\
\midrule
\multicolumn{4}{@{}l}{\textit{State aggregation, matching, and hybrid control}}\\
\midrule
\(\mathfrak G_{\mathrm{adm}}\) & Relabeling group for transient identifiers. &
\(\sigma(\varsigma),\Sigma\) & Feasibility signature and signature set.\\
\(\mathcal Q_K,\Sigma_K\) & Prototype set and represented signatures. &
\(\epsilon_K\) & Same-signature feature covering radius.\\
\(M_{\varsigma\rightarrow\varsigma'}\) & Role-aware matched-action correspondence. &
\(Q_K\) & Cached prototype action-value table.\\
\(\phi_v,\phi_e\) & GNN node and pair encoders. &
\(\pi_{\mathrm G},\pi_{\mathrm P},\pi_{\mathrm H}\) & GNN, planner, and hybrid policies.\\
\(\alpha_t\) & GNN trust coefficient in policy fusion. &
\(\beta_t^{\mathrm G}\) & GNN responsibility in the semi-gradient update.\\
\(\mathcal D_t^{\mathrm{ind}}\) & Independent-worker transition batch. &
\(\Ent(\cdot)\), \(\KL(\cdot\|\cdot)\) & Shannon entropy and Kullback--Leibler divergence.\\
\bottomrule
\end{tabularx}
\end{table*}

We consider a quantum communication network with fixed physical connectivity
\begin{equation}
\mathcal G^{\mathrm{phy}}
=
\bigl(\mathcal V,\mathcal E^{\mathrm{phy}},\boldsymbol\ell\bigr),
\label{eq:physical_network_graph}
\end{equation}
where \(\mathcal V=\{v_1,\ldots,v_n\}\) is the set of quantum-capable nodes, \(\mathcal E^{\mathrm{phy}}\subseteq\mathcal V\times\mathcal V\) is the set of physical optical links, and \(\boldsymbol\ell=[\ell_e]_{e\in\mathcal E^{\mathrm{phy}}}\) contains link lengths. Fixed node and link attributes, including memory capacities, hardware roles, directionality, attenuation parameters, and demand roles, are intrinsic properties of the physical network and remain unchanged when transient runtime objects are relabeled.

At decision epoch \(t\), stored entanglement is represented by the pair-instance multigraph
\begin{equation}
\mathcal G_t=(\mathcal V,\mathcal P_t,\omega_t),
\qquad
\omega_t(p)=\bigl(\rho_{b,t}^{p},\mathbf e_{p,t}\bigr),
\label{eq:quantum_network_graph}
\end{equation}
where \(\mathcal P_t\) is the finite collection of stored elementary or swapped Bell-pair instances. Each pair instance \(p\in\mathcal P_t\) has endpoints
\begin{equation}
\partial p=\{u_p,v_p\}\subseteq\mathcal V,
\label{eq:pair_endpoint_map}
\end{equation}
posterior reduced state \(\rho_{b,t}^{p}\in\mathcal D(\mathbb C^2\otimes\mathbb C^2)\), and identifier-free record \(\mathbf e_{p,t}\in\mathbb R^{d_e}\). The record contains age, generation history, purification depth, endpoint roles, delivery eligibility, and an operational-error summary \(\varepsilon_{p,t}^{\mathrm{op}}\in[0,1]\). Multiple pair instances may have the same endpoints. Transient pair identifiers distinguish runtime instances only; they are not numerical features and are not used for value transfer. The graph \(\mathcal G_t\) is a controller-side representation and does not assume that the hidden global quantum state factorizes across stored pairs.

\begin{revblock}

\subsection{Atomic Micro-Epoch Timing}
\label{subsec:atomic_timing_model}

Time is divided into routing micro-epochs \(t=0,1,\ldots\). At the beginning of epoch \(t\), the controller observes the information state \((b_t,c_t)\) and selects a feasible elementary action \(a_t\); the idle action is always feasible. The events within epoch \(t\) occur in the following order:
\begin{equation}
\label{eq:system_event_order}
\small
\begin{aligned}
\mathsf{Obs}
&\to \mathsf{Sel}
\to \mathsf{Res}
\to \mathsf{Evol} \\
&\to \mathsf{Exec}
\to \mathsf{Out}
\to \mathsf{Upd}
\to \mathsf{Rew}
\to \mathsf{Bel}.
\end{aligned}
\end{equation}
Here, \(\mathsf{Obs}\) denotes observation, \(\mathsf{Sel}\) action selection, \(\mathsf{Res}\) reservation, \(\mathsf{Evol}\) passive evolution, \(\mathsf{Exec}\) execution of the selected instrument, \(\mathsf{Out}\) outcome realization, \(\mathsf{Upd}\) inventory and memory update, \(\mathsf{Rew}\) reward realization, and \(\mathsf{Bel}\) belief update.

No operation, pair lock, memory reservation, or unfinished heralding attempt persists across decision boundaries. A generation action may contain multiple parallel optical attempts, but all attempts occur on one selected physical link and are internal to that one atomic action. Operations involving different links or disjoint groups of pairs require separate micro-epochs.

\end{revblock}

\subsection{Finite Quantum and Classical Memory}
\label{subsec:memory_model}

Node \(v\) has \(m_v^{\max}\) quantum-memory cells and memory Hilbert space \(\mathcal M_v\) with \(\dim(\mathcal M_v)=2^{m_v^{\max}}\). The global memory space is
\begin{equation}
\mathcal H^q=\bigotimes_{v\in\mathcal V}\mathcal M_v.
\label{eq:global_hilbert_space}
\end{equation}
Qubits supporting pair instances are subregisters of their endpoint memories and are not represented by an additional tensor factor.

Let \(\mu_v(t)\) be the number of occupied quantum-memory cells immediately before action selection at node \(v\). Synchronous completion implies that the decision-boundary reservation count is zero. If action \(a_t\) requires \(r_v(a_t,c_t)\) new within-epoch reservations at node \(v\), feasibility is checked before execution:
\begin{equation}
0\le \mu_v(t)+r_v(a_t,c_t)\le m_v^{\max},
\qquad v\in\mathcal V.
\label{eq:memory_capacity_constraint}
\end{equation}
During the epoch, \(\Delta\mu_{v,t}^{\mathrm{conv}}\) and \(\Delta\mu_{v,t}^{\mathrm{rel}}\) denote reservation conversions and reservation releases, whereas \(N_{v,t}^{+}\) and \(N_{v,t}^{-}\) denote other occupied-cell creations and removals. Reservation resolution requires
\begin{equation}
\Delta\mu_{v,t}^{\mathrm{conv}}+\Delta\mu_{v,t}^{\mathrm{rel}}
=
r_v(a_t,c_t),
\label{eq:reservation_resolution}
\end{equation}
and the occupied-cell update is
\begin{equation}
\mu_v(t+1)
=
\mu_v(t)+\Delta\mu_{v,t}^{\mathrm{conv}}+N_{v,t}^{+}-N_{v,t}^{-}.
\label{eq:memory_occupancy_update}
\end{equation}
The pair inventory and memory ledger are updated atomically and satisfy, at every decision boundary,
\begin{equation}
\mu_v(t)
=
\sum_{p\in\mathcal P_t}\mathbf 1_{\{v\in\partial p\}},
\qquad v\in\mathcal V.
\label{eq:memory_pair_inventory_invariant}
\end{equation}
Thus, every stored pair occupies one memory cell at each endpoint, and every pair consumption, release, delivery, or expiration removes one occupied cell at each endpoint.

The observable classical state \(c_t\) contains the authenticated pair inventory, occupied-memory counters, request queues, heralding and measurement records, timestamps, posterior summaries, hardware status, and calibration estimates. If \(\chi_v(t)\) is the number of stored classical records at node \(v\), then \(0\le\chi_v(t)\le\chi_v^{\max}\). Classical memory is finite but is not the limiting resource in the theoretical routing formulation; when a classical buffer is full, new classical records are rejected by admission control or compressed into finite posterior summaries already included in \(c_t\).

Thus, memory is treated as a finite hardware resource rather than an abundant background assumption; the simulations explicitly sweep \(m_v^{\max}\in\{4,8,16\}\) to expose memory-limited and memory-abundant regimes, while classical information is compressed into bounded counters, records, and posterior summaries because classical buffering is not the scarce resource in the routing model.

\begin{revblock}

\subsection{On-Demand Heralded Entanglement Generation}
\label{subsec:entanglement_generation}

Elementary entanglement generation is on demand. A generation action \(a_t=(\mathsf G,(e,g_e(t)))\) selects one physical link \(e=(u,v)\) and an attempt count \(g_e(t)\in\mathcal G_e\), where \(\mathcal G_e\subset\mathbb Z_+\) is finite and also memory-bounded by \cref{eq:memory_capacity_constraint}. It reserves \(g_e(t)\) cells at each endpoint, so \(r_u=r_v=g_e(t)\) and \(r_w=0\) for \(w\notin\{u,v\}\). Conditional on latent environment \(Z_t\) and physical parameter vector \(\boldsymbol\vartheta_t\), the heralded success count is
\begin{equation}
G_e(t)\mid g_e(t),Z_t,\boldsymbol\vartheta_t
\sim
\operatorname{Binomial}\!\left(g_e(t),
p_e^{\mathrm{gen}}(Z_t,\boldsymbol\vartheta_t)\right),
\label{eq:heralded_generation}
\end{equation}
where
\begin{equation}
p_e^{\mathrm{gen}}(Z_t,\boldsymbol\vartheta_t)
=
p_e^{\mathrm{sys}}(\boldsymbol\vartheta_t)
10^{-\alpha_{\mathrm{att}}(\boldsymbol\vartheta_t)\ell_e/10}
\xi_e(Z_t,\boldsymbol\vartheta_t).
\label{eq:generation_probability}
\end{equation}
Here \(p_e^{\mathrm{sys}}\in[0,1]\) aggregates source, coupling, detection, and heralding efficiencies; \(\alpha_{\mathrm{att}}\) is the attenuation coefficient; and \(\xi_e\in[0,1]\) captures short-term excess loss, alignment, or availability. Each success creates one pair instance and converts one reserved cell into one occupied cell at both endpoints, while each failure releases its reservation:
\begin{equation}
\label{eq:generation_reservation_outcome}
\begin{aligned}
\Delta\mu_{u,t}^{\mathrm{conv}}
=\Delta\mu_{v,t}^{\mathrm{conv}}
&= G_e(t), \\
\Delta\mu_{u,t}^{\mathrm{rel}}
=\Delta\mu_{v,t}^{\mathrm{rel}}
&= g_e(t)-G_e(t).
\end{aligned}
\end{equation}
The binomial count in \cref{eq:heralded_generation} is the classical success-count marginal of the generation instrument introduced in \cref{subsec:atomic_quantum_operations}; it is not an independent random draw outside the quantum operation model.

\subsection{Passive Dynamics, Decoherence, and Markovian Interpretation}
\label{subsec:quantum_dynamics}

During one micro-epoch, occupied memories undergo passive open-system evolution before the selected operation outcome is registered. For \(\tau\in[0,\Delta t)\),
\begin{equation}
\label{eq:network_master_equation}
\begin{aligned}
\dot\rho(\tau)
={}&
-i\bigl[H_{\mathrm{mem}}(\tau),\rho(\tau)\bigr]
\\
&+
\sum_{v\in\mathcal V}
\mathcal L_{v,t}^{\mathrm{mem},\neg\phi}\bigl(\rho(\tau)\bigr)
\\
&+
\sum_{p\in\mathcal P_t}
\mathcal L_{p,t}^{\mathrm{phase}}\bigl(\rho(\tau)\bigr).
\end{aligned}
\end{equation}
The local non-phase generator is
\begin{equation}
\label{eq:memory_lindbladian}
\begin{aligned}
\mathcal L_{v,t}^{\mathrm{mem},\neg\phi}(\rho)
={}&
\sum_{j=1}^{m_v^{\max}}
\sum_{\kappa\in\mathcal N_v^{\neg\phi}}
\Bigl(
L_{v,j,t}^{(\kappa)}\rho L_{v,j,t}^{(\kappa)\dagger}
\\
&\qquad\qquad
-\frac12
\bigl\{
L_{v,j,t}^{(\kappa)\dagger}
L_{v,j,t}^{(\kappa)},\rho
\bigr\}
\Bigr).
\end{aligned}
\end{equation}
where \(\mathcal N_v^{\neg\phi}\) may include amplitude damping, leakage, and other calibrated mechanisms.

Independent phase dephasing at node \(v\) is parameterized by
\begin{equation}
\Gamma_{v,t}^{\phi}=\frac{1}{T_{2,v}(t)}.
\label{eq:memory_dephasing_rate}
\end{equation}
The symbol \(\Gamma\) is reserved for physical decoherence and dephasing rates; the symbol \(\gamma\) is used only for the discount factor. For pair \(p\) with endpoints \(\{u_p,v_p\}\), independent and correlated phase noise is represented by the GKSL generator
\begin{equation}
\label{eq:correlated_dephasing}
\begin{aligned}
\mathcal L_{p,t}^{\mathrm{phase}}(\rho)
={}&
\sum_{i,j\in\{u_p,v_p\}}
K_{ij,t}^{p,\phi}
\Bigl[
Z_p^{(i)}\rho Z_p^{(j)}
-\frac12\{Z_p^{(j)}Z_p^{(i)},\rho\}
\Bigr],
\\
&\mathbf K_t^{p,\phi}\succeq 0 .
\end{aligned}
\end{equation}
with
\begin{equation}
\label{eq:correlated_phase_matrix}
\begin{aligned}
\mathbf K_t^{p,\phi}
&=
\frac12
\begin{pmatrix}
\Gamma_{u_p,t}^{\phi} & \eta_{p,t}^{\phi} \\
\eta_{p,t}^{\phi} & \Gamma_{v_p,t}^{\phi}
\end{pmatrix},
\\
\eta_{p,t}^{\phi}
&=
\kappa_{p,t}^{\mathrm{corr}}
\sqrt{\Gamma_{u_p,t}^{\phi}\Gamma_{v_p,t}^{\phi}},
\qquad
\mathbf K_t^{p,\phi}\succeq0 .
\end{aligned}
\end{equation}
where \(|\kappa_{p,t}^{\mathrm{corr}}|\le1\). The \(\ket{00}\!\bra{11}\) Bell coherence then decays at rate
\begin{equation}
\Gamma_{p,t}^{\Phi}
=
\Gamma_{u_p,t}^{\phi}+\Gamma_{v_p,t}^{\phi}
+
2\kappa_{p,t}^{\mathrm{corr}}
\sqrt{\Gamma_{u_p,t}^{\phi}\Gamma_{v_p,t}^{\phi}}.
\label{eq:pair_dephasing_rate}
\end{equation}
The generator in \cref{eq:correlated_dephasing} replaces, rather than duplicates, independent phase terms for the two supporting cells. Bell-state fidelity decay is not modeled by mere \(Z\otimes Z\) conjugation, because Bell projectors are eigenprojectors of that operation.

The latent state \(Z_t\in\mathcal Z\) captures stochastic short-term effects such as alignment, excess loss, detector state, temperature, and correlated operating conditions. The vector \(\boldsymbol\vartheta_t\in\Theta\) indexes slowly varying physical-model parameters. For fixed \(\boldsymbol\vartheta\),
\begin{equation}
Z_{t+1}\sim P_{Z,\boldsymbol\vartheta}(\cdot\mid Z_t),
\label{eq:environment_transition}
\end{equation}
and the controlled process \((X_t,C_t)\) is Markovian after \(X_t\) is augmented with \(Z_t\) and any latent finite-memory environmental variables. The observable inventory, queues, reports, and memory counters remain in \(C_t\); they are not silently absorbed into \(X_t\). Thus, decoherence does not by itself imply non-Markovian dynamics. Non-Markovian effects are present only if the chosen finite latent state is insufficient to make \((X_t,C_t)\) Markov.

Conditional on \((z,c,\boldsymbol\vartheta)\), passive evolution over one epoch is the CPTP map
\begin{equation}
\mathcal E_{z,c,\boldsymbol\vartheta}^{\Delta t}
=
\exp\!\left(\Delta t\,\mathcal L_{z,c,\boldsymbol\vartheta}\right).
\label{eq:conditional_continuous_channel}
\end{equation}
The controller uses calibration estimates \(\widehat{\boldsymbol\vartheta}_t\), which may differ from the true parameter sequence.

\end{revblock}

\subsection{Atomic Quantum Operations}
\label{subsec:atomic_quantum_operations}

The elementary actions are generation, purification, swapping, delivery, release, and idle. For action \(a\), let \(\mathcal R(a)\) be its action-support register set and \(\overline{\mathcal R(a)}\) the complementary registers. Pair creation and consumption are represented on fixed input and output support spaces by reset/vacuum cell conventions. Thus, all selective branches for a given action are maps between common finite-dimensional spaces, which makes the flagged-instrument and nonselective-channel distances below well-defined.

The evolved action-support state is defined directly from the global passive channel:
\begin{equation}
\rho_{a}^{\mathrm{evo}}(x,c)
=
\Tr_{\overline{\mathcal R(a)}}
\left[
\mathcal E_{z(x),c,\boldsymbol\vartheta}^{\Delta t}(\rho(x))
\right],
\label{eq:evolved_action_support_state}
\end{equation}
where \(\rho(x)\) is the global hidden density operator contained in \(x\). For each outcome \(y\in\mathcal Y_a\), let \(\mathcal K_{a,y}^{z,\boldsymbol\vartheta}\) be a completely positive trace-nonincreasing map on the action-support space, with
\begin{equation}
\sum_{y\in\mathcal Y_a}\mathcal K_{a,y}^{z,\boldsymbol\vartheta}
\quad\text{CPTP}.
\label{eq:instrument_completeness}
\end{equation}
The operation branch probability is
\begin{equation}
p_{\boldsymbol\vartheta}^{\mathrm{op}}(y\mid x,c,a)
=
\Tr\!\left[
\mathcal K_{a,y}^{z(x),\boldsymbol\vartheta}
\left(\rho_{a}^{\mathrm{evo}}(x,c)\right)
\right].
\label{eq:quantum_operation_probability}
\end{equation}
Passive evolution and the selected instrument are each applied exactly once. For generation, the marginal distribution of the success-count component of \(y\) is the binomial law in \cref{eq:heralded_generation}; purification, swapping, and delivery outcomes may depend on the hidden input pair states through \(\rho_a^{\mathrm{evo}}(x,c)\).

For outcome-law and model-error analysis, define an action-support passive channel \(\mathcal E_{z,c,\boldsymbol\vartheta}^{\Delta t,a}\) on the fixed support Hilbert space \(\mathcal H_a\). This local channel is the calibrated support-level reduction used for channel comparison; branch probabilities for the actual global state remain defined by \cref{eq:evolved_action_support_state,eq:quantum_operation_probability}. The reduced flagged instrument is
\begin{equation}
\mathfrak I_{a,z,c,\boldsymbol\vartheta}^{\mathrm{red}}(\sigma_a)
=
\sum_{y\in\mathcal Y_a}
\ket{y}\!\bra{y}
\otimes
\mathcal K_{a,y}^{z,\boldsymbol\vartheta}
\left(
\mathcal E_{z,c,\boldsymbol\vartheta}^{\Delta t,a}(\sigma_a)
\right).
\label{eq:flagged_action_instrument}
\end{equation}
The corresponding nonselective local channel is
\begin{equation}
\mathcal E_{a,z,c,\boldsymbol\vartheta}^{\mathrm{ns}}(\sigma_a)
=
\sum_{y\in\mathcal Y_a}
\mathcal K_{a,y}^{z,\boldsymbol\vartheta}
\left(
\mathcal E_{z,c,\boldsymbol\vartheta}^{\Delta t,a}(\sigma_a)
\right).
\label{eq:realized_local_action_channel}
\end{equation}
Thus, the nonselective channel includes passive evolution exactly once. The flagged instrument retains the classical outcome law and is used for observation-law, filtering, and transition-kernel calibration; the nonselective channel is used for the robust reward penalty. Nonselective channel error alone is not assumed to control classical outcome probabilities.

For the global post-branch physical update, let \(\widetilde{\mathcal K}_{a,y}^{z,\boldsymbol\vartheta}\) denote the selected branch extended to the global memory, including register reset, output-pair placement, and tracing of optical or measurement ancillas. In this paragraph, \(y\in\mathcal Y_{a_t}\) denotes the raw branch label; the q-POMDP observation stores the tagged outcome \((a_t,y)\). For positive-probability \(y\),
\begin{equation}
\rho_{t+1}
=
\frac{
\widetilde{\mathcal K}_{a_t,y}^{z(x_t),\boldsymbol\vartheta_t}
\left(
\mathcal E_{z(x_t),c_t,\boldsymbol\vartheta_t}^{\Delta t}(\rho(x_t))
\right)
}{
p_{\boldsymbol\vartheta_t}^{\mathrm{op}}(y\mid x_t,c_t,a_t)
}.
\label{eq:global_post_branch_state}
\end{equation}
Pair creation and consumption are reflected simultaneously in the inventory update below.

Let \(\operatorname{Cons}_t(a,y)\subseteq\mathcal P_t\) be consumed input pairs and \(\operatorname{New}_t(a,y)\) be newly created pair instances. The branch-wise pair-inventory update is
\begin{equation}
\mathcal P_{t+1}
=
\bigl(\mathcal P_t\setminus\operatorname{Cons}_t(a_t,y_{t+1})\bigr)
\uplus
\operatorname{New}_t(a_t,y_{t+1}),
\label{eq:action_inventory_update}
\end{equation}
where \(\uplus\) is disjoint multiset union. The notation avoids overloading \(\mathcal C\), which is reserved for the classical-state space. The memory ledger in \cref{eq:memory_occupancy_update} is required to agree with \cref{eq:action_inventory_update} and with the invariant in \cref{eq:memory_pair_inventory_invariant}.

The default abstraction is that a purification or swapping output pair reuses endpoint cells freed by consumed inputs. Hence \(r_v(a,c)=0\) for purification and swapping unless an implementation requires additional temporary ancilla or output reservations. If such cells are required, they are included in \(r_v(a,c)\); a final output placed in a reserved cell is counted through \(\Delta\mu_{v,t}^{\mathrm{conv}}\), with the corresponding \(N_{v,t}^{+}\) adjusted to avoid double-counting. In all cases, the invariant in \cref{eq:memory_pair_inventory_invariant} is the authoritative boundary condition.

For generation on \(e=(u,v)\), the outcome is \(y=(g_{\mathrm s},\eta_{\mathrm{her}})\), where \(g_{\mathrm s}=G_e(t)\in\{0,\ldots,g_e(t)\}\) and \(\eta_{\mathrm{her}}\) denotes other heralding reports. Then \(\operatorname{Cons}_t=\varnothing\), \(\operatorname{New}_t\) consists of \(g_{\mathrm s}\) new pairs with endpoints \(\{u,v\}\), and
\begin{equation}
\label{eq:generation_ledger}
\begin{aligned}
\Delta\mu_{x,t}^{\mathrm{conv}}
&= g_{\mathrm s},
\\
\Delta\mu_{x,t}^{\mathrm{rel}}
&= g_e(t)-g_{\mathrm s},
\qquad x\in\{u,v\},
\\
N_{w,t}^{+}
&= N_{w,t}^{-}=0 .
\end{aligned}
\end{equation}

For purification of distinct \(p_1,p_2\) with \(\partial p_1=\partial p_2=\{u,v\}\), one action is one DEJMPS recurrence round. Outcomes are \(y\in\{\mathrm s,\mathrm f\}\). Both input pairs are consumed in all branches. If \(y=\mathrm s\), one output pair \(p^+\) with endpoints \(\{u,v\}\) is created; if \(y=\mathrm f\), no output pair is created:
\begin{equation}
\begin{aligned}
\operatorname{Cons}_t&=\{p_1,p_2\},&
\operatorname{New}_t&=
\begin{cases}
\{p^+\},&y=\mathrm s,\\
\varnothing,&y=\mathrm f,
\end{cases}
\\
N_{u,t}^{-}&=N_{v,t}^{-}=2,&
N_{u,t}^{+}&=N_{v,t}^{+}=\mathbf 1_{\{y=\mathrm s\}},
\end{aligned}
\label{eq:purification_ledger}
\end{equation}
under the default reuse convention. Thus, success changes occupancy by \(-1\) at each endpoint and failure by \(-2\). Additional purification rounds require additional micro-epochs and fresh input pairs.

For swapping of \(p_{\mathrm L}\) with endpoints \(\{u,v\}\) and \(p_{\mathrm R}\) with endpoints \(\{v,w\}\), \(v\) is the intermediate node. Both inputs are consumed in all branches. If \(y=\mathrm s\), a new pair \(p^{uw}\) with endpoints \(\{u,w\}\) is created; if \(y=\mathrm f\), no output pair is created:
\begin{equation}
\label{eq:swapping_ledger}
\begin{aligned}
\operatorname{Cons}_t
&= \{p_{\mathrm L},p_{\mathrm R}\},
\\
\operatorname{New}_t(y)
&=
\begin{cases}
\{p^{uw}\}, & y=\mathrm s,\\
\varnothing, & y=\mathrm f,
\end{cases}
\\
\bigl(N_{u,t}^{-},N_{v,t}^{-},N_{w,t}^{-}\bigr)
&= (1,2,1),
\\
\bigl(N_{u,t}^{+},N_{v,t}^{+},N_{w,t}^{+}\bigr)
&=
\bigl(
\mathbf 1_{\{y=\mathrm s\}},
0,
\mathbf 1_{\{y=\mathrm s\}}
\bigr).
\end{aligned}
\end{equation}
Under the default reuse convention, success leaves outer-node occupancies unchanged and frees two cells at the intermediate node; failure removes all four occupied input cells.

For delivery of pair \(p\) with \(\partial p=\{u,v\}\) and release of pair \(p\), the pair is consumed in every branch. Delivery can serve a request only when its demand class has endpoints matching \(\partial p\); release never serves a request:
\begin{equation}
\label{eq:delivery_release_ledger}
\begin{aligned}
\operatorname{Cons}_t
&= \{p\},
&
\operatorname{New}_t
&= \varnothing,
\\
N_{u,t}^{-}
&= N_{v,t}^{-}=1,
&
N_{u,t}^{+}
&= N_{v,t}^{+}=0.
\end{aligned}
\end{equation}
Idle has \(\operatorname{Cons}_t=\operatorname{New}_t=\varnothing\) and zero reservation, creation, and removal terms.

For a stored pair \(p\) and target Bell projector \(\Pi_\Phi\), define its belief-induced decision-boundary fidelity as
\begin{equation}
F_b^p(t)=\Tr[\Pi_\Phi\rho_{b,t}^{p}],
\label{eq:belief_pair_fidelity}
\end{equation}
and its hidden-state fidelity as \(F_p(x)=\Tr[\Pi_\Phi\rho_p(x)]\). For a delivery action \(a\) selecting pair \(p\), the default delivery operation is a classical service/handoff that does not add additional quantum degradation beyond passive evolution. The completion-time state used for service is
\begin{equation}
\label{eq:completion_time_fidelity}
\begin{aligned}
\rho_{p}^{\mathrm{comp}}(x,c,a)
&=
\Tr_{\mathcal R(a)\setminus p}
\!\left[\rho_a^{\mathrm{evo}}(x,c)\right],
\\
F_p^{\mathrm{comp}}(x,c,a)
&=
\Tr\!\left[
\Pi_\Phi\rho_{p}^{\mathrm{comp}}(x,c,a)
\right].
\end{aligned}
\end{equation}
If a hardware implementation includes additional delivery-channel degradation, the delivery instrument is extended so that \(F_p^{\mathrm{comp}}\) is defined after the successful delivery branch. In all cases, delivery feasibility and reward use completion-time fidelity, not stale pre-action fidelity.

\subsection{Bounded State and Observation Perturbations}
\label{subsec:adversarial_perturbations}

For pair \(p\), bounded perturbations of the posterior pair state and nonsafety-critical metadata are modeled as
\begin{align}
\widetilde\rho_{b,t}^{p}
&=
(1-\epsilon_{p,t}^{q})\rho_{b,t}^{p}
+
\epsilon_{p,t}^{q}\Phi_{p,t}^{\mathrm{adv}}(\rho_{b,t}^{p}),
\label{eq:quantum_perturbation}
\\
\widetilde{\mathbf e}_{p,t}
&=
\mathbf e_{p,t}+\boldsymbol\delta_{p,t}^{c},
\qquad
\|\boldsymbol\delta_{p,t}^{c}\|_2\le\epsilon_{p,t}^{c},
\label{eq:classical_perturbation}
\end{align}
where \(\Phi_{p,t}^{\mathrm{adv}}\) is CPTP and \(0\le\epsilon_{p,t}^{q}\le1\). For any Bell projector,
\begin{equation}
\left|
\Tr[\Pi_\Phi(\widetilde\rho_{b,t}^{p}-\rho_{b,t}^{p})]
\right|
\le
\frac12\|\widetilde\rho_{b,t}^{p}-\rho_{b,t}^{p}\|_1
\le
\epsilon_{p,t}^{q}.
\label{eq:fidelity_perturbation_bound}
\end{equation}
Perturbed observations may affect inference and policy scores but cannot create, reserve, occupy, or remove memory cells. The execution layer revalidates every selected action against authenticated inventory and hard memory constraints.

\subsection{Belief and Feature Representation}
\label{subsec:belief_representation}

Let \(X_t\in\mathcal X\) contain the hidden global quantum state, \(Z_t\), and latent variables required for Markovian evolution under fixed \(\boldsymbol\vartheta\). Let \(C_t\in\mathcal C\) be the observable classical state. With history
\begin{equation}
h_t=(c_0,a_0,y_1,c_1,\ldots,a_{t-1},y_t,c_t),
\label{eq:action_observation_history}
\end{equation}
the online belief is the classical probability measure
\begin{equation}
b_t(dx)
=
\Pr\!\left(X_t\in dx\mid h_t,\widehat{\boldsymbol\vartheta}_{0:t-1}\right).
\label{eq:classical_belief_state}
\end{equation}
For pair instance \(p\), let \(\rho_p(x)\) be its hidden reduced density operator under hidden state \(x\). The posterior pair state in \cref{eq:quantum_network_graph} is induced by the belief:
\begin{equation}
\rho_{b,t}^{p}
=
\int_{\mathcal X}\rho_p(x)\,b_t(dx).
\label{eq:belief_induced_state}
\end{equation}
Thus, \(b_t\) is a classical distribution over hidden physical states, whereas \(\rho_{b,t}^{p}\) is a quantum density operator. A particle representation may be written as \(b_t^{(N_b)}=\sum_{i=1}^{N_b}w_t^{(i)}\delta_{x_t^{(i)}}\), with nonnegative normalized weights.

The controller embeds the information state through one global feature map
\begin{equation}
\mathbf f_t=\phi(b_t,c_t)\in\mathbb R^{d_f},
\qquad
\phi:\mathcal B\times\mathcal C\rightarrow\mathbb R^{d_f}.
\label{eq:belief_feature_map}
\end{equation}
Observable inputs may include normalized memory occupancy, queues, request classes, physical-link attributes, pair ages, stored posterior summaries, operational-error summaries, purification depths, recent outcomes, and calibration estimates. Reservations and pending-operation features are absent because no such state persists across decision boundaries.

\subsection{GNN State Encoder}
\label{subsec:system_gnn_policy}

Let \(\mathbf x_{v,t}\) collect identifier-free node features and let \(\mathbf x_{p,t}=(\rho_{b,t}^{p},\mathbf e_{p,t})\) collect pair features. The density operator is supplied through a fixed real representation, such as independent matrix components or selected pair-local observables. Distinct encoders initialize
\begin{equation}
\mathbf h_v^{(0)}=\phi_v(\mathbf x_{v,t}),
\qquad
\mathbf g_{p,t}=\phi_e(\mathbf x_{p,t}),
\label{eq:initial_graph_embeddings}
\end{equation}
where \(\phi_v\) and \(\phi_e\) are not the global information-state map \(\phi\). For \(p\in\mathcal P_t(v)\), let \(u(p,v)\) be its other endpoint. The canonical message-passing rule is
\begin{align}
\mathbf m_{p\rightarrow v}^{(k)}
&=
\operatorname{MLP}_{\theta}^{(k)}
\left(
\mathbf h_{u(p,v)}^{(k)}
\oplus
\mathbf h_v^{(k)}
\oplus
\mathbf g_{p,t}
\right),
\label{eq:gnn_message}
\\
\mathbf h_v^{(k+1)}
&=
\operatorname{GRU}_{\theta}^{(k)}
\left(
\mathbf h_v^{(k)},
\sum_{p\in\mathcal P_t(v)}\mathbf m_{p\rightarrow v}^{(k)}
\right),
\label{eq:gnn_update}
\end{align}
for \(k=0,\ldots,L-1\). The sum is invariant to transient pair identifiers while preserving separate messages from multiple pairs having the same endpoints. The final embeddings are combined through operation-specific, role-aware action representations in \cref{subsec:graph_action_representation}.

\section{Problem Formulation}
\label{sec:problem}

The controller observes the information state
\begin{equation}
\varsigma_t=(b_t,c_t)\in\mathcal S,
\qquad
\mathcal S\coloneqq\mathcal B\times\mathcal C,
\label{eq:problem_information_state}
\end{equation}
selects one feasible atomic action, observes its completed outcome, and updates the physical state, classical state, reward, and posterior belief before epoch \(t+1\).

\subsection{Demand Model and Atomic Actions}
\label{subsec:routing_actions}

Let
\begin{equation}
\mathbf D=[\lambda_{sd}]\in\mathbb R_+^{n\times n}
\label{eq:demand_matrix}
\end{equation}
be the entanglement-demand matrix, where \(\lambda_{sd}\) is the mean request-arrival rate from \(s\) to \(d\). The active demand classes are
\begin{equation}
\mathcal J
=
\{(s,d)\in\mathcal V^2:s\neq d,\ \lambda_{sd}>0\}.
\label{eq:active_demand_classes}
\end{equation}
For \(k=(s_k,d_k)\in\mathcal J\), let \(q_k(t)\) be its request backlog and \(F_k^{\min}\in(0,1]\) its minimum delivery fidelity. Each class has finite admission buffer \(q_k^{\max}>0\):
\begin{equation}
0\le q_k(t)\le q_k^{\max}.
\label{eq:finite_request_buffer}
\end{equation}
Requests arriving to a full buffer are blocked or deferred outside the modeled system.

The elementary operation alphabet is
\begin{equation}
\mathcal O=\{\mathsf G,\mathsf P,\mathsf S,\mathsf D,\mathsf R,\mathsf I\},
\label{eq:operation_alphabet}
\end{equation}
representing generation, purification, swapping, delivery, release, and idle. An action is \(a=(o,\iota)\), where \(o\in\mathcal O\) and \(\iota\) identifies its physical resources. The nominal action space is the disjoint union
\begin{equation}
\mathcal A=\biguplus_{o\in\mathcal O}\mathcal A^{(o)}.
\label{eq:routing_action_space}
\end{equation}
A generation action \(a=(\mathsf G,(e,g))\) selects one physical link and one optical-attempt count. A purification action \(a=(\mathsf P,(p_1,p_2))\) selects two distinct stored pairs with identical endpoints. A swapping action \(a=(\mathsf S,(p_{\mathrm L},p_{\mathrm R},v))\) selects two pairs sharing intermediate node \(v\), while preserving left and right pair roles. A delivery action \(a=(\mathsf D,(p,k))\) assigns one stored pair to one demand class, and a release action \(a=(\mathsf R,p)\) deletes one stored pair. The idle action performs no deliberate routing operation but permits passive evolution.

Because exactly one action completes per micro-epoch, realized delivery indicators satisfy
\begin{equation}
n_k^{\mathrm{del}}(x,c,a,y)\in\{0,1\},
\qquad
\sum_{k\in\mathcal J}n_k^{\mathrm{del}}(x,c,a,y)\le1.
\label{eq:atomic_delivery_count}
\end{equation}

\subsection{State-Dependent Feasibility}
\label{subsec:routing_feasibility}

The controller acts on the nonempty feasible set
\begin{equation}
\mathcal A_{\mathrm f}(\varsigma)
=
\mathcal A_{\mathrm f}(b,c)
\subseteq\mathcal A,
\label{eq:feasible_action_set}
\end{equation}
which always contains \(\mathsf I\). For an observable classical state \(c\), let \(\mathcal P(c)\), \(\mu_v(c)\), and \(q_k(c)\) denote the authenticated pair inventory, occupied memory count at node \(v\), and demand backlog for class \(k\) encoded in \(c\). Memory feasibility requires
\begin{equation}
\mu_v(c)+r_v(a,c)\le m_v^{\max},
\qquad v\in\mathcal V.
\label{eq:action_memory_feasibility}
\end{equation}
For \(a=(\mathsf G,(e,g))\), \(e=(u,v)\),
\begin{equation}
r_w(a,c)
=
\begin{cases}
g, & w\in\{u,v\},\\
0, & \text{otherwise}.
\end{cases}
\label{eq:generation_action_reservation}
\end{equation}
Purification requires distinct \(p_1,p_2\in\mathcal P(c)\) with \(\partial p_1=\partial p_2\). Swapping requires distinct pairs satisfying \(\partial p_{\mathrm L}=\{u,v\}\), \(\partial p_{\mathrm R}=\{v,w\}\), and \(u\neq w\). Delivery \(a=(\mathsf D,(p,k))\), \(k=(s_k,d_k)\), requires
\begin{equation}
p\in\mathcal P(c),
\qquad
\partial p=\{s_k,d_k\},
\qquad
q_k(c)>0.
\label{eq:delivery_endpoint_queue_feasibility}
\end{equation}
Release requires the selected pair to be present in the authenticated inventory. Existing occupied cells containing purification or swapping inputs are not counted as new reservations; hardware-specific ancillas or temporary output cells are included in \(r_v(a,c)\).

For hidden state \(x\), let \(F_p^{\mathrm{comp}}(x,c,a)\) denote completion-time fidelity as in \cref{eq:completion_time_fidelity}, with the environmental component extracted from \(x\). Delivery of pair \(p\) to demand class \(k\) is feasible only if
\begin{equation}
\Pr_{X\sim b}
\left(
F_p^{\mathrm{comp}}(X,c,a)\ge F_k^{\min}
\right)
\ge1-\epsilon_k,
\quad
0<\epsilon_k<1.
\label{eq:delivery_chance_constraint}
\end{equation}
Thus, feasibility uses the posterior probability of meeting the fidelity target at action completion, not the pre-action posterior mean. Actions violating physical-link requirements, inventory conditions, memory or classical-buffer capacities, hardware roles, endpoint compatibility, nonempty-queue delivery, or delivery-risk thresholds are excluded from \(\mathcal A_{\mathrm f}(\varsigma)\).

\begin{revblock}

\subsection{Completed-Action Reward, Queue Service, and Objective}
\label{subsec:routing_objective}

For a delivery action \(a=(\mathsf D,(p,k))\), define the service indicator
\begin{equation}
\label{eq:delivery_service_indicator}
\begin{aligned}
n_k^{\mathrm{del}}(x,c,a,y)
={}&
\mathbf 1_{\{a=(\mathsf D,(p,k))\}}
\mathbf 1_{\{q_k(c)>0\}}
\mathbf 1_{\{\partial p=\{s_k,d_k\}\}}
\\
&\times
\mathbf 1_{\{y\in\mathcal Y_a^{\mathrm{succ}}\}}
\mathbf 1_{\{F_p^{\mathrm{comp}}(x,c,a)\ge F_k^{\min}\}} .
\end{aligned}
\end{equation}
and set \(n_{\ell}^{\mathrm{del}}(x,c,a,y)=0\) for all \(\ell\neq k\). Here \(\mathcal Y_a^{\mathrm{succ}}\subseteq\mathcal Y_a\) is the successful handoff branch set of the delivery instrument. This definition is a defensive service check: feasibility already requires endpoint and nonempty-queue compatibility, but the realized service indicator still verifies them.

For outcome \(y\), let \(\mathcal P^{\mathrm{handoff}}(a,y)\) be the set of pairs physically handed off by action \(a\). By atomicity it has cardinality at most one. A below-threshold handoff may appear in \(\mathcal P^{\mathrm{handoff}}(a,y)\) and then contributes a negative fidelity-margin term, but it does not count as successful service through \(n_k^{\mathrm{del}}\). For handed-off pair \(p\), let \(\kappa(p)\in\mathcal J\) denote its assigned demand class. The realized base reward is
\begin{equation}
\label{eq:routing_reward}
\begin{aligned}
R(x,c,a,y,c')
={}&
\sum_{k\in\mathcal J}
w_k n_k^{\mathrm{del}}(x,c,a,y)
-
\lambda_q
\sum_{k\in\mathcal J}
\frac{q_k(c)}{q_k^{\max}}
\\
&+
\lambda_F
\sum_{\mathclap{p\in\mathcal P^{\mathrm{handoff}}(a,y)}}
\Bigl(
F_p^{\mathrm{comp}}(x,c,a)-F_{\kappa(p)}^{\min}
\Bigr)
\\
&-
\lambda_o C_{\mathrm{op}}(x,c,a,y,c') .
\end{aligned}
\end{equation}
where \(w_k>0\) and \(\lambda_q,\lambda_F,\lambda_o\ge0\). The operation cost includes only resources used by the selected atomic action: optical attempts, consumed pairs, gates, measurements, reservations, holding time, and classical signaling. We assume
\begin{equation}
0\le C_{\mathrm{op}}(x,c,a,y,c')\le C_{\mathrm{op}}^{\max}.
\label{eq:bounded_operation_cost}
\end{equation}

Let \(A_{k,t}\in\{0,\ldots,A_k^{\max}\}\) be the exogenous arrivals admitted to demand class \(k\) during epoch \(t\), before buffer capping. The queue component of the branch update satisfies
\begin{equation}
q_k(c')
=
\min\left\{
q_k^{\max},
q_k(c)-n_k^{\mathrm{del}}(x,c,a,y)+A_{k,t}
\right\},
\label{eq:queue_update}
\end{equation}
where \(n_k^{\mathrm{del}}=1\) is possible only when \(q_k(c)>0\). Arrivals beyond \(q_k^{\max}\) are blocked or deferred outside the modeled queue, preserving the finite-buffer condition.

For hidden state \(x\), let \(z(x)\) denote its environmental component and abbreviate the realized nonselective channel by
\begin{equation}
\mathcal E_{a,x,c,\boldsymbol\vartheta}^{\mathrm{ns}}
\coloneqq
\mathcal E_{a,z(x),c,\boldsymbol\vartheta}^{\mathrm{ns}}.
\label{eq:abbreviated_nonselective_channel}
\end{equation}
Let \(\mathcal E_a^{0,\mathrm{ns}}\) be the calibrated nominal nonselective channel on the same support. For \(a\neq\mathsf I\), define
\begin{equation}
d_{\diamond}
\left(
\mathcal E_{a,x,c,\boldsymbol\vartheta}^{\mathrm{ns}},
\mathcal E_a^{0,\mathrm{ns}}
\right)
=
\frac12
\left\|
\mathcal E_{a,x,c,\boldsymbol\vartheta}^{\mathrm{ns}}
-
\mathcal E_a^{0,\mathrm{ns}}
\right\|_{\diamond},
\label{eq:normalized_channel_deviation}
\end{equation}
which lies in \([0,1]\). For idle,
\begin{equation}
d_{\diamond}
\left(
\mathcal E_{\mathsf I,x,c,\boldsymbol\vartheta}^{\mathrm{ns}},
\mathcal E_{\mathsf I}^{0,\mathrm{ns}}
\right)=0.
\label{eq:problem_idle_channel_convention}
\end{equation}
Passive storage evolution remains part of the physical transition but is not penalized as a deliberate operation-channel deviation.

The realized robust reward is
\begin{align}
R_{\mathrm{rob},\boldsymbol\vartheta}(x,c,a,y,c')
={}&
R(x,c,a,y,c')
\nonumber\\
&-
\lambda_{\mathrm{rob}}
 d_{\diamond}
\left(
\mathcal E_{a,x,c,\boldsymbol\vartheta}^{\mathrm{ns}},
\mathcal E_a^{0,\mathrm{ns}}
\right),
\label{eq:robust_reward}
\end{align}
where \(\lambda_{\mathrm{rob}}\ge0\). Let \(w_{\max}=\max_{k\in\mathcal J}w_k\). Since at most one pair is delivered per epoch, queue penalties are normalized, fidelities and thresholds lie in \([0,1]\), operation cost is bounded, and \(d_\diamond\le1\),
\begin{equation}
\left|R_{\mathrm{rob},\boldsymbol\vartheta}\right|
\le
R_{\max}
\coloneqq
w_{\max}+\lambda_q|\mathcal J|+\lambda_F+
\lambda_o C_{\mathrm{op}}^{\max}+\lambda_{\mathrm{rob}}.
\label{eq:bounded_robust_reward}
\end{equation}

For fixed \(\boldsymbol\vartheta\), stationary feasible policy \(\pi\), and initial information-state distribution \(\nu_0\), the discounted objective is
\begin{equation}
\label{eq:discounted_routing_objective}
\begin{aligned}
J_{\boldsymbol\vartheta}(\pi;\nu_0)
=
\EX_{\nu_0,\pi,\boldsymbol\vartheta}
\!\left[
\sum_{t=0}^{\infty}\gamma^t
R_{\mathrm{rob},\boldsymbol\vartheta}
\!\left(
X_t,C_t,A_t,Y_{t+1},C_{t+1}
\right)
\right],
\end{aligned}
\end{equation}
where \(0<\gamma<1\). The corresponding pointwise value is denoted \(V_{\boldsymbol\vartheta}^{\pi}(\varsigma)\). The frozen-model optimization problem is
\begin{equation}
\max_{\pi}\ J_{\boldsymbol\vartheta}(\pi;\nu_0)
\quad
\text{s.t.}\quad
\pi(\mathcal A_{\mathrm f}(\varsigma)\mid\varsigma)=1
\quad
\forall\varsigma\in\mathcal S_{\mathrm r},
\label{eq:routing_optimization_problem}
\end{equation}
with \(\mathcal S_{\mathrm r}\subseteq\mathcal S\) the reachable information-state domain.

\end{revblock}

\subsection{q-POMDP Construction}
\label{subsec:qpomdp_formulation}

For fixed \(\boldsymbol\vartheta\), define the q-POMDP
\begin{equation}
\mathfrak Q_{\boldsymbol\vartheta}
=
\left(
\mathcal X,\mathcal C,\mathcal Y,\mathcal A,
\mathsf Q_{\boldsymbol\vartheta},
R_{\mathrm{rob},\boldsymbol\vartheta},
\gamma
\right).
\label{eq:qpomdp_definition}
\end{equation}
The global outcome space is the action-tagged union
\begin{equation}
\mathcal Y
=
\biguplus_{a\in\mathcal A}\bigl(\{a\}\times\mathcal Y_a\bigr).
\label{eq:global_action_tagged_outcome_space}
\end{equation}
Inside action-conditioned formulas, \(y\in\mathcal Y_a\) denotes the raw branch label; inside the joint kernel, belief filter, and information-state transition, \(y\) denotes the corresponding action-tagged outcome \((a,y)\in\mathcal Y\).

The joint controlled transition--observation kernel is constructed from two components. First, the action outcome kernel \(\mathsf O_{\boldsymbol\vartheta}\) satisfies
\begin{equation}
\mathsf O_{\boldsymbol\vartheta}(\{(a,y)\}\mid x,c,a)
=
p_{\boldsymbol\vartheta}^{\mathrm{op}}(y\mid x,c,a),
\label{eq:outcome_kernel}
\end{equation}
for finite branch spaces. Generation outcomes have the binomial success-count marginal in \cref{eq:heralded_generation}. Second, the branch update kernel
\begin{equation}
\mathsf U_{\boldsymbol\vartheta}(dx',dc'\mid x,c,a,y)
\label{eq:branch_update_kernel}
\end{equation}
contains the global post-branch quantum update in \cref{eq:global_post_branch_state}, the inventory update in \cref{eq:action_inventory_update}, the memory ledger in \cref{eq:memory_occupancy_update}, the latent transition \(P_{Z,\boldsymbol\vartheta}\), request arrivals, queue recursion in \cref{eq:queue_update}, classical reports, and finite classical-memory updates. The joint kernel is
\begin{equation}
\label{eq:joint_kernel_operational_factorization}
\begin{aligned}
\mathsf Q_{\boldsymbol\vartheta}
(dx',dy,dc'\mid x,c,a)
={}&
\mathsf O_{\boldsymbol\vartheta}
(dy\mid x,c,a)
\\
&\quad
\mathsf U_{\boldsymbol\vartheta}
(dx',dc'\mid x,c,a,y).
\end{aligned}
\end{equation}
Thus, passive evolution and the selected instrument are used once in the physical branch, and all classical and memory consequences of that branch are folded into \(\mathsf U_{\boldsymbol\vartheta}\).

Assume that \(\mathsf Q_{\boldsymbol\vartheta}\) has a regular conditional density representation with respect to a fixed \(\sigma\)-finite observation reference measure \(m_{\mathrm{obs}}(dy,dc')\):
\begin{align}
&\mathsf Q_{\boldsymbol\vartheta}(dx',dy,dc'\mid x,c,a)
\nonumber\\
&\quad=
\mathsf P_{\boldsymbol\vartheta}(dx'\mid x,c,a)
 o_{\boldsymbol\vartheta}(y,c'\mid x,x',c,a)m_{\mathrm{obs}}(dy,dc'),
\label{eq:joint_kernel_factorization}
\end{align}
where \(\mathsf P_{\boldsymbol\vartheta}(dx'\mid x,c,a)=\mathsf Q_{\boldsymbol\vartheta}(dx',\mathcal Y,\mathcal C\mid x,c,a)\). This density is used only to write Bayes' rule; \cref{eq:joint_kernel_operational_factorization} is the operational construction of the kernel.

At epoch \(t\), the online filter uses \(\widehat{\boldsymbol\vartheta}_t\). After action \(a_t\) and observation \((y_{t+1},c_{t+1})\), the posterior is given by \cref{eq:qpomdp_belief_update} for every measurable \(B\subseteq\mathcal X\). The denominator in \cref{eq:qpomdp_belief_update} is assumed positive for every realized observation. Because \(y_{t+1}\) is action-tagged, identical raw labels such as ``success'' for different instruments are not confused.

\begin{figure*}[!b]
\hrule
\vspace{0.6\baselineskip}
\begin{equation}
\label{eq:qpomdp_belief_update}
\begin{aligned}
b_{t+1}(B)
=
\frac{
\displaystyle
\int_{\mathcal X}\int_B
o_{\widehat{\boldsymbol\vartheta}_t}
\!\left(
y_{t+1},c_{t+1}
\mid x,x',c_t,a_t
\right)
\mathsf P_{\widehat{\boldsymbol\vartheta}_t}
\!\left(dx'\mid x,c_t,a_t\right)
b_t(dx)
}{
\displaystyle
\int_{\mathcal X}\int_{\mathcal X}
o_{\widehat{\boldsymbol\vartheta}_t}
\!\left(
y_{t+1},c_{t+1}
\mid x,\widetilde x,c_t,a_t
\right)
\mathsf P_{\widehat{\boldsymbol\vartheta}_t}
\!\left(d\widetilde x\mid x,c_t,a_t\right)
b_t(dx)
}.
\end{aligned}
\end{equation}
\vspace{-0.4\baselineskip}
\end{figure*}

Define the Bayesian update map \(\Psi_{\boldsymbol\vartheta}(b,c,a,y,c')\) by \cref{eq:qpomdp_belief_update}. The belief-averaged joint predictive law is
\begin{equation}
\overline{\mathsf Q}_{b,\boldsymbol\vartheta}(dx',dy,dc'\mid c,a)
=
\int_{\mathcal X}\mathsf Q_{\boldsymbol\vartheta}(dx',dy,dc'\mid x,c,a)b(dx),
\label{eq:belief_joint_predictive_kernel}
\end{equation}
and the induced information-state kernel is
\begin{equation}
\label{eq:information_state_transition_kernel}
\begin{aligned}
\mathsf T_{\boldsymbol\vartheta}(B\mid\varsigma,a)
={}&
\int
\mathbf 1_B\!\left(
\Psi_{\boldsymbol\vartheta}(b,c,a,y,c'),c'
\right)
\\
&\qquad
\overline{\mathsf Q}_{b,\boldsymbol\vartheta}
(dx',dy,dc'\mid c,a).
\end{aligned}
\end{equation}
for measurable \(B\subseteq\mathcal S\).

The belief-expected one-step reward is
\begin{equation}
\label{eq:belief_expected_reward}
\begin{aligned}
\overline R_{\boldsymbol\vartheta}(\varsigma,a)
={}&
\int_{\mathcal X}\int
R_{\mathrm{rob},\boldsymbol\vartheta}(x,c,a,y,c')
\\
&\qquad
\mathsf Q_{\boldsymbol\vartheta}
(dx',dy,dc'\mid x,c,a)\,b(dx).
\end{aligned}
\end{equation}
This separates the realized completed-action reward from the conditional reward used in Bellman equations. For a frozen model, the optimal value satisfies
\begin{align}
V_{\boldsymbol\vartheta}^*(\varsigma)
=
\max_{a\in\mathcal A_{\mathrm f}(\varsigma)}
\biggl\{
&\overline R_{\boldsymbol\vartheta}(\varsigma,a)
\nonumber\\
&+
\gamma\int_{\mathcal S}V_{\boldsymbol\vartheta}^*(\varsigma')
\mathsf T_{\boldsymbol\vartheta}(d\varsigma'\mid\varsigma,a)
\biggr\}.
\label{eq:quantum_bellman_equation}
\end{align}
All Bellman equations and policy-improvement statements use maximization, consistently with \cref{eq:discounted_routing_objective}.

\subsection{Feature-Based Value Representation}
\label{subsec:feature_policy_value}

To retain pair-local quantum information, let \(\{O_i\}_{i=1}^{m_O}\) be fixed Hermitian observables on a two-qubit pair space and define
\begin{equation}
\mathbf o_p(b)
=
\left[
\Tr(O_i\rho_b^p),
\;
\Tr(O_i^2\rho_b^p)-\Tr^2(O_i\rho_b^p)
\right]_{i=1}^{m_O}.
\label{eq:pair_quantum_observables}
\end{equation}
A permutation-invariant operator pools these features:
\begin{equation}
\mathbf o_Q(b,c)
=
\operatorname{Pool}\left(\{\mathbf o_p(b):p\in\mathcal P(c)\}\right),
\label{eq:pooled_quantum_observables}
\end{equation}
with empty-set output defined as zero. From \cref{eq:bounded_robust_reward},
\begin{equation}
V_{\max}=\frac{R_{\max}}{1-\gamma}.
\label{eq:problem_value_bound}
\end{equation}
The critic is parameterized as
\begin{equation}
V_{\psi}(\varsigma)
=
V_{\max}
\tanh\left(
f_{\psi}^{V}\left(\phi(\varsigma)\oplus\mathbf o_Q(b,c)\right)
\right),
\label{eq:hybrid_value_approximation}
\end{equation}
and therefore satisfies
\begin{equation}
\|V_{\psi}\|_{\infty}\le V_{\max}.
\label{eq:bounded_critic_output}
\end{equation}
This representation preserves the pair reduced states and pair-local observables supplied to the critic; it does not claim to preserve arbitrary multipartite correlations across different pair instances.

\section{Feature Selection and Belief Aggregation Guarantees}
\label{sec:feature_selection_guarantees}

Planning directly on \(\mathcal S=\mathcal B\times\mathcal C\) is generally intractable. We approximate the reachable information-state domain by feasibility-stratified prototypes while preserving physical attributes, hard resource constraints, and concrete action structure. Only transient runtime identifiers may be removed by canonicalization.

\subsection{Joint Information-State Metric}
\label{subsec:joint_information_state}

For beliefs \(b,b'\in\mathcal B\), define
\begin{equation}
d_{\mathcal B}(b,b')
=
\TV(b,b')
=
\sup_{A\in\mathfrak B(\mathcal X)}|b(A)-b'(A)|,
\label{eq:belief_metric}
\end{equation}
where \(\mathfrak B(\mathcal X)\) is the measurable sigma-field of \(\mathcal X\). For a finite hypothesis set, \(d_{\mathcal B}(b,b')=\frac12\|b-b'\|_1\).

The observable-state metric uses only quantities stored in \(c\). Let
\begin{equation}
\overline{\boldsymbol\mu}(c)=\left[\frac{\mu_v(c)}{m_v^{\max}}\right]_{v\in\mathcal V},
\qquad
\overline{\mathbf q}(c)=\left[\frac{q_k(c)}{q_k^{\max}}\right]_{k\in\mathcal J},
\label{eq:normalized_memory_queue_vectors}
\end{equation}
and \(\mathbf u^{\mathrm{req}}(c)=[\mathbf 1_{\{q_k(c)>0\}}]_{k\in\mathcal J}\). For \(\mathbf z\in\mathbb R^d\), write \(\|\mathbf z\|_{1,\mathrm{av}}=\|\mathbf z\|_1/\max\{1,d\}\). For each \(p\in\mathcal P(c)\), let \(\boldsymbol\eta_p(c)\in[0,1]^{d_p}\) be its normalized identifier-free record.

Finite quantum memory gives the global pair-count bound
\begin{equation}
M_{\mathrm P}^{\max}
=
\max\left\{1,\left\lfloor\frac12\sum_{v\in\mathcal V}m_v^{\max}\right\rfloor\right\}.
\label{eq:maximum_pair_count}
\end{equation}
Every pair-record multiset is padded to cardinality \(M_{\mathrm P}^{\max}\) with a dummy record \(\bot\). Define
\begin{equation}
d_0(\mathbf u,\mathbf u')=
\min\{1,\|\mathbf u-\mathbf u'\|_{1,\mathrm{av}}\},
\quad
d_0(\mathbf u,\bot)=1,
\quad
d_0(\bot,\bot)=0.
\label{eq:record_ground_metric}
\end{equation}
For padded multisets \(\widetilde{\mathcal U}\) and \(\widetilde{\mathcal U}'\), define
\begin{equation}
d_{\mathrm{match}}(\mathcal U,\mathcal U')
=
\frac{1}{M_{\mathrm P}^{\max}}
\min_{\pi\in\mathfrak S_{M_{\mathrm P}^{\max}}}
\sum_{i=1}^{M_{\mathrm P}^{\max}}
 d_0(\widetilde{\mathbf u}_i,\widetilde{\mathbf u}'_{\pi(i)}).
\label{eq:multiset_matching_metric}
\end{equation}
The fixed padding cardinality makes the distance permutation invariant and comparable across different inventory sizes.

Let \(\mathcal U_{\mathrm P}(c)=\{\boldsymbol\eta_p(c):p\in\mathcal P(c)\}\). Define
\begin{align}
d_{\mu}(c,c')&=\|\overline{\boldsymbol\mu}(c)-\overline{\boldsymbol\mu}(c')\|_{1,\mathrm{av}},
\label{eq:occupied_memory_metric}
\\
d_q(c,c')&=\|\overline{\mathbf q}(c)-\overline{\mathbf q}(c')\|_{1,\mathrm{av}},
\label{eq:queue_metric}
\\
d_{\mathrm P}(c,c')&=d_{\mathrm{match}}(\mathcal U_{\mathrm P}(c),\mathcal U_{\mathrm P}(c')),
\label{eq:pair_inventory_metric}
\\
d_{\mathrm{req}}(c,c')&=\|\mathbf u^{\mathrm{req}}(c)-\mathbf u^{\mathrm{req}}(c')\|_{1,\mathrm{av}}.
\label{eq:request_metric}
\end{align}
The bounded observable-state metric is
\begin{equation}
\label{eq:observable_state_metric}
\begin{aligned}
d_{\mathcal C}(c,c')
={}&
\beta_{\mu}d_{\mu}(c,c')
+
\beta_qd_q(c,c')
\\
&\quad+
\beta_{\mathrm P}d_{\mathrm P}(c,c')
+
\beta_{\mathrm{req}}d_{\mathrm{req}}(c,c') .
\end{aligned}
\end{equation}
where the nonnegative coefficients sum to one. The joint metric is
\begin{equation}
d_{\mathcal S}(\varsigma,\varsigma')
=
d_{\mathcal B}(b,b')+\lambda_c d_{\mathcal C}(c,c'),
\qquad
\lambda_c>0.
\label{eq:joint_information_metric}
\end{equation}
We assume the global encoder satisfies
\begin{equation}
\|\phi(\varsigma)-\phi(\varsigma')\|_2
\le
L_{\phi}d_{\mathcal S}(\varsigma,\varsigma').
\label{eq:feature_lipschitz}
\end{equation}
If the same pair instance \(p\) is represented under \(b\) and \(b'\), then
\begin{equation}
\|\rho_b^p-\rho_{b'}^p\|_1\le2d_{\mathcal B}(b,b').
\label{eq:belief_to_pair_trace_bound}
\end{equation}

\begin{revblock}

\subsection{Admissible Relabeling and Signatures}
\label{subsec:admissible_relabelings}

Let \(\mathfrak G_{\mathrm{adm}}\) be the admissible relabeling group. Its elements may permute transient pair identifiers and request-instance identifiers within the same demand class. They may not relabel physical nodes, physical links, endpoint roles, hardware classes, memory capacities, demand classes, action roles, or calibration attributes. For \(g\in\mathfrak G_{\mathrm{adm}}\), let \(g\varsigma\) and \(ga\) denote consistently relabeled state and action. We assume
\begin{align}
\mathcal A_{\mathrm f}(g\varsigma)&=g\mathcal A_{\mathrm f}(\varsigma),
\label{eq:relabeling_feasibility_invariance}
\\
\overline R(g\varsigma,ga)&=\overline R(\varsigma,a),
\label{eq:relabeling_reward_invariance}
\\
\mathsf T(gB\mid g\varsigma,ga)&=\mathsf T(B\mid\varsigma,a),
\label{eq:relabeling_transition_invariance}
\\
\phi(g\varsigma)&=\phi(\varsigma).
\label{eq:relabeling_feature_invariance}
\end{align}
With \(\operatorname{Can}:\mathcal S\to\mathcal S/\mathfrak G_{\mathrm{adm}}\) a deterministic canonical-representative map, these invariances imply \(V^*(g\varsigma)=V^*(\varsigma)\).

Each state is assigned an identifier-free feasibility signature
\begin{equation}
\sigma(\varsigma)\in\Sigma,
\label{eq:feasibility_signature}
\end{equation}
where \(\Sigma\) is the finite set of structural feasibility classes for the fixed network. The signature preserves every discrete feature that affects hard feasibility: operation types, eligible generation links and attempt-count classes, purification groups, swapping motifs, free-capacity classes, active demand classes, delivery-eligibility classes, hardware roles, and the multiplicity of each action-template class. Continuous values remain in feature and action descriptors unless they determine a hard threshold, such as a memory, buffer, request-nonemptiness, generation-attempt, or delivery-fidelity chance constraint.

For \(\zeta\in\Sigma\), let \(\mathcal T_{\mathrm a}(\zeta)\) be its finite action-template set and \(m_{\tau}(\zeta)\) the multiplicity of template \(\tau\). The canonical action-label set is
\begin{equation}
\Lambda(\zeta)
=
\biguplus_{\tau\in\mathcal T_{\mathrm a}(\zeta)}
\{(\tau,j):j=1,\ldots,m_{\tau}(\zeta)\}.
\label{eq:canonical_action_label_set}
\end{equation}
Canonicalization provides a bijection
\begin{equation}
\ell_{\varsigma}:\mathcal A_{\mathrm f}(\varsigma)
\leftrightarrow
\Lambda(\sigma(\varsigma)).
\label{eq:canonical_action_label_map}
\end{equation}
Its inverse instantiates the current physical action, including concrete link, pair instances, node roles, demand class, and reserved resources.

\subsection{Role-Aware Action Correspondence}
\label{subsec:canonical_action_matching}

Canonical multiplicity indices alone are insufficient for value transfer because structurally identical actions may involve pairs with different fidelities, ages, or error estimates. For feasible \(a\) in state \(\varsigma\), let
\begin{equation}
\mathbf d_{\varsigma}(a)\in\mathbb R^{d_a}
\label{eq:action_matching_descriptor}
\end{equation}
be a normalized descriptor containing operation type, ordered roles, fixed node and link attributes, pair summaries, ages, purification depths, operation errors, reservation requirements, demand class, fidelity threshold, and expected output role. Transient identifiers are excluded. The descriptor is chosen to include every role-aware continuous variable on which matched-action reward and transition regularity may depend.

For states \(\varsigma,\varsigma'\) with \(\sigma(\varsigma)=\sigma(\varsigma')=\zeta\), let
\begin{equation}
\mathcal A_{\tau}(\varsigma)
=
\{a\in\mathcal A_{\mathrm f}(\varsigma):a\text{ has template }\tau\}.
\label{eq:template_action_subset}
\end{equation}
Equal signatures imply \(|\mathcal A_{\tau}(\varsigma)|=|\mathcal A_{\tau}(\varsigma')|=m_{\tau}(\zeta)\). For actions in the same template class, define
\begin{equation}
c_{\mathrm{act}}(a,a')
=
\|\mathbf d_{\varsigma}(a)-\mathbf d_{\varsigma'}(a')\|_{\mathbf W_{\mathrm{act}}},
\quad
\|\mathbf z\|_{\mathbf W}=\sqrt{\mathbf z^{\mathsf T}\mathbf W\mathbf z},
\label{eq:action_matching_cost}
\end{equation}
where \(\mathbf W_{\mathrm{act}}\succeq0\). The template-level correspondence is the minimum-cost bijection
\begin{equation}
M_{\varsigma\rightarrow\varsigma'}^{\tau}
\in
\argmin_{M}
\sum_{a\in\mathcal A_{\tau}(\varsigma)}
c_{\mathrm{act}}(a,M(a)),
\label{eq:template_action_matching}
\end{equation}
where \(M\) ranges over bijections from \(\mathcal A_{\tau}(\varsigma)\) to \(\mathcal A_{\tau}(\varsigma')\). The complete correspondence is
\begin{equation}
M_{\varsigma\rightarrow\varsigma'}
=
\biguplus_{\tau\in\mathcal T_{\mathrm a}(\zeta)}
M_{\varsigma\rightarrow\varsigma'}^{\tau}.
\label{eq:complete_action_matching}
\end{equation}
Deterministic role-aware tie breaking is used when several minimum-cost matchings exist. Actions that remain indistinguishable are treated as physically symmetric, with invariant reward and transition laws. A cached prototype action \(a_j\) is instantiated at current state \(\varsigma\) as
\begin{equation}
a=M_{\varsigma_j\rightarrow\varsigma}(a_j).
\label{eq:prototype_to_current_action_transfer}
\end{equation}

\end{revblock}

\subsection{Feasibility-Stratified Prototypes}
\label{subsec:feasibility_stratified_prototypes}

Let
\begin{equation}
\mathcal Q_K=\{\varsigma_1,\ldots,\varsigma_K\}\subseteq\mathcal S_{\mathrm r}
\label{eq:joint_prototype_set}
\end{equation}
be reachable prototypes and
\begin{equation}
\Sigma_K=\{\sigma(\varsigma_j):j=1,\ldots,K\}
\label{eq:represented_signature_set}
\end{equation}
be represented signatures. The represented reachable domain is
\begin{equation}
\mathcal S_{\mathrm r}^{K}
=
\{\varsigma\in\mathcal S_{\mathrm r}:\sigma(\varsigma)\in\Sigma_K\}.
\label{eq:represented_reachable_set}
\end{equation}
The projected Bellman analysis assumes represented-domain transition closure:
\begin{equation}
\mathsf T(\mathcal S_{\mathrm r}^{K}\mid\varsigma,a)=1
\label{eq:represented_domain_transition_closure}
\end{equation}
for every \(\varsigma\in\mathcal S_{\mathrm r}^{K}\) and \(a\in\mathcal A_{\mathrm f}(\varsigma)\). If a prototype table is not closed, it must be expanded or supplied with an explicit boundary-state construction before the aggregation theorem is applied.

For \(\varsigma\in\mathcal S_{\mathrm r}^{K}\), define the same-signature nearest prototype
\begin{equation}
j_K(\varsigma)
\in
\argmin_{\substack{1\le j\le K\\ \sigma(\varsigma_j)=\sigma(\varsigma)}}
\|\phi(\varsigma)-\phi(\varsigma_j)\|_2,
\label{eq:feasibility_aware_assignment}
\end{equation}
with deterministic tie breaking. The covering radius is
\begin{equation}
\epsilon_K
=
\sup_{\varsigma\in\mathcal S_{\mathrm r}^{K}}
\|\phi(\varsigma)-\phi(\varsigma_{j_K(\varsigma)})\|_2.
\label{eq:joint_covering_radius}
\end{equation}
If \(\sigma(\varsigma)\notin\Sigma_K\), neither \(j_K(\varsigma)\) nor prototype action matching is defined. Such states are handled by the unseen-signature fallback in \cref{sec:hybrid_pomdp_gnn} and excluded from the aggregation guarantee until their strata are inserted.

\subsection{Stratified Covering and Sampling}
\label{subsec:covering_guarantees}

For \(\zeta\in\Sigma_K\), define
\begin{equation}
\mathcal S_{\zeta}
=
\{\varsigma\in\mathcal S_{\mathrm r}^{K}:\sigma(\varsigma)=\zeta\}.
\label{eq:feasibility_strata}
\end{equation}

\begin{proposition}[Stratified Covering and Sampling Complexity]
\label{prop:covering_complexity}
Suppose each feature image \(\phi(\mathcal S_{\zeta})\) is compact and contained in a Euclidean ball of radius \(R_{\phi}\). Then a signature-preserving \(\epsilon\)-cover exists with at most
\begin{equation}
K_{\epsilon}
\le
|\Sigma_K|\left(1+\frac{2R_{\phi}}{\epsilon}\right)^{d_f}
\label{eq:stratified_covering_bound}
\end{equation}
prototypes.

Let \(\mathbb P_{\mathrm{sam}}\) be a probability distribution on \(\mathcal S_{\mathrm r}^{K}\). For each \(\zeta\in\Sigma_K\), assume
\begin{equation}
\label{eq:sampling_stratum_definitions}
\begin{aligned}
p_{\zeta}
&= \mathbb P_{\mathrm{sam}}(\mathcal S_{\zeta}) > 0,
\\
\mathbb P_{\mathrm{sam},\zeta}
&= \mathbb P_{\mathrm{sam}}(\cdot\mid\mathcal S_{\zeta}),
\\
\nu_{\zeta}
&= \phi_{\#}\mathbb P_{\mathrm{sam},\zeta}.
\end{aligned}
\end{equation}
Suppose there exist \(c_{\zeta}>0\) and \(d_{\zeta}>0\) such that
\begin{equation}
\nu_{\zeta}(B_r(\mathbf z))
\ge
c_{\zeta}\left(\frac{r}{R_{\phi}}\right)^{d_{\zeta}}
\label{eq:stratified_lower_mass}
\end{equation}
for every \(\mathbf z\in\supp(\nu_{\zeta})\) and \(0<r\le R_{\phi}\). For \(0<r\le2R_{\phi}\), define
\begin{equation}
\underline m(r)
=
\min_{\zeta\in\Sigma_K}
 p_{\zeta}c_{\zeta}\left(\frac{r}{2R_{\phi}}\right)^{d_{\zeta}}.
\label{eq:minimum_unconditional_ball_mass}
\end{equation}
If \(N\) states are sampled independently from \(\mathbb P_{\mathrm{sam}}\), their feature vectors form an \(r\)-cover of every \(\supp(\nu_{\zeta})\) with probability at least \(1-\delta\) whenever
\begin{equation}
N
\ge
\frac{
\log|\Sigma_K|+
d_f\log\!\left(1+\frac{4R_{\phi}}{r}\right)+
\log(1/\delta)
}{\underline m(r)}.
\label{eq:stratified_sampling_bound}
\end{equation}
If \(\supp(\nu_{\zeta})=\phi(\mathcal S_{\zeta})\) for every represented stratum, the sampled features form an \(r\)-cover of the complete represented feature domain.
\end{proposition}

The proof is provided in Appendix~\ref{app:covering_complexity_full}.

\subsection{Matched-Action Regularity and Projected Bellman Approximation}
\label{subsec:value_regularity}

For states \(\varsigma,\varsigma'\in\mathcal S_{\mathrm r}^{K}\) with the same signature, compare \(a\in\mathcal A_{\mathrm f}(\varsigma)\) with \(M_{\varsigma\rightarrow\varsigma'}(a)\). We assume
\begin{align}
&\left|
\overline R(\varsigma,a)
-
\overline R\left(\varsigma',M_{\varsigma\rightarrow\varsigma'}(a)\right)
\right|
\nonumber\\
&\qquad\le
L_R\|\phi(\varsigma)-\phi(\varsigma')\|_2
\label{eq:reward_feature_regular}
\end{align}
and
\begin{align}
&\TV\Bigl(
\operatorname{Can}_{\#}\mathsf T(\cdot\mid\varsigma,a),
\operatorname{Can}_{\#}\mathsf T\bigl(\cdot\mid\varsigma',M_{\varsigma\rightarrow\varsigma'}(a)\bigr)
\Bigr)
\nonumber\\
&\qquad\le
L_T\|\phi(\varsigma)-\phi(\varsigma')\|_2.
\label{eq:transition_feature_regular}
\end{align}

\begin{assumption}[Threshold-margin regularity]
\label{ass:threshold_margin_regular}
Within each represented signature stratum, posterior distributions assign no atom exactly to delivery thresholds \(F_k^{\min}\), and the probability mass in a width-\(\eta\) neighborhood of any such threshold is at most \(C_{\mathrm{thr}}\eta\). This excludes discontinuities caused solely by probability mass sitting exactly on hard fidelity boundaries and is the regularity condition under which \cref{eq:reward_feature_regular,eq:transition_feature_regular} are applied.
\end{assumption}

\begin{lemma}[Feature-Space Regularity of the Optimal Value]
\label{lem:value_feature_regular}
Under \cref{eq:bounded_robust_reward,eq:reward_feature_regular,eq:transition_feature_regular}, states with the same signature satisfy
\begin{equation}
|V^*(\varsigma)-V^*(\varsigma')|
\le
L_V\|\phi(\varsigma)-\phi(\varsigma')\|_2,
\label{eq:derived_value_lipschitz}
\end{equation}
where
\begin{equation}
L_V=L_R+2\gamma V_{\max}L_T
=L_R+\frac{2\gamma R_{\max}L_T}{1-\gamma}.
\label{eq:derived_value_constant}
\end{equation}
\end{lemma}

The proof is given in Appendix~\ref{app:value_feature_regular_full}.

Define
\begin{equation}
\label{eq:value_classes}
\begin{aligned}
\mathbb V_{\max}^{K}
&=
\bigl\{\,V:\|V\|_{\infty,\mathcal S_{\mathrm r}^{K}}\le V_{\max}\,\bigr\},
\\
\mathbb V_{\max}^{\mathrm{full}}
&=
\bigl\{\,V:\|V\|_{\infty,\mathcal S_{\mathrm r}}\le V_{\max}\,\bigr\}.
\end{aligned}
\end{equation}
For \(V\in\mathbb V_{\max}^{K}\), the same-signature prototype projection is
\begin{equation}
(\mathfrak P_KV)(\varsigma)=V(\varsigma_{j_K(\varsigma)}).
\label{eq:value_projection_operator}
\end{equation}
It is nonexpansive. Under transition closure, the exact frozen-model Bellman operator on the represented domain is
\begin{equation}
\label{eq:canonical_quantum_bellman_operator}
\begin{aligned}
(\mathcal T_QV)(\varsigma)
=
\max_{a\in\mathcal A_{\mathrm f}(\varsigma)}
\Biggl\{
&\overline R(\varsigma,a)
\\
&\quad+
\gamma
\int_{\mathcal S_{\mathrm r}^{K}}
V(\varsigma')
\mathsf T(d\varsigma'\mid\varsigma,a)
\Biggr\}.
\end{aligned}
\end{equation}
Because \(\mathcal S_{\mathrm r}^{K}\) is transition closed and the represented restriction uses the same rewards and feasible actions as the full model on that domain, the restriction of the full-domain optimal value to \(\mathcal S_{\mathrm r}^{K}\) is the represented-domain optimal value. The exact projected operator is
\begin{equation}
\mathcal T_K=\mathfrak P_K\mathcal T_Q.
\label{eq:exact_projected_operator}
\end{equation}
It is a \(\gamma\)-contraction on \(\mathbb V_{\max}^{K}\). Let \(V_K\) be its unique fixed point:
\begin{equation}
V_K=\mathfrak P_K\mathcal T_QV_K.
\label{eq:projected_quantum_bellman}
\end{equation}

\begin{theorem}[Feature-Based Aggregation Bound]
\label{thm:feature_selection}
Suppose \(\mathcal S_{\mathrm r}^{K}\) is transition closed, represented states use the same reward and feasible action sets as the full model, assignment is restricted to same-signature prototypes, the covering radius is \(\epsilon_K\), and \cref{eq:reward_feature_regular,eq:transition_feature_regular} hold under \cref{ass:threshold_margin_regular}. Then,
\begin{equation}
\|V^*-V_K\|_{\infty,\mathcal S_{\mathrm r}^{K}}
\le
\frac{L_V\epsilon_K}{1-\gamma}.
\label{eq:feature_aggregation_bound}
\end{equation}
\end{theorem}

\begin{corollary}[Information-State Estimation Error]
\label{cor:error_propagation}
Let \(\widehat\varsigma=(\widehat b,\widehat c)\) satisfy
\begin{equation}
d_{\mathcal S}(\varsigma,\widehat\varsigma)\le\epsilon_{\mathrm{est}},
\qquad
\sigma(\varsigma)=\sigma(\widehat\varsigma)\in\Sigma_K,
\label{eq:estimation_same_signature_condition}
\end{equation}
with both states in \(\mathcal S_{\mathrm r}^{K}\). Then,
\begin{equation}
|V^*(\varsigma)-V_K(\widehat\varsigma)|
\le
L_VL_{\phi}\epsilon_{\mathrm{est}}
+
\frac{L_V\epsilon_K}{1-\gamma}.
\label{eq:information_state_estimation_bound}
\end{equation}
\end{corollary}

The proofs of Theorem~\ref{thm:feature_selection} and Corollary~\ref{cor:error_propagation} are provided in Appendix~\ref{app:feature_selection_full}.    

Theorem~\ref{thm:feature_selection} bounds the projected value function \(V_K\). It does not establish near-optimality of the finite-temperature planner policy introduced below; such a policy guarantee must additionally account for action-value transfer, and planner temperature.   

\subsection{Aggregation and Matching Complexity}
\label{subsec:aggregation_design_implications}

For \(N\) sampled states, \(K\) prototypes, feature dimension \(d_f\), and \(I\) clustering iterations, signature-stratified Lloyd clustering costs \(\mathcal O(INKd_f)\), whereas greedy farthest-point selection costs \(\mathcal O(NKd_f)\) after feature computation. For one current state and one assigned prototype, exact matching within template classes costs
\begin{equation}
\mathcal O\left(
\sum_{\tau\in\mathcal T_{\mathrm a}(\sigma(\varsigma))}
 m_{\tau}(\sigma(\varsigma))^3
\right)
\label{eq:action_matching_complexity}
\end{equation}
under a cubic assignment algorithm.

\section{Hybrid q-POMDP--GNN Routing Framework}
\label{sec:hybrid_pomdp_gnn}

The hybrid controller combines graph-based action scoring with the feature-based q-POMDP planner. Both components act on the same information state \(\varsigma_t=(b_t,c_t)\) and the same feasible atomic action set \(\mathcal A_{\mathrm f}(\varsigma_t)\). We write their policies as \(\pi_{\mathrm G}\), \(\pi_{\mathrm P}\), and \(\pi_{\mathrm H}\).

\subsection{Role-Aware Graph Action Representation}
\label{subsec:graph_action_representation}

The node and pair embeddings \(\{\mathbf h_v^{(L)}\}\) and \(\{\mathbf g_{p,t}\}\) are computed using \cref{eq:initial_graph_embeddings,eq:gnn_message,eq:gnn_update}. For operation type \(o\in\mathcal O\), let \(\mathsf L_o^v\) and \(\mathsf L_o^p\) denote node-role and pair-role labels. For action \(a=(o,\iota)\), let \(\mathcal V_r(a)\) and \(\mathcal P_r(a)\) contain nodes and pairs assigned role \(r\).

Generation retains endpoint roles, purification retains endpoints and two input-pair roles, and swapping distinguishes left endpoint, intermediate node, right endpoint, and the two ordered input pairs. Delivery retains source, destination, and delivered-pair roles; release retains endpoint roles and released pair. For undirected hardware, endpoint order is determined by a fixed physical-node ordering, not by transient pair identifiers.

Let \(\mathbf q_a\) contain normalized action attributes such as operation type, physical-link properties, attempt count, duration, reservation demand, pair-consumption pattern, demand class, and fidelity threshold. The role-aware action embedding is
\begin{align}
\mathbf z_a
={}&
\bigoplus_{r\in\mathsf L_o^v}
\left(\sum_{v\in\mathcal V_r(a)}\mathbf h_v^{(L)}\right)
\nonumber\\
&\oplus
\bigoplus_{r\in\mathsf L_o^p}
\left(\sum_{p\in\mathcal P_r(a)}\mathbf g_{p,t}\right)
\oplus\mathbf q_a.
\label{eq:action_representation}
\end{align}
An empty role sum is the zero vector. Pooling is permutation invariant within a role but does not merge distinct operational roles. The action score is
\begin{equation}
s_{\theta}(a\mid\varsigma_t)
=
f_{\theta}^{\mathrm{act}}(\mathbf z_a\oplus\phi(\varsigma_t)).
\label{eq:gnn_action_score}
\end{equation}
The feasibility-masked GNN policy is
\begin{equation}
\pi_{\mathrm G}(a\mid\varsigma_t)
=
\frac{\mathbf 1_{\{a\in\mathcal A_{\mathrm f}(\varsigma_t)\}}
\exp(s_{\theta}(a\mid\varsigma_t))}{
\sum_{a'\in\mathcal A_{\mathrm f}(\varsigma_t)}
\exp(s_{\theta}(a'\mid\varsigma_t))}.
\label{eq:hybrid_gnn_policy}
\end{equation}
Since idle is feasible,
\begin{equation}
\pi_{\mathrm G}(\mathcal A_{\mathrm f}(\varsigma_t)\mid\varsigma_t)=1.
\label{eq:gnn_feasible_support}
\end{equation}

\begin{revblock}

\subsection{Cached q-POMDP Planner}
\label{subsec:feature_pomdp_planner}

Let \(\tau_{\mathrm{ref}}(t)\le t\) denote the latest planner-refresh epoch, and let \(\widehat{\boldsymbol\vartheta}_{\tau_{\mathrm{ref}}(t)}\) be the calibration model used by the current cached table. At refresh epoch \(\tau\), let \(V_{K,\tau}\) be the fixed point of the projected Bellman operator built from \(\widehat{\boldsymbol\vartheta}_{\tau}\). The planner table is keyed by prototype index and canonical prototype action label \((j,\ell_{\varsigma_j}(a_j))\). For prototype \(\varsigma_j\) and feasible prototype action \(a_j\), define
\begin{align}
Q_{K,\tau}(\varsigma_j,a_j)
={}&
\overline R_{\widehat{\boldsymbol\vartheta}_{\tau}}(\varsigma_j,a_j)
\nonumber\\
&+
\gamma\int_{\mathcal S_{\mathrm r}^{K}}
V_{K,\tau}(\varsigma')
\mathsf T_{\widehat{\boldsymbol\vartheta}_{\tau}}
(d\varsigma'\mid\varsigma_j,a_j).
\label{eq:prototype_action_value}
\end{align}

For represented current state \(\varsigma\), set \(j=j_K(\varsigma)\) and \(M_{j\rightarrow\varsigma}=M_{\varsigma_j\rightarrow\varsigma}\). The transferred value of current action \(a\) is
\begin{equation}
Q_{K,t}^{\rightarrow}(\varsigma,a)
=
Q_{K,\tau_{\mathrm{ref}}(t)}
\left(\varsigma_j,M_{j\rightarrow\varsigma}^{-1}(a)\right).
\label{eq:transferred_planner_action_value}
\end{equation}
The planner policy is
\begin{equation}
\pi_{\mathrm P}(a\mid\varsigma)
=
\frac{\exp\left(Q_{K,t}^{\rightarrow}(\varsigma,a)/\tau_{\mathrm P}\right)}{
\sum_{a'\in\mathcal A_{\mathrm f}(\varsigma)}
\exp\left(Q_{K,t}^{\rightarrow}(\varsigma,a')/\tau_{\mathrm P}\right)},
\label{eq:pomdp_planner_policy}
\end{equation}
where \(\tau_{\mathrm P}>0\). As \(\tau_{\mathrm P}\to0\), it approaches a greedy transferred-value policy. The aggregation theorem concerns \(V_{K,\tau}\), not directly this finite-temperature policy.

If \(\sigma(\varsigma)\notin\Sigma_K\), no nearest-prototype query, table lookup, or action matching is performed. For bookkeeping only,
\begin{equation}
\pi_{\mathrm P}(\cdot\mid\varsigma)
\coloneqq
\pi_{\mathrm G}(\cdot\mid\varsigma),
\qquad
\sigma(\varsigma)\notin\Sigma_K.
\label{eq:online_planner_fallback}
\end{equation}
The state is added to the prototype-insertion buffer and becomes eligible for planner guarantees only after its stratum is represented. A planner refresh occurs periodically, when a new signature is inserted, or when \(\widehat{\boldsymbol\vartheta}_t\) changes beyond a prescribed tolerance. Ordinary stochastic evolution of \(Z_t\) does not trigger a refresh by itself.

\end{revblock}

\subsection{Trust-Adaptive Policy Fusion}
\label{subsec:trust_adaptive_fusion}

Define the represented-signature indicator
\begin{equation}
I_t^K=\mathbf 1_{\{\sigma(\varsigma_t)\in\Sigma_K\}}.
\label{eq:represented_signature_indicator}
\end{equation}
For \(I_t^K=1\), let \(u_t\) be uniform on \(\mathcal A_{\mathrm f}(\varsigma_t)\) and define the smoothed policies
\begin{equation}
\overline\pi_{i,t}
=
(1-\varepsilon_{\mathrm{KL}})\pi_i(\cdot\mid\varsigma_t)+\varepsilon_{\mathrm{KL}}u_t,
\qquad
i\in\{\mathrm G,\mathrm P\},
\label{eq:smoothed_component_policies}
\end{equation}
where \(0<\varepsilon_{\mathrm{KL}}<1\). Smoothing is used only in the trust computation. The planner-referenced disagreement is
\begin{equation}
D_t
=
\KL\left(\overline\pi_{\mathrm P,t}\,\middle\|\,\overline\pi_{\mathrm G,t}\right).
\label{eq:policy_kl_disagreement}
\end{equation}
The normalized GNN uncertainty is
\begin{equation}
U_t
=
\begin{cases}
\dfrac{\Ent(\overline\pi_{\mathrm G,t})}{
\log|\mathcal A_{\mathrm f}(\varsigma_t)|},
& |\mathcal A_{\mathrm f}(\varsigma_t)|>1,\\[2mm]
0,& |\mathcal A_{\mathrm f}(\varsigma_t)|=1.
\end{cases}
\label{eq:normalized_policy_uncertainty}
\end{equation}
For represented states, the GNN trust coefficient is
\begin{equation}
\alpha_t
=
\alpha_{\min}+(
\alpha_{\max}-\alpha_{\min})
\exp(-\kappa_DD_t-\kappa_UU_t),
\label{eq:adaptive_trust_coefficient}
\end{equation}
where \(0\le\alpha_{\min}<\alpha_{\max}\le1\) and \(\kappa_D,\kappa_U\ge0\). Agreement and low GNN uncertainty increase \(\alpha_t\); disagreement or uncertainty shifts weight toward the planner. The executed represented-state policy is
\begin{equation}
\pi_{\mathrm H}(a\mid\varsigma_t)
=
\alpha_t\pi_{\mathrm G}(a\mid\varsigma_t)
+(1-\alpha_t)\pi_{\mathrm P}(a\mid\varsigma_t).
\label{eq:hybrid_policy_composition}
\end{equation}

For unseen signatures, the controller enforces
\begin{equation}
\alpha_t=1,
\qquad
\pi_{\mathrm H}(\cdot\mid\varsigma_t)
=
\pi_{\mathrm G}(\cdot\mid\varsigma_t)
=
\pi_{\mathrm P}(\cdot\mid\varsigma_t).
\label{eq:forced_unseen_fallback}
\end{equation}
The distillation loss is omitted, and the GNN receives full responsibility for the executed action. Therefore, for every reachable state,
\begin{equation}
\pi_{\mathrm H}(\mathcal A_{\mathrm f}(\varsigma_t)\mid\varsigma_t)=1.
\label{eq:hybrid_support_identity}
\end{equation}

\subsection{Bounded Critic and Semi-Gradient Learning}
\label{subsec:hybrid_policy_learning}

The bounded critic and target critic use \cref{eq:hybrid_value_approximation}. Hence \(|V_{\psi}(\varsigma)|\le V_{\max}\) and \(|V_{\psi^-}(\varsigma)|\le V_{\max}\). For completed transition \((\varsigma_t,a_t,r_t,\varsigma_{t+1})\), where
\begin{equation}
r_t=R_{\mathrm{rob},\boldsymbol\vartheta_t}(X_t,C_t,A_t,Y_{t+1},C_{t+1}),
\label{eq:realized_training_reward}
\end{equation}
define
\begin{align}
y_t^{\mathrm{TD}}&=r_t+\gamma V_{\psi^-}(\varsigma_{t+1}),
\label{eq:hybrid_td_target}
\\
\widehat A_t&=y_t^{\mathrm{TD}}-V_{\psi}(\varsigma_t).
\label{eq:hybrid_advantage}
\end{align}
The critic loss is
\begin{equation}
\mathcal L_V(\psi)
=
\frac12\EX\left[(V_{\psi}(\varsigma_t)-y_t^{\mathrm{TD}})^2\right].
\label{eq:hybrid_critic_loss}
\end{equation}
By \cref{eq:bounded_robust_reward,eq:problem_value_bound},
\begin{equation}
|\widehat A_t|
\le
R_{\max}+(1+\gamma)V_{\max}
\le
\frac{2R_{\max}}{1-\gamma}.
\label{eq:bounded_hybrid_advantage}
\end{equation}

For represented states, actions are sampled from \(\pi_{\mathrm H}\), whereas only \(\pi_{\mathrm G}\) is parameterized by \(\theta\). Holding \(\alpha_t\) fixed during one update, define the GNN responsibility
\begin{equation}
\beta_t^{\mathrm G}
=
\frac{\alpha_t\pi_{\mathrm G}(a_t\mid\varsigma_t)}{
\pi_{\mathrm H}(a_t\mid\varsigma_t)},
\qquad
I_t^K=1.
\label{eq:gnn_responsibility}
\end{equation}
Then \(0\le\beta_t^{\mathrm G}\le1\) and
\begin{equation}
\nabla_{\theta}\log\pi_{\mathrm H}(a_t\mid\varsigma_t)
=
\beta_t^{\mathrm G}
\nabla_{\theta}\log\pi_{\mathrm G}(a_t\mid\varsigma_t)
\label{eq:hybrid_score_identity}
\end{equation}
for the fixed-\(\alpha_t\) mixture. Under the unseen fallback, \(\beta_t^{\mathrm G}=1\) and \(I_t^K=0\).

Let \(\operatorname{sg}[\cdot]\) denote stop-gradient. The actor loss is
\begin{align}
\mathcal L_{\pi}(\theta)
={}&
-
\EX\left[
\operatorname{sg}[\beta_t^{\mathrm G}\widehat A_t]
\log\pi_{\mathrm G}(a_t\mid\varsigma_t)
\right]
\nonumber\\
&+
\lambda_{\mathrm{IL}}
\EX\left[
I_t^K
\KL\left(
\operatorname{sg}[\pi_{\mathrm P}(\cdot\mid\varsigma_t)]
\,\middle\|\,
\pi_{\mathrm G}(\cdot\mid\varsigma_t)
\right)
\right]
\nonumber\\
&-
\lambda_{\mathrm H}
\EX\left[\Ent\left(\pi_{\mathrm G}(\cdot\mid\varsigma_t)\right)\right],
\label{eq:hybrid_actor_loss}
\end{align}
where \(\lambda_{\mathrm{IL}},\lambda_{\mathrm H}\ge0\). The distillation term uses unsmoothed planner-to-GNN KL and is evaluated only on represented states. Smoothing is used only for trust. The trust coefficient, prototype assignment, matching, and planner table are recomputed before action selection but treated as constants during the corresponding actor update; \cref{eq:hybrid_actor_loss} is therefore a semi-gradient surrogate.

\subsection{Conditionally Independent Worker Batches}
\label{subsec:independent_worker_batches}

Let \(\mathcal F_t\) contain actor and critic parameters, target critic, current prototype table, cached planner values, training history, worker information states before action sampling, and all shared randomness realized before the current transitions. Conditional on \(\mathcal F_t\), workers use independent action samples and simulator randomness.

With \(W\ge B\) workers, the theoretical update batch contains at most one current transition from each of \(B\) distinct workers:
\begin{equation}
\mathcal D_t^{\mathrm{ind}}
=
\{\mathcal Z_t^{(i)}\}_{i\in\mathcal I_t},
\qquad
|\mathcal I_t|=B,
\label{eq:independent_worker_batch}
\end{equation}
where \(\mathcal Z_t^{(i)}=(\varsigma_t^{(i)},a_t^{(i)},r_t^{(i)},\varsigma_{t+1}^{(i)})\). Successive transitions from one worker remain temporally dependent; the variance result in \cref{lem:var_bound} applies only to batches satisfying this distinct-worker construction.

\subsection{Hybrid Routing Algorithm}
\label{subsec:hybrid_algorithm}

\begin{algorithm}[t]
\caption{Synchronous Hybrid q-POMDP--GNN Routing}
\label{alg:hybrid_quantum_pomdp_gnn_routing}
\KwIn{Workers \(1{:}W\), prototypes \(\mathcal Q_K\), parameters \((\theta,\psi,\psi^-)\), refresh period \(H_{\mathrm P}\), batch size \(B\).}
\KwOut{Stationary policy snapshots and online hybrid routing decisions.}
Initialize planner table and new-signature buffer \(\mathcal U_{\mathrm{new}}\leftarrow\varnothing\)\;
\For{\(t=0,1,\ldots\)}{
    Define shared filtration \(\mathcal F_t\)\;
    \For{\(i=1,\ldots,W\)}{
        Observe \(\varsigma_t^{(i)}=(b_t^{(i)},c_t^{(i)})\) and construct \(\mathcal A_{\mathrm f}(\varsigma_t^{(i)})\)\;
        Compute graph embeddings, role-aware action scores, and masked \(\pi_{\mathrm G}\)\;
        \eIf{\(\sigma(\varsigma_t^{(i)})\in\Sigma_K\)}{
            Set \(I_{t,i}^{K}\leftarrow1\), assign \(j\leftarrow j_K(\varsigma_t^{(i)})\), construct \(M_{\varsigma_j\rightarrow\varsigma_t^{(i)}}\), transfer cached \(Q_{K,\tau_{\mathrm{ref}}(t)}\), and form \(\pi_{\mathrm P}\)\;
            Compute \(D_t^{(i)}\), \(U_t^{(i)}\), \(\alpha_t^{(i)}\), and \(\pi_{\mathrm H}\)\;
        }{
            Set \(I_{t,i}^{K}\leftarrow0\), \(\alpha_t^{(i)}\leftarrow1\), \(\pi_{\mathrm P}\leftarrow\pi_{\mathrm G}\), \(\pi_{\mathrm H}\leftarrow\pi_{\mathrm G}\), and \(\beta_{t,i}^{\mathrm G}\leftarrow1\)\;
            Add \(\varsigma_t^{(i)}\) to \(\mathcal U_{\mathrm{new}}\) and skip distillation for this sample\;
        }
        Sample \(a_t^{(i)}\sim\pi_{\mathrm H}(\cdot\mid\varsigma_t^{(i)})\), revalidate feasibility, and reserve required memory\;
        Apply passive evolution and the selected instrument; observe \(Y_{t+1}^{(i)}\)\;
        Resolve reservations and update pair inventory, memory ledger, queues, and \(c_{t+1}^{(i)}\); enforce \cref{eq:memory_pair_inventory_invariant}\;
        Compute \(r_t^{(i)}\) and update \(b_{t+1}^{(i)}\) by \cref{eq:qpomdp_belief_update}\;
    }
    Form \(\mathcal D_t^{\mathrm{ind}}\) from \(B\) distinct workers and update \((\psi,\theta,\psi^-)\) using \cref{eq:hybrid_critic_loss,eq:hybrid_actor_loss}\;
    \If{\((t+1)\bmod H_{\mathrm P}=0\), significant calibration drift, or \(\mathcal U_{\mathrm{new}}\neq\varnothing\)}{
        Insert new signatures, update \(\Sigma_K\), expand represented model as required for transition closure, refresh \(V_{K,t+1}\) and \(Q_{K,t+1}\), and clear \(\mathcal U_{\mathrm{new}}\)\;
    }
}
\end{algorithm}

All stationary-policy value results below are policy-snapshot guarantees. For such a result, actor parameters, prototype set, cached planner table, calibration model used at the latest refresh, matching rule, and trust map are held fixed. The resulting \(\pi_{\mathrm G}\), \(\pi_{\mathrm P}\), and \(\pi_{\mathrm H}\) are stationary Markov policies on common feasible action sets.

\section{Analysis of the Hybrid Framework}
\label{sec:hybrid_analysis}

By \cref{eq:bounded_robust_reward,eq:problem_value_bound}, every stationary feasible policy satisfies
\begin{equation}
\|V^{\pi}\|_{\infty,\mathcal S_{\mathrm r}}
\le
V_{\max}.
\label{eq:policy_value_bound}
\end{equation}
Prototype and critic approximation results are stated on transition-closed \(\mathcal S_{\mathrm r}^{K}\). Robustness and nonstationarity results use the full reachable domain \(\mathcal S_{\mathrm r}\). Unless stated otherwise, all policy values refer to stationary policy snapshots. Whenever two frozen models are compared, they share the measurable information-state domain \(\mathcal S_{\mathrm r}\), hard-feasibility correspondence \(\mathcal A_{\mathrm f}(\varsigma)\), discount \(\gamma\), and reward bound \(R_{\max}\).

\subsection{Hybrid-Policy Performance}
\label{subsec:hybrid_performance}

For represented states, define
\begin{equation}
\Delta(\varsigma)
=
\frac{1}{1-\varepsilon_{\mathrm{KL}}}
\sqrt{\frac12
\KL\left(\overline\pi_{\mathrm P}(\cdot\mid\varsigma)\,\middle\|\,
\overline\pi_{\mathrm G}(\cdot\mid\varsigma)\right)}.
\label{eq:policy_disagreement}
\end{equation}
For unseen signatures, set \(\Delta(\varsigma)=0\). Define
\begin{align}
\varepsilon_{\mathrm P}
&=
\sup_{\varsigma\in\mathcal S_{\mathrm r}}
\alpha(\varsigma)\Delta(\varsigma),
\label{eq:planner_hybrid_deviation}
\\
\varepsilon_{\mathrm G}
&=
\sup_{\varsigma\in\mathcal S_{\mathrm r}}
(1-\alpha(\varsigma))\Delta(\varsigma).
\label{eq:gnn_hybrid_deviation}
\end{align}

\begin{theorem}[Hybrid-Policy Performance Bound]
\label{thm:hybrid_bound}
Assume \(\pi_{\mathrm G}\), \(\pi_{\mathrm P}\), and \(\pi_{\mathrm H}\) are stationary policy snapshots on common feasible action sets, that represented states use the same feasible support for the two component policies, and that unseen states use the forced fallback in \cref{eq:forced_unseen_fallback}. Then, for every reachable information state \(\varsigma\),
\begin{align}
V^{\pi_{\mathrm H}}(\varsigma)
\ge
\max\bigl\{&
V^{\pi_{\mathrm P}}(\varsigma)-C_{\gamma}\varepsilon_{\mathrm P},
\nonumber\\
&V^{\pi_{\mathrm G}}(\varsigma)-C_{\gamma}\varepsilon_{\mathrm G}
\bigr\},
\label{eq:hybrid_performance_bound}
\end{align}
where
\begin{equation}
C_{\gamma}=\frac{2R_{\max}}{(1-\gamma)^2}.
\label{eq:hybrid_performance_constant}
\end{equation}
\end{theorem}

The proof is provided in Appendix~\ref{app:hybrid_bound_full}

The theorem does not assert that mixing improves on both components. It bounds the loss relative to each component through the corresponding trust-weighted disagreement, using the same \(\KL(\pi_{\mathrm P}\|\pi_{\mathrm G})\) direction as the trust rule.

\subsection{Conditional Semi-Gradient Variance}
\label{subsec:learning_guarantees}

Assume
\begin{equation}
\|\nabla_{\theta}\log\pi_{\mathrm G}(a\mid\varsigma;\theta)\|_2\le G_{\pi}
\label{eq:bounded_policy_score}
\end{equation}
for every reachable feasible state--action pair. For worker \(i\), define
\begin{equation}
\mathbf g_t^{(i)}
=
\beta_{t,i}^{\mathrm G}
\nabla_{\theta}\log\pi_{\mathrm G}(a_t^{(i)}\mid\varsigma_t^{(i)};\theta)
\widehat A_t^{(i)},
\label{eq:single_worker_semigradient}
\end{equation}
and \(\widehat{\mathbf g}_t=B^{-1}\sum_{i\in\mathcal I_t}\mathbf g_t^{(i)}\). For a random vector \(\mathbf Z\), let
\begin{equation}
\operatorname{Var}(\mathbf Z\mid\mathcal F_t)
=
\EX\left[
\left\|\mathbf Z-\EX[\mathbf Z\mid\mathcal F_t]\right\|_2^2
\middle| \mathcal F_t
\right].
\label{eq:conditional_vector_variance}
\end{equation}

\begin{lemma}[Conditional Independent-Worker Variance]
\label{lem:var_bound}
Suppose \(\mathcal D_t^{\mathrm{ind}}\) contains one current transition from each of \(B\) distinct workers and these current transitions are conditionally independent given \(\mathcal F_t\). If \(|\widehat A_t^{(i)}|\le A_{\max}\), then
\begin{equation}
\operatorname{Var}(\widehat{\mathbf g}_t\mid\mathcal F_t)
\le
\frac{G_{\pi}^2A_{\max}^2}{B}.
\label{eq:hybrid_gradient_variance}
\end{equation}
Using \cref{eq:bounded_hybrid_advantage},
\begin{equation}
\operatorname{Var}(\widehat{\mathbf g}_t\mid\mathcal F_t)
\le
\frac{4G_{\pi}^2R_{\max}^2}{B(1-\gamma)^2}.
\label{eq:reward_gradient_variance}
\end{equation}
\end{lemma}

See Appendix~\ref{app:var_bound_full} for the proof.

The result is conditional on shared parameters, prototype table, planner cache, worker states, and training history. It does not apply to arbitrary replay samples or a contiguous rollout from one worker.

\begin{revblock}

\subsection{Convex Final-Head Distillation}
\label{subsec:convex_distillation}

For represented sample \(s\), define
\begin{equation}
\ell_s(\theta)
=
\KL\left(\pi_{\mathrm P,s}\,\middle\|\,\pi_{\mathrm G,s}(\theta)\right).
\label{eq:distillation_loss}
\end{equation}
The target and encoded action features are held fixed during this subproblem.

\begin{theorem}[Regret of the Convex Distillation Subproblem]
\label{thm:il_converge}
Suppose graph and information-state encoders are fixed, the feasible action set is finite, final GNN logits are affine in \(\theta\), the masked softmax assigns positive probability to every feasible action, and \(\Theta\) is convex. If \(\|\theta_1-\theta^*\|_2\le B_{\theta}\) and \(\|\nabla\ell_s(\theta)\|_2\le G_{\ell}\), then projected online gradient descent with
\begin{equation}
\nu_N=\frac{B_{\theta}}{G_{\ell}\sqrt N}
\label{eq:distillation_step_size}
\end{equation}
satisfies
\begin{equation}
\frac1N\sum_{s=1}^{N}\left[\ell_s(\theta_s)-\ell_s(\theta^*)\right]
\le
\frac{B_{\theta}G_{\ell}}{\sqrt N}.
\label{eq:distillation_convergence}
\end{equation}
\end{theorem}

The proof is provided in Appendix~\ref{app:il_converge_full}

For fixed encoded features and affine logits, \(\ell_s\) equals planner-to-GNN cross-entropy plus a target-dependent constant and is convex. The theorem concerns only this final-head subproblem.

\end{revblock}

\subsection{Bellman-Residual Approximation}
\label{subsec:approximation_error}

Let \(\widehat{\mathcal T}_K=\mathfrak P_K\widehat{\mathcal T}_Q\) be the projected Bellman operator under an estimated frozen model. Both \(\mathcal T_K\) and \(\widehat{\mathcal T}_K\) are \(\gamma\)-contractive on \(\mathbb V_{\max}^{K}\). Let \(\widetilde V_{\psi}\in\mathbb V_{\max}^{K}\) satisfy
\begin{equation}
\|\widetilde V_{\psi}-\widehat{\mathcal T}_K\widetilde V_{\psi}\|_{\infty,\mathcal S_{\mathrm r}^{K}}
\le
\epsilon_{\mathrm{BR}}.
\label{eq:critic_bellman_residual}
\end{equation}
Assume the exact and estimated projected operators act on the same transition-closed represented domain and the same represented feasible action sets. On that domain, suppose
\begin{equation}
\sup_{\varsigma,a}
\left|\overline R(\varsigma,a)-\widehat{\overline R}(\varsigma,a)\right|
\le
\zeta_Q,
\label{eq:quantum_reward_model_error}
\end{equation}
and for every \(V\in\mathbb V_{\max}^{K}\),
\begin{equation}
\sup_{\varsigma,a}
\left|
\int V\,d\mathsf T(\cdot\mid\varsigma,a)
-
\int V\,d\widehat{\mathsf T}(\cdot\mid\varsigma,a)
\right|
\le
\delta_Q.
\label{eq:quantum_transition_value_error}
\end{equation}

\begin{theorem}[Bellman-Residual Quantum Value Approximation]
\label{thm:q_error}
Let \(V_K\) be the fixed point of \(\mathcal T_K\). Then
\begin{equation}
\|V_K-\widetilde V_{\psi}\|_{\infty,\mathcal S_{\mathrm r}^{K}}
\le
\frac{\epsilon_{\mathrm{BR}}+\zeta_Q+\gamma\delta_Q}{1-\gamma}.
\label{eq:quantum_value_approximation_bound}
\end{equation}
Together with \cref{thm:feature_selection},
\begin{equation}
\|V^*-\widetilde V_{\psi}\|_{\infty,\mathcal S_{\mathrm r}^{K}}
\le
\frac{L_V\epsilon_K+\epsilon_{\mathrm{BR}}+\zeta_Q+\gamma\delta_Q}{1-\gamma}.
\label{eq:total_value_approximation_bound}
\end{equation}
\end{theorem}

\begin{corollary}[Entanglement-Sensitive Feature Requirement]
\label{cor:ent_error}
Suppose a bipartite state is represented only by its local marginals and reconstructed as \(\widehat\rho_{AB}=\rho_A\otimes\rho_B\). For the maximally entangled state \(\rho_{AB}=\ket{\Phi_D}\!\bra{\Phi_D}\),
\begin{equation}
\frac12\left\|\rho_{AB}-\rho_A\otimes\rho_B\right\|_1
=
1-\frac1{D^2}.
\label{eq:entanglement_compression_error}
\end{equation}
\end{corollary}
Thus, local marginals alone may incur order-one error. The proposed representation retains pair-reduced states and selected pair-local observables but does not claim to preserve arbitrary multipartite correlations across separate pair instances.

The proofs of Theorem~\ref{thm:q_error} and Corollary~\ref{cor:ent_error} are provided in Appendix~\ref{app:q_error_full}.

\begin{revblock}

\subsection{Amortized Computational Complexity}
\label{subsec:approximation_scalability}

Let \(n=|\mathcal V|\), \(m_t=|\mathcal P_t|\), \(L\) be GNN layers, \(h\) hidden dimension, \(K\) prototypes, and \(\overline A\) average feasible actions. Let \(d_K\) be average nonzero successor prototypes per prototype--action pair, \(N_b\) belief particles, and \(c_{\mathrm q}\) average cost of propagating one particle through channel, instrument, and likelihood calculation. Belief filtering costs \(C_{\mathrm{filter}}=\mathcal O(N_bc_{\mathrm q})\). Let \(C_{\mathrm{sig},t}\) be signature and action-label construction cost, and let
\begin{equation}
C_{\mathrm{match},t}
=
\mathcal O\left(
\sum_{\tau\in\mathcal T_{\mathrm a}(\sigma(\varsigma_t))}
 m_{\tau}(\sigma(\varsigma_t))^3
\right).
\label{eq:online_action_matching_cost}
\end{equation}
A complete projected Bellman refresh costs \(C_{\mathrm{backup}}=\mathcal O(K\overline A d_K)\). Let \(\eta_{\mathrm{rep}}\), \(\eta_{\mathrm P}\), and \(\eta_{\mathrm{new}}\) denote fractions of represented-state epochs, complete planner-refresh epochs, and unseen-signature epochs. Insertion of one prototype costs \(C_{\mathrm{ins}}=\mathcal O(d_f+\overline A d_K)\).

\begin{theorem}[Amortized Online Complexity]
\label{thm:scalability}
One role-aware GNN evaluation requires
\begin{equation}
C_{\mathrm{GNN}}
=
\mathcal O\left(L(n+m_t)h^2+\overline A h^2\right).
\label{eq:gnn_evaluation_complexity}
\end{equation}
With direct prototype search, the amortized per-worker, per-epoch cost is
\begin{equation}
\label{eq:hybrid_amortized_complexity}
\begin{aligned}
C_{\mathrm{hybrid}}
=
\mathcal O\Bigl(
&N_bc_{\mathrm q}
+C_{\mathrm{sig},t}
\\
&+
L(n+m_t)h^2
+\overline A h^2
\\
&+
\eta_{\mathrm{rep}}
\bigl[
Kd_f
+C_{\mathrm{match},t}
+\overline A
\bigr]
\\
&+
\eta_{\mathrm P}K\overline A d_K
+
\eta_{\mathrm{new}}C_{\mathrm{ins}}
\Bigr).
\end{aligned}
\end{equation}
For \(W\) synchronized workers, simulation and policy evaluation scale linearly in \(W\), whereas a shared planner-table refresh is performed once.
\end{theorem}

The proof is provided in Appendix~\ref{app:scalability_full}.

\end{revblock}

\subsection{Sensitivity to Pair-State Perturbations}
\label{subsec:policy_sensitivity}

For states \(\varsigma=(b,c)\) and \(\varsigma'=(b',c)\) with the same inventory, signature, and feasible action set, define
\begin{equation}
d_Q(\varsigma,\varsigma')
=
\max_{p\in\mathcal P(c)}\|\rho_b^p-\rho_{b'}^p\|_1,
\label{eq:pair_state_metric}
\end{equation}
with value zero when \(\mathcal P(c)=\varnothing\).

\begin{proposition}[GNN Sensitivity to Pair-Local Noise]
\label{prop:policy_sensitivity}
Assume
\begin{equation}
\|\pi_{\mathrm G}(\cdot\mid\varsigma)-\pi_{\mathrm G}(\cdot\mid\varsigma')\|_1
\le
L_{\pi}d_Q(\varsigma,\varsigma').
\label{eq:gnn_pair_lipschitz}
\end{equation}
If each pair state is perturbed by a CPTP map \(\mathcal E_p\), while \(c\), signature, and feasible action set remain unchanged, then
\begin{equation}
\|\pi_{\mathrm G}(\cdot\mid\varsigma^{\mathcal E})-\pi_{\mathrm G}(\cdot\mid\varsigma)\|_1
\le
L_{\pi}\max_{p\in\mathcal P(c)}\|\mathcal E_p-\mathcal I\|_{\diamond}.
\label{eq:policy_noise_sensitivity}
\end{equation}
\end{proposition}

\subsection{Policy Improvement under Static Model Error}
\label{subsec:static_robustness}

Let \(\mathcal M_{\boldsymbol\vartheta}\) be the true frozen model and \(\widehat{\mathcal M}_{\boldsymbol\vartheta}\) the calibrated model used for planning. Both include endogenous \(Z_t\) and share \(\mathcal S_{\mathrm r}\), \(\mathcal A_{\mathrm f}(\varsigma)\), \(\gamma\), and \(R_{\max}\). Assume first that both models use the same nominal nonselective channel \(\mathcal E_a^{0,\mathrm{ns}}\). Define
\begin{equation}
\epsilon_{\mathrm{ch}}
=
\sup_{\varsigma,a,x\in\supp(b)}
\frac12
\left\|
\mathcal E_{a,x,c,\boldsymbol\vartheta}^{\mathrm{ns}}
-
\widehat{\mathcal E}_{a,x,c,\boldsymbol\vartheta}^{\mathrm{ns}}
\right\|_{\diamond}.
\label{eq:local_channel_calibration_error}
\end{equation}
If \(\epsilon_r^{\mathrm{base}}\) bounds non-channel reward-model errors, then
\begin{equation}
\epsilon_r\le\epsilon_r^{\mathrm{base}}+\lambda_{\mathrm{rob}}\epsilon_{\mathrm{ch}}.
\label{eq:channel_to_reward_error}
\end{equation}

\begin{assumption}[Flagged-instrument and filter stability]
\label{ass:filter_stability}
The nonselective-channel error in \cref{eq:local_channel_calibration_error} controls the robust reward penalty but not the classical outcome law. For transition and filtering error, define
\begin{equation}
\epsilon_{\mathrm{inst}}
=
\sup_{\varsigma,a,x\in\supp(b)}
\frac12
\left\|
\mathfrak I_{a,x,c,\boldsymbol\vartheta}^{\mathrm{red}}
-
\widehat{\mathfrak I}_{a,x,c,\boldsymbol\vartheta}^{\mathrm{red}}
\right\|_{\diamond},
\label{eq:flagged_instrument_calibration_error}
\end{equation}
where \(\mathfrak I_{a,x,c,\boldsymbol\vartheta}^{\mathrm{red}}\) abbreviates \(\mathfrak I_{a,z(x),c,\boldsymbol\vartheta}^{\mathrm{red}}\). Let \(\epsilon_{\mathrm{cl}}\) bound additional errors in latent-environment, request, memory, and classical-report kernels. There exists \(L_{\mathrm B}<\infty\) such that
\begin{equation}
\epsilon_{\mathrm T}
\le
L_{\mathrm B}(\epsilon_{\mathrm{inst}}+\epsilon_{\mathrm{cl}}).
\label{eq:instrument_to_transition_error}
\end{equation}
A sufficient regularity condition is that all realized observation normalizers are uniformly bounded away from zero on the analyzed domain.
\end{assumption}

Assume uniformly over reachable feasible \((\varsigma,a)\) that
\begin{align}
|\overline R_{\boldsymbol\vartheta}(\varsigma,a)-\widehat{\overline R}_{\boldsymbol\vartheta}(\varsigma,a)|&\le\epsilon_r,
\label{eq:static_reward_model_error}
\\
\TV\left(\mathsf T_{\boldsymbol\vartheta}(\cdot\mid\varsigma,a),
\widehat{\mathsf T}_{\boldsymbol\vartheta}(\cdot\mid\varsigma,a)\right)&\le\epsilon_{\mathrm T}.
\label{eq:static_transition_model_error}
\end{align}

\begin{theorem}[Policy Improvement under Static Model Error]
\label{thm:noise_improve}
Suppose stationary policy snapshots satisfy
\begin{equation}
V_{\widehat{\mathcal M}_{\boldsymbol\vartheta}}^{\pi_{j+1}}(\varsigma)
\ge
V_{\widehat{\mathcal M}_{\boldsymbol\vartheta}}^{\pi_j}(\varsigma)
\label{eq:estimated_model_policy_improvement}
\end{equation}
for every \(\varsigma\in\mathcal S_{\mathrm r}\). Then
\begin{align}
V_{\mathcal M_{\boldsymbol\vartheta}}^{\pi_{j+1}}(\varsigma)
\ge{}&
V_{\mathcal M_{\boldsymbol\vartheta}}^{\pi_j}(\varsigma)
-
\frac{2\epsilon_r}{1-\gamma}
\nonumber\\
&-
\frac{4\gamma R_{\max}\epsilon_{\mathrm T}}{(1-\gamma)^2}.
\label{eq:static_noise_policy_improvement}
\end{align}
\end{theorem}

See Appendix~\ref{app:noise_improve_full} for the full proof.

If the nominal channel is also estimated, define \(\epsilon_{\mathrm{nom}}=\sup_a\frac12\|\mathcal E_a^{0,\mathrm{ns}}-\widehat{\mathcal E}_a^{0,\mathrm{ns}}\|_{\diamond}\) and replace \cref{eq:channel_to_reward_error} by \(\epsilon_r\le\epsilon_r^{\mathrm{base}}+\lambda_{\mathrm{rob}}(\epsilon_{\mathrm{ch}}+\epsilon_{\mathrm{nom}})\).

\subsection{Frozen-Model Stability and Dynamic Value-Regret}
\label{subsec:nonstationary_stability}

The endogenous state \(Z_t\) belongs to every frozen q-POMDP. External nonstationarity is represented by \(\{\boldsymbol\vartheta_t\}_{t\ge0}\). Let \(\mathcal M_t=\mathcal M_{\boldsymbol\vartheta_t}\) denote the model obtained by holding \(\boldsymbol\vartheta_t\) fixed over the infinite horizon. Let \(\mathcal T_t^*\) and \(V_t^*\) be its optimal Bellman operator and fixed point on \(\mathbb V_{\max}^{\mathrm{full}}\). Define
\begin{equation}
d_t^*
=
\sup_{V\in\mathbb V_{\max}^{\mathrm{full}}}
\|\mathcal T_t^*V-\mathcal T_{t-1}^*V\|_{\infty,\mathcal S_{\mathrm r}}.
\label{eq:optimal_operator_drift}
\end{equation}

\begin{theorem}[Frozen-Model Value Stability]
\label{thm:nonstationary_stability}
If every \(\mathcal T_t^*\) is a \(\gamma\)-contraction, then
\begin{equation}
\|V_t^*-V_{t-1}^*\|_{\infty,\mathcal S_{\mathrm r}}
\le
\frac{d_t^*}{1-\gamma}.
\label{eq:general_value_stability}
\end{equation}
If \(d_t^*\le L_{\mathrm{env}}\|\boldsymbol\vartheta_t-\boldsymbol\vartheta_{t-1}\|\), then
\begin{equation}
\|V_t^*-V_{t-1}^*\|_{\infty,\mathcal S_{\mathrm r}}
\le
\frac{L_{\mathrm{env}}}{1-\gamma}
\|\boldsymbol\vartheta_t-\boldsymbol\vartheta_{t-1}\|.
\label{eq:parameter_value_stability}
\end{equation}
\end{theorem}

The proof is provided in Appendix~\ref{app:nonstationary_full}.

For stationary policy \(\pi\), let \(\mathcal T_t^{\pi}\) and \(V_t^{\pi}\) denote its Bellman operator and infinite-horizon value in frozen model \(\mathcal M_t\). Define
\begin{equation}
\overline d_t
=
\max\left\{
 d_t^*,
\sup_{\pi}\sup_{V\in\mathbb V_{\max}^{\mathrm{full}}}
\|\mathcal T_t^{\pi}V-\mathcal T_{t-1}^{\pi}V\|_{\infty}
\right\}.
\label{eq:uniform_operator_drift}
\end{equation}
A concrete sufficient route for bounding \(\overline d_t\) is to bound one-step reward variation, flagged-instrument variation, nonselective-channel variation where it enters the robust penalty, and classical-kernel variation, then invoke the same model-perturbation logic used in \cref{subsec:static_robustness}.

Let \(\pi_t\) be the stationary snapshot used at epoch \(t\), and define frozen-model dynamic value-regret
\begin{equation}
\mathfrak R_T^{\mathrm{val}}
=
\sum_{t=2}^{T}\left[V_t^*(\varsigma_t)-V_t^{\pi_t}(\varsigma_t)\right].
\label{eq:dynamic_regret_definition}
\end{equation}
Assume
\begin{equation}
\|V_{t-1}^*-V_{t-1}^{\pi_t}\|_{\infty,\mathcal S_{\mathrm r}}
\le
\epsilon_t^{\mathrm{trk}}.
\label{eq:policy_tracking_error}
\end{equation}

\begin{theorem}[Dynamic Value-Regret under Nonstationarity]
\label{thm:nonstationary_regret}
The frozen-model dynamic value-regret satisfies
\begin{equation}
\mathfrak R_T^{\mathrm{val}}
\le
\sum_{t=2}^{T}\epsilon_t^{\mathrm{trk}}
+
\frac{2}{1-\gamma}\sum_{t=2}^{T}\overline d_t.
\label{eq:nonstationary_regret_bound}
\end{equation}
If \(\overline d_t\le\overline L_{\mathrm{env}}\|\boldsymbol\vartheta_t-\boldsymbol\vartheta_{t-1}\|\) and
\begin{equation}
B_T^{\vartheta}
=
\sum_{t=2}^{T}\|\boldsymbol\vartheta_t-\boldsymbol\vartheta_{t-1}\|,
\label{eq:environment_variation_budget}
\end{equation}
then
\begin{equation}
\mathfrak R_T^{\mathrm{val}}
\le
\sum_{t=2}^{T}\epsilon_t^{\mathrm{trk}}
+
\frac{2\overline L_{\mathrm{env}}B_T^{\vartheta}}{1-\gamma}.
\label{eq:variation_budget_regret}
\end{equation}
Consequently,
\begin{equation}
\frac{\mathfrak R_T^{\mathrm{val}}}{T}\rightarrow0
\quad
\text{whenever}\quad
\sum_{t=2}^{T}\epsilon_t^{\mathrm{trk}}=o(T)
\;\text{and}\;
B_T^{\vartheta}=o(T).
\label{eq:sublinear_tracking_and_variation}
\end{equation}
\end{theorem}

The proof is provided in Appendix~\ref{app:nonstationary_full}

This is a frozen-model dynamic value-regret bound, not a sample-path cumulative-reward regret bound. Sublinear environmental variation alone is insufficient without sublinear cumulative tracking error.

\subsection{Feasibility}
\label{subsec:feasibility_sample}

\begin{theorem}[Hybrid-Policy Feasibility]
\label{thm:feasibility}
Assume the GNN is feasibility masked as in \cref{eq:gnn_feasible_support}. For represented signatures, suppose the assigned prototype has the same signature, \(M_{\varsigma_j\rightarrow\varsigma}\) is a bijection between prototype and current feasible action sets, and the planner normalizes only over instantiated actions in \(\mathcal A_{\mathrm f}(\varsigma)\). Then
\begin{equation}
\pi_{\mathrm P}(\mathcal A_{\mathrm f}(\varsigma)\mid\varsigma)=1.
\label{eq:planner_component_feasibility}
\end{equation}
Under the unseen-signature fallback in \cref{eq:forced_unseen_fallback}, the hybrid policy satisfies
\begin{equation}
\pi_{\mathrm H}(\mathcal A_{\mathrm f}(\varsigma)\mid\varsigma)=1
\label{eq:hybrid_policy_feasibility}
\end{equation}
for every reachable information state.
\end{theorem}

See Appendix~\ref{app:feasibility_full} for the proof.

The feasibility result applies immediately to unseen signatures because execution falls back to the masked GNN. Covering, aggregation, and planner-transfer guarantees apply only after the corresponding stratum becomes represented. Theorem~\ref{thm:il_converge} gives average convex-head distillation regret at most \(\epsilon_{\mathrm{IL}}>0\) whenever
\begin{equation}
N\ge\left(\frac{B_{\theta}G_{\ell}}{\epsilon_{\mathrm{IL}}}\right)^2.
\label{eq:distillation_iteration_requirement}
\end{equation}
This is not a global sample-complexity guarantee for the complete hybrid algorithm.

\section{Simulation and Numerical Results}
\label{sec:simulation}

This section evaluates the revised atomic q-POMDP--GNN routing framework under finite memory, on-demand heralded generation, completion-time delivery fidelity, belief-state filtering, role-aware action matching, and time-varying physical conditions. The evaluation is designed to stress the modeling components introduced in \cref{sec:system_model,sec:problem,sec:feature_selection_guarantees,sec:hybrid_pomdp_gnn}: pair-instance inventories, stochastic memory ledgers, branch-dependent purification and swapping costs, planner staleness, represented and unseen signatures, and robust belief-state control. The objective of the experiments is not only to compare final goodput, but also to verify that the operation semantics assumed by the theory are reflected in the behavior of the implemented controller.

\subsection{Simulation Setup}
\label{subsec:simulation_setup}

Each experiment uses discrete micro-epochs of duration \(\Delta t=1\) ms. The 1 ms value is the physical model's routing granularity; the wall-clock latency reported below is offline simulator and policy-evaluation time on the software platform, not a real-time control deadline imposed on hardware. At every epoch, the controller observes \(\varsigma_t=(b_t,c_t)\), selects one feasible atomic action from \(\mathcal A_{\mathrm f}(\varsigma_t)\), reserves any required quantum memory, applies passive evolution, executes the selected instrument, observes the outcome, updates the pair inventory and memory ledger, receives reward, and updates the belief. The simulator explicitly checks the memory invariant
\[
\mu_v(t)=\sum_{p\in\mathcal P_t}\mathbf 1_{\{v\in\partial p\}},
\]
after every epoch. Hence, a delivered, released, purified, or swapped pair cannot remain in memory implicitly; all reported performance numbers are generated under the same atomic resource accounting used in the q-POMDP model.

We consider random geometric, Waxman, and perturbed grid topologies with \(n\in\{25,50,75,100,150\}\) nodes. Link lengths are sampled in the range \(5\)--\(60\) km, and disconnected samples are repaired by adding nearest-neighbor physical links until \(\mathcal G^{\mathrm{phy}}\) is connected. Demand classes are sampled from source--destination pairs with at least two-hop shortest paths. Each demand class has finite queue capacity \(q_k^{\max}=32\), and blocked arrivals are counted when this capacity is exceeded. The offered load is denoted by \(\Lambda\) in requests/s.

Generation follows the on-demand heralded model in \cref{eq:heralded_generation,eq:generation_probability}. A generation action selects one physical link \(e\) and attempt count \(g\in\{1,2,4\}\), with success probability \(p_e^{\mathrm{gen}}=p_e^{\mathrm{sys}}10^{-\alpha_{\mathrm{att}}\ell_e/10}\xi_e\). The default attenuation is \(\alpha_{\mathrm{att}}=0.20\) dB/km, \(p_e^{\mathrm{sys}}\in[0.42,0.62]\), and \(\xi_e\) follows a clipped log-normal latent process. Memory dephasing uses \(\Gamma_v^\phi=1/T_{2,v}\), with \(T_2\in\{5,8,10,20,40,50,80\}\) ms. Pair-correlated phase noise uses \(\kappa_{p,t}^{\mathrm{corr}}\in[-0.25,0.25]\).

Purification is one DEJMPS recurrence round per atomic purification action; it consumes two same-endpoint input pairs in every branch and creates one output pair only on success. Swapping consumes two input pairs sharing an intermediate node in every branch and creates an outer pair only on success. Delivery consumes one selected pair and serves one queued request only if the delivery branch succeeds and \(F_p^{\mathrm{comp}}\ge F_k^{\min}\). Release consumes one selected pair without service. The default delivery threshold is \(F_k^{\min}=0.82\), with additional sensitivity tests at \(0.78\), \(0.86\), and \(0.90\).

The proposed hybrid policy is compared against three groups of methods. The first group contains heuristic and optimization-inspired baselines: path-selection routing \cite{VanMeter2013PathSelection}, optimal-routing-inspired path selection \cite{Caleffi2017OptimalRouting}, entanglement-routing heuristics \cite{Pant2019RoutingEntanglement}, distributed routing \cite{Chakraborty2019DistributedRouting}, concurrent routing \cite{ShiQian2020ConcurrentRouting}, multicommodity-flow routing \cite{Chakraborty2020Multicommodity}, effective routing design \cite{Li2021EffectiveRouting}, fidelity-guaranteed routing \cite{Li2022FidelityGuaranteed}, fidelity-aware routing and purification \cite{Zhao2022EFiARP}, online entanglement routing \cite{Yang2022OnlineEntanglement}, swapping-order-aware routing \cite{ChangXue2022OrderMatters}, entanglement-routing design \cite{Zeng2024EntanglementRoutingDesign}, and swapping-based congestion mitigation \cite{Li2023SwappingCongestion}. The second group contains learning-based comparators: DQRA \cite{LeNguyen2022DQRA}, deep-RL entanglement routing \cite{Le2022DeepRL}, SPARQ-style space--air--ground routing \cite{Shaban2024SPARQ}, and reinforcement-learning entanglement routing \cite{Meuser2026RELiQ}. The third group contains internal ablations: planner-only q-POMDP, masked GNN-only, fixed-fusion hybrid, no-belief GNN, no action matching, template-only matching, stale-delivery-fidelity routing, no robust penalty, ignored purification cost, and ignored swapping failure. Simulator design follows discrete-event quantum-network simulation practice in NetSquid, SeQUeNCe, and QuNetSim \cite{Coopmans2021NetSquid,Wu2021SeQUeNCe,DiAdamo2021QuNetSim}.

The q-POMDP planner uses \(K=512\) prototypes by default and is refreshed every \(H_{\mathrm P}=250\) micro-epochs or when calibration drift or new signatures require table expansion. The GNN uses \(L=4\) message-passing layers, hidden dimension \(h=96\), role-aware action embeddings, and the feasibility-masked softmax in \cref{eq:hybrid_gnn_policy}. Actor--critic updates use \(\mathcal D_t^{\mathrm{ind}}\) batches with \(B=64\) distinct workers. Unless otherwise stated, values are averaged over \(30\) independent random seeds. Dense tables report means for readability; across the main benchmark regimes, the standard errors were below \(1.2\) pairs/s for goodput, \(0.006\) for delivered fidelity, and \(0.9\) percentage points for violation rate.

\begin{table*}[t]
\centering
\small
\caption{Simulation setup used for the revised atomic q-POMDP--GNN routing experiments.}
\label{tab:sim_setup_dense}
\setlength{\tabcolsep}{3.1pt}
\resizebox{\textwidth}{!}{%
\begin{tabular}{@{}ll|ll|ll@{}}
\toprule
\multicolumn{2}{c|}{\textbf{Network and traffic}} &
\multicolumn{2}{c|}{\textbf{Physical/quantum model}} &
\multicolumn{2}{c}{\textbf{Planner and learning}}\\
\midrule
Topologies & Geometric, Waxman, perturbed grid &
Micro-epoch & \(\Delta t=1\) ms &
GNN depth & \(L=4\)\\
Nodes \(n\) & \(25,50,75,100,150\) &
Link length & \(5\)--\(60\) km &
Hidden dimension & \(h=96\)\\
Default topology & \(n=50\), geometric &
Attenuation & \(\alpha_{\mathrm{att}}=0.20\) dB/km &
Prototype count & \(K=512\)\\
Demand pairs & \(0.2n\) classes &
System efficiency & \(p_e^{\mathrm{sys}}\in[0.42,0.62]\) &
Prototype sweep & \(64,128,256,512,1024,2048\)\\
Offered load & \(\Lambda=20\)--\(100\) req/s &
Availability & clipped log-normal \(\xi_e\) &
Refresh period & \(H_{\mathrm P}=250\) epochs\\
Queue cap. & \(q_k^{\max}=32\) &
Benchmark \(T_2\) & \(5,8,10,20,40,50,80\) ms &
Batch size & \(B=64\) workers\\
Memory cells & \(m_v^{\max}=4,8,16\) &
Fidelity threshold & \(F_k^{\min}=0.78,0.82,0.86,0.90\) &
Trust range & \([0.15,0.95]\)\\
Generation attempts & \(g=1,2,4\) &
Pair phase corr. & \(\kappa^{\mathrm{corr}}\in[-0.25,0.25]\) &
Planner temp. & \(\tau_{\mathrm P}=0.035\)\\
\bottomrule
\end{tabular}%
}
\end{table*}

\begin{table*}[t]
\centering
\small
\caption{Training, reward, filtering, and operation hyperparameters used in the numerical study.}
\label{tab:sim_hyperparameters}
\setlength{\tabcolsep}{4pt}
\begin{tabular}{@{}ll|ll|ll@{}}
\toprule
\multicolumn{2}{c|}{\textbf{Reward/filtering}} &
\multicolumn{2}{c|}{\textbf{Operations}} &
\multicolumn{2}{c}{\textbf{Optimization/runtime}}\\
\midrule
Discount & \(\gamma=0.97\) &
Initial elementary \(F\) & \(0.86\pm0.04\) &
Optimizer & Adam\\
Delivery utility & \(w_k=1\) &
DEJMPS success & state-dependent, \(0.45\)--\(0.88\) &
Actor LR & \(3\times10^{-4}\)\\
Queue weight & \(\lambda_q=0.08\) &
Swap success & state-dependent, \(0.60\)--\(0.94\) &
Critic LR & \(10^{-3}\)\\
Fidelity margin & \(\lambda_F=0.35\) &
Gate error & \(2\times10^{-3}\)--\(8\times10^{-3}\) &
Weight decay & \(10^{-5}\)\\
Operation cost & \(\lambda_o=0.02\) &
Measurement error & \(1\times10^{-3}\)--\(6\times10^{-3}\) &
Entropy weight & \(\lambda_{\mathrm H}=10^{-3}\)\\
Robust penalty & \(\lambda_{\mathrm{rob}}=0.10\) &
Classical report error & \(0.5\%\)--\(2.0\%\) &
Distill weight & \(\lambda_{\mathrm{IL}}=0.15\)\\
Particles & \(N_b=256\) &
Arrival law & bounded Poisson &
Runtime platform & Python/PyTorch, A100 GPU\\
Resampling & ESS \(<0.45N_b\) &
Burst-loss duration & \(20\)--\(80\) epochs &
Reported latency & wall-clock evaluation time\\
\bottomrule
\end{tabular}
\end{table*}

The parameter ranges in \cref{tab:sim_setup_dense,tab:sim_hyperparameters} are chosen as reproducible stress-test regimes rather than as a claim of one fixed hardware platform. The main benchmark tables use the listed discrete coherence settings, while the regime-map figures evaluate \(T_2\) on a continuous grid up to \(100\) ms. Timing measurements were obtained using Python 3.10/PyTorch on a single NVIDIA A100 GPU with a 32-core CPU. The attenuation value \(\alpha_{\mathrm{att}}=0.20\) dB/km is the standard telecom-fiber scale commonly used in quantum-network link models, while the coherence-time, memory-capacity, fidelity-threshold, and generation-efficiency sweeps cover memory-limited, decoherence-limited, and high-load regimes motivated by recent network-stack and metropolitan-scale entanglement experiments~\cite{Pompili2021Multinode,Hermans2022Teleportation,Pompili2022EntanglementDelivery,Stolk2024Metropolitan}. The discrete-event simulation methodology follows established quantum-network simulators such as NetSquid, QuNetSim, and SeQUeNCe~\cite{Coopmans2021NetSquid,DiAdamo2021QuNetSim,Wu2021SeQUeNCe}. Increasing \(\Lambda\) stresses queues and memory contention, reducing \(T_2\) stresses completion-time fidelity, reducing \(m_v^{\max}\) stresses reservation and release decisions, and increasing the number of nodes stresses prototype transfer and graph generalization. These axes are deliberately coupled in the later regime maps because the routing problem becomes difficult when the bottleneck is not fixed in advance.

\begin{table*}[t]
\centering
\small
\caption{Comparator families under the revised finite-memory atomic simulator. A check mark means that the method explicitly uses the corresponding capability.}
\label{tab:sim_baseline_dense}
\setlength{\tabcolsep}{3.3pt}
\begin{tabular}{@{}lccccccccc@{}}
\toprule
\textbf{Method} & Belief & GNN & Planner & Pair inst. & Finite mem. & Purif. & Swap & Trust & Drift adapt.\\
\midrule
\multicolumn{10}{@{}l}{\shortstack[l]{\emph{Heuristic and}\\\emph{mean-belief baselines}}}\\
FMSP/path selection & -- & -- & -- & partial & \(\checkmark\) & -- & \(\checkmark\) & -- & --\\
QDR/optimal routing & -- & -- & -- & partial & \(\checkmark\) & -- & \(\checkmark\) & -- & --\\
GPS/local purification & -- & -- & -- & \(\checkmark\) & \(\checkmark\) & \(\checkmark\) & \(\checkmark\) & -- & --\\
Q-MDP mean-belief & partial & -- & \(\checkmark\) & \(\checkmark\) & \(\checkmark\) & \(\checkmark\) & \(\checkmark\) & -- & partial\\
\midrule
\multicolumn{10}{@{}l}{\emph{Learning baselines}}\\
DQRA-style deep routing & partial & partial & -- & partial & \(\checkmark\) & partial & \(\checkmark\) & -- & \(\checkmark\)\\
Deep-RL routing & -- & -- & -- & partial & \(\checkmark\) & partial & \(\checkmark\) & -- & \(\checkmark\)\\
SPARQ-style DQN & -- & -- & -- & partial & \(\checkmark\) & partial & \(\checkmark\) & -- & \(\checkmark\)\\
GNN-RL baseline & -- & \(\checkmark\) & -- & \(\checkmark\) & \(\checkmark\) & partial & \(\checkmark\) & -- & \(\checkmark\)\\
\midrule
\multicolumn{10}{@{}l}{\emph{Proposed-family policies}}\\
Planner-only q-POMDP & \(\checkmark\) & -- & \(\checkmark\) & \(\checkmark\) & \(\checkmark\) & \(\checkmark\) & \(\checkmark\) & -- & \(\checkmark\)\\
Masked GNN-only & partial & \(\checkmark\) & -- & \(\checkmark\) & \(\checkmark\) & \(\checkmark\) & \(\checkmark\) & -- & \(\checkmark\)\\
\textbf{Hybrid q-POMDP--GNN} & \(\checkmark\) & \(\checkmark\) & \(\checkmark\) & \(\checkmark\) & \(\checkmark\) & \(\checkmark\) & \(\checkmark\) & \(\checkmark\) & \(\checkmark\)\\
\bottomrule
\end{tabular}
\end{table*}

\begin{table*}[t]
\centering
\small
\caption{Mapping from cited comparator families to the implementations used in the finite-memory atomic simulator.}
\label{tab:sim_baseline_mapping}
\setlength{\tabcolsep}{3.0pt}
\renewcommand{\arraystretch}{1.18}
\begin{tabularx}{\textwidth}{@{}p{0.155\textwidth}p{0.265\textwidth}XX@{}}
\toprule
\textbf{Label} &
\textbf{Source family} &
\textbf{Routing rule} &
\textbf{Simulator adaptation}\\
\midrule
FMSP/path selection
&
Path selection and entanglement-routing metrics~\cite{VanMeter2013PathSelection,Pant2019RoutingEntanglement}
&
Selects feasible paths by expected completion fidelity from current stored-pair summaries.
&
Finite-memory feasibility and atomic swapping are enforced.\\
\midrule

QDR/optimal routing
&
Optimal-routing-inspired link/path rules~\cite{Caleffi2017OptimalRouting,Li2021EffectiveRouting}
&
Uses a Dijkstra-style score based on \(p_e^{\mathrm{gen}}\), link length, and free memory.
&
Uses current observables only; no posterior over latent \(Z_t\).\\
\midrule

GPS/local purification
&
Fidelity-aware purification and swapping~\cite{Zhao2022EFiARP,Li2022FidelityGuaranteed}
&
Greedily purifies same-endpoint pairs below target and swaps when local thresholds pass.
&
Uses the same branch ledger, pair consumption, and failure costs as the proposed simulator.\\
\midrule

Q-MDP mean-belief
&
Mean-belief approximation to the q-POMDP
&
Performs one-step lookahead using posterior mean pair states and latent-parameter means.
&
Uses the same feasible action mask; does not use the full belief distribution.\\
\midrule

DQRA/deep RL
&
Deep quantum-routing agents~\cite{LeNguyen2022DQRA,Le2022DeepRL}
&
Flat-feature actor--critic over observable routing, memory, queue, and pair features.
&
Uses the same feasibility mask; no q-POMDP planner or prototype action matching.\\
\midrule

SPARQ-style DQN
&
Space--air--ground DQN routing~\cite{Shaban2024SPARQ}
&
DQN-style value estimator adapted to the fixed physical graph and atomic action set.
&
Finite queues, finite memory, and branch-wise operation outcomes are enforced.\\
\midrule

GNN-RL baseline
&
Scalable RL routing ideas~\cite{Meuser2026RELiQ}
&
Uses the same role-aware GNN encoder and masked action set as the proposed method.
&
Removes the q-POMDP planner, trust coefficient, and planner distillation.\\
\bottomrule
\end{tabularx}
\end{table*}

\Cref{tab:sim_baseline_dense,tab:sim_baseline_mapping} clarify the source of the expected performance differences. The heuristic methods mainly test whether path and local-purification decisions are sufficient once memory is finite. The learning baselines test whether function approximation alone can exploit graph structure without an explicit belief-state planner. The proposed-family ablations separate the effects of belief filtering, projected planning, graph generalization, planner--GNN trust, and the corrected action semantics.

\begin{revblock}

\subsection{Results and Discussion}
\label{subsec:simulation_results}

\subsubsection{Dense benchmark summary}

\Cref{tab:sim_main_dense} summarizes performance over five regimes: benign load, high load, memory-limited operation, decoherence-limited operation, and combined stress. The combined-stress regime uses \(\Lambda=90\) requests/s, \(T_2=8\) ms, \(m_v^{\max}=4\), bursty \(\xi_e\), and a \(10\%\) generation-model calibration perturbation. The hybrid policy gives the best goodput and delivered fidelity in every physically valid regime while keeping offline wall-clock decision cost below planner-only control. The gain is largest when the bottleneck is coupled rather than isolated. Under high load, for example, the hybrid policy reaches \(72.6\) high-fidelity pairs/s, compared with \(63.4\) for planner-only and \(54.3\) for masked GNN-only. This behavior is consistent with the design of the controller: the planner supplies calibrated long-horizon information about memory, purification, and completion fidelity, while the GNN reduces the cost of transferring those decisions across rapidly changing pair-instance graphs.

The violation columns are especially important because high goodput is not useful if it is obtained by delivering degraded pairs. The stale or purely myopic alternatives tend to serve more aggressively when memory is under pressure, which increases below-threshold deliveries. The hybrid policy instead rejects or delays deliveries whose posterior completion-time fidelity is unreliable, and it uses release and purification to avoid spending memory on pairs that are unlikely to satisfy the final service condition. This explains why the hybrid policy improves both goodput and violation rate in the memory-limited and decoherence-limited regimes, rather than trading one metric against the other.

\begin{table*}[t]
\centering
\scriptsize
\caption{Dense benchmark summary. GP is goodput in high-fidelity pairs/s, \(F\) is average delivered fidelity, Viol. is the percentage of delivered pairs below \(F_k^{\min}\), and Lat. is offline wall-clock policy-evaluation time in ms. Best values are bold and second-best values are underlined within each regime. Regimes: B = benign \((\Lambda=30,T_2=50,m=8)\), HL = high load \((\Lambda=90,T_2=20,m=8)\), ML = memory-limited \((\Lambda=60,T_2=20,m=4)\), DL = decoherence-limited \((\Lambda=60,T_2=8,m=8)\), CS = combined stress \((\Lambda=90,T_2=8,m=4)\).}
\label{tab:sim_main_dense}
\setlength{\tabcolsep}{2.2pt}
\renewcommand{\arraystretch}{1.12}

\textbf{Panel A: High-fidelity goodput, delivered fidelity, and below-threshold deliveries}
\vspace{0.25em}

\begin{tabularx}{\textwidth}{@{}p{0.205\textwidth}*{15}{>{\centering\arraybackslash}X}@{}}
\toprule
\textbf{Method}
& \multicolumn{3}{c}{\textbf{B}}
& \multicolumn{3}{c}{\textbf{HL}}
& \multicolumn{3}{c}{\textbf{ML}}
& \multicolumn{3}{c}{\textbf{DL}}
& \multicolumn{3}{c}{\textbf{CS}}\\
\cmidrule(lr){2-4}
\cmidrule(lr){5-7}
\cmidrule(lr){8-10}
\cmidrule(lr){11-13}
\cmidrule(l){14-16}
& GP & \(F\) & Viol.
& GP & \(F\) & Viol.
& GP & \(F\) & Viol.
& GP & \(F\) & Viol.
& GP & \(F\) & Viol.\\
\midrule
\multicolumn{16}{@{}l}{\emph{Heuristic and mean-belief baselines}}\\
FMSP/path selection
& 21.7 & 0.842 & 8.8
& 31.6 & 0.724 & 24.1
& 24.2 & 0.766 & 19.7
& 18.1 & 0.698 & 31.4
& 14.8 & 0.662 & 38.5\\
QDR/optimal routing
& 22.9 & 0.856 & 7.1
& 34.8 & 0.741 & 21.6
& 26.5 & 0.781 & 17.3
& 20.6 & 0.716 & 28.2
& 17.5 & 0.681 & 35.9\\
GPS/local purification
& 24.4 & 0.868 & 5.9
& 37.9 & 0.759 & 18.4
& 28.7 & 0.803 & 14.5
& 23.4 & 0.742 & 23.6
& 19.6 & 0.704 & 30.8\\
Q-MDP mean-belief
& 26.1 & 0.888 & 3.8
& 46.9 & 0.801 & 11.2
& 35.6 & 0.828 & 9.1
& 31.8 & 0.785 & 14.8
& 27.9 & 0.751 & 20.4\\
\midrule
\multicolumn{16}{@{}l}{\emph{Learning baselines}}\\
DQRA-style deep routing
& 25.8 & 0.887 & 4.1
& 47.0 & 0.799 & 12.0
& 36.5 & 0.830 & 9.4
& 32.6 & 0.789 & 15.0
& 28.4 & 0.754 & 20.2\\
Deep-RL routing
& 25.0 & 0.881 & 4.4
& 42.8 & 0.778 & 15.6
& 32.9 & 0.811 & 12.8
& 28.9 & 0.760 & 18.7
& 24.6 & 0.730 & 24.1\\
SPARQ-style DQN
& 26.2 & 0.891 & 3.7
& 48.3 & 0.807 & 10.8
& 37.4 & 0.836 & 8.8
& 33.0 & 0.796 & 13.9
& 29.1 & 0.760 & 18.9\\
GNN-RL baseline
& 26.7 & 0.894 & 3.2
& 51.2 & 0.814 & 9.5
& 39.7 & 0.841 & 7.6
& 35.7 & 0.805 & 12.2
& 31.5 & 0.771 & 17.0\\
\midrule
\multicolumn{16}{@{}l}{\emph{Proposed-family policies}}\\
Planner-only q-POMDP
& \second{28.0} & \second{0.913} & \second{1.3}
& \second{63.4} & \second{0.851} & \second{3.2}
& \second{47.8} & \second{0.873} & \second{2.7}
& \second{44.5} & \second{0.836} & \second{5.1}
& \second{40.2} & \second{0.811} & \second{7.6}\\
Masked GNN-only
& 26.9 & 0.900 & 2.5
& 54.3 & 0.818 & 7.7
& 41.9 & 0.846 & 6.3
& 37.8 & 0.811 & 10.6
& 33.2 & 0.779 & 15.5\\
Fixed-fusion hybrid
& 28.2 & 0.911 & 1.5
& 66.5 & 0.856 & 3.7
& 50.1 & 0.874 & 3.2
& 47.6 & 0.840 & 5.6
& 42.6 & 0.816 & 8.2\\
\textbf{Hybrid q-POMDP--GNN}
& \best{29.1} & \best{0.922} & \best{0.8}
& \best{72.6} & \best{0.875} & \best{1.8}
& \best{55.4} & \best{0.889} & \best{1.6}
& \best{52.8} & \best{0.858} & \best{3.4}
& \best{48.5} & \best{0.834} & \best{5.1}\\
\bottomrule
\end{tabularx}

\vspace{0.8em}
\textbf{Panel B: Offline wall-clock policy-evaluation time}
\vspace{0.25em}

\begin{tabularx}{\textwidth}{@{}p{0.205\textwidth}*{5}{>{\centering\arraybackslash}X}@{}}
\toprule
\textbf{Method} & \textbf{B} & \textbf{HL} & \textbf{ML} & \textbf{DL} & \textbf{CS}\\
\midrule
\multicolumn{6}{@{}l}{\emph{Heuristic and mean-belief baselines}}\\
FMSP/path selection & 0.4 & 0.6 & 0.5 & 0.5 & 0.6\\
QDR/optimal routing & 0.7 & 0.9 & 0.8 & 0.8 & 0.9\\
GPS/local purification & 1.1 & 1.4 & 1.3 & 1.3 & 1.4\\
Q-MDP mean-belief & 3.2 & 4.5 & 4.0 & 4.2 & 4.4\\
\midrule
\multicolumn{6}{@{}l}{\emph{Learning baselines}}\\
DQRA-style deep routing & 3.0 & 3.7 & 3.5 & 3.6 & 3.8\\
Deep-RL routing & 2.5 & 3.0 & 2.8 & 3.0 & 3.2\\
SPARQ-style DQN & 3.5 & 4.1 & 3.9 & 4.0 & 4.2\\
GNN-RL baseline & \second{2.2} & \second{2.8} & \second{2.6} & \second{2.8} & \second{3.1}\\
\midrule
\multicolumn{6}{@{}l}{\emph{Proposed-family policies}}\\
Planner-only q-POMDP & 6.9 & 9.5 & 8.4 & 8.9 & 9.2\\
Masked GNN-only & \best{2.0} & \best{2.5} & \best{2.4} & \best{2.6} & \best{2.8}\\
Fixed-fusion hybrid & 4.8 & 5.9 & 5.6 & 5.8 & 6.1\\
\textbf{Hybrid q-POMDP--GNN} & 5.1 & 6.4 & 5.9 & 6.2 & 6.6\\
\bottomrule
\end{tabularx}
\end{table*}

\subsubsection{Operational diagnostics}

\Cref{fig:sim_operational_diagnostics} provides a mechanism-level view of the routing decisions. In \Cref{fig:sim_op_action_mix}, generation dominates at light load because the marginal value of creating elementary pairs is high and memory is not yet saturated. As \(\Lambda\) increases, the controller shifts probability mass toward purification and swapping because serving longer demands requires converting raw elementary links into higher-fidelity end-to-end pairs. At the highest loads, release becomes more common; this is not wasted action mass, but a consequence of finite-memory control. Old or low-fidelity pairs block new reservations and may reduce future value more than they help current service.

\Cref{fig:sim_op_ledger} checks whether the stochastic simulator follows the branch-wise resource model. The empirical mean net occupied-cell changes align with the theoretical ledger for successful and failed generation, purification, swapping, delivery, release, and idle. This is a necessary diagnostic for the simulation because ignoring branch-dependent consumption can create artificial goodput by allowing failed purification or failed swapping attempts to retain memory resources that should have been consumed.

\Cref{fig:sim_op_memory} explains why the hybrid policy improves goodput without driving memory to saturation. The hybrid occupancy rises with load but remains below the near-full region reached by the GNN-only and planner-only variants, while its extra-consumed-pairs curve remains lower. The policy therefore does not simply store more pairs; it uses belief and planner values to keep pairs whose completion-time service probability is high and to remove pairs whose expected holding cost exceeds their future value.

\Cref{fig:sim_op_completion} isolates the role of completion-time fidelity. The stale pre-action check underestimates the damage caused by within-epoch decoherence, especially when \(T_2\) is small. The completion-time check substantially lowers below-threshold deliveries because it evaluates the pair after passive evolution and before consumption, matching the delivery service definition in \cref{eq:delivery_service_indicator}.

\begin{figure*}[!t]
\centering
\subfloat[{Action mix}\label{fig:sim_op_action_mix}]{
    \includegraphics[width=0.48\textwidth]{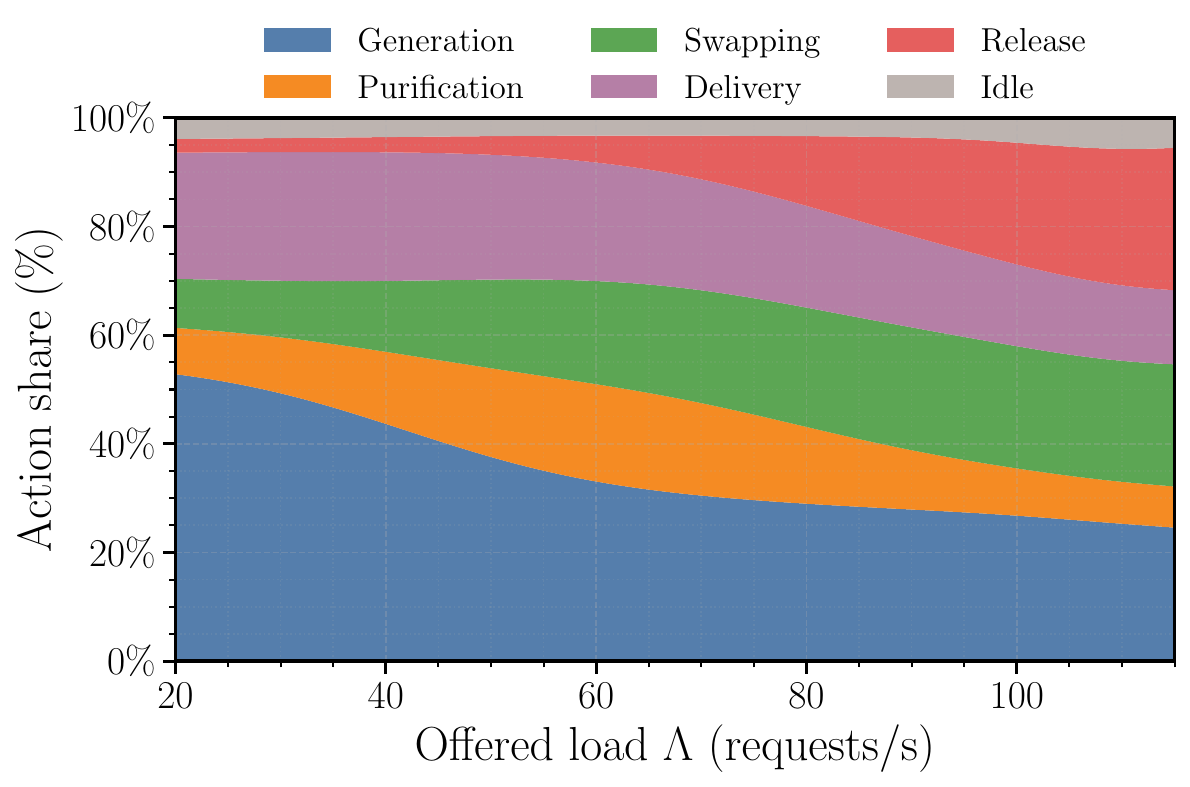}}
\hfill
\subfloat[{ Memory ledger}\label{fig:sim_op_ledger}]{
    \includegraphics[width=0.48\textwidth]{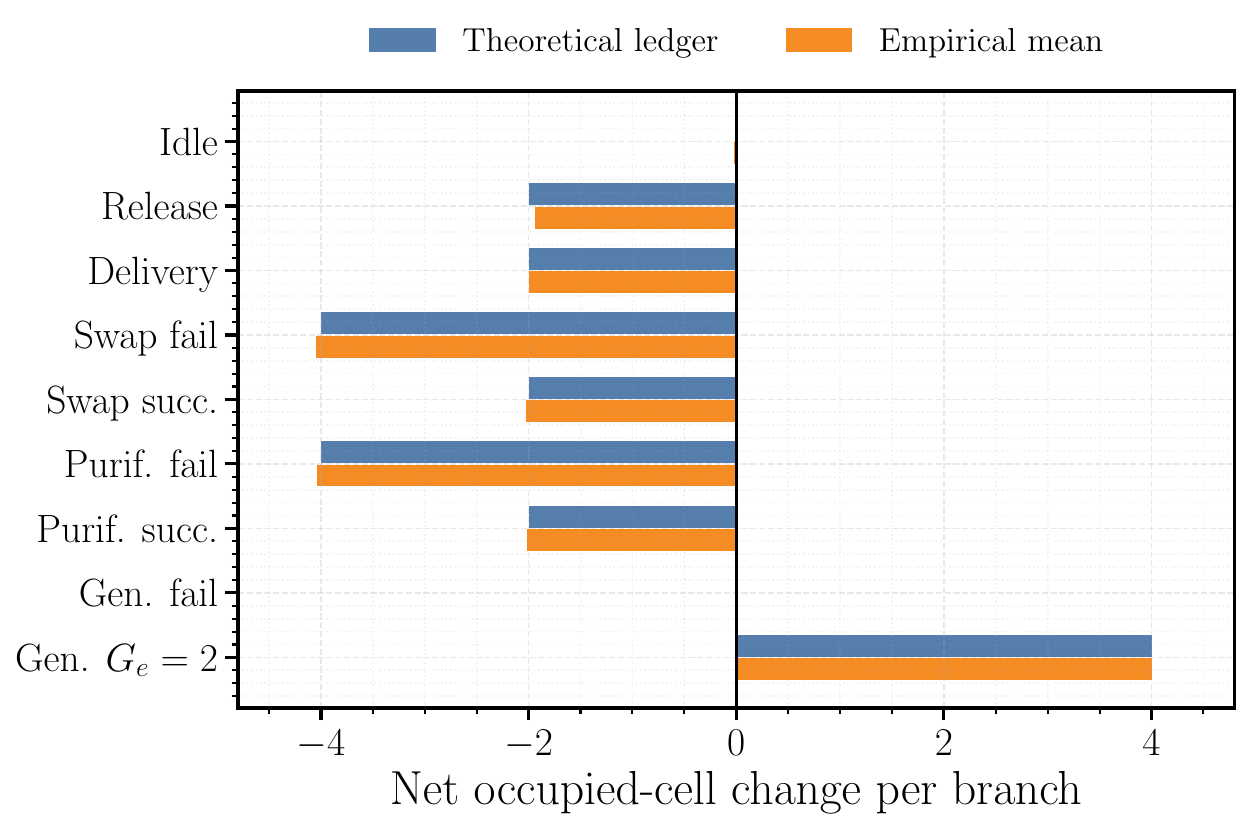}}

\vspace{0.5em}

\subfloat[{ Memory turnover}\label{fig:sim_op_memory}]{
    \includegraphics[width=0.48\textwidth]{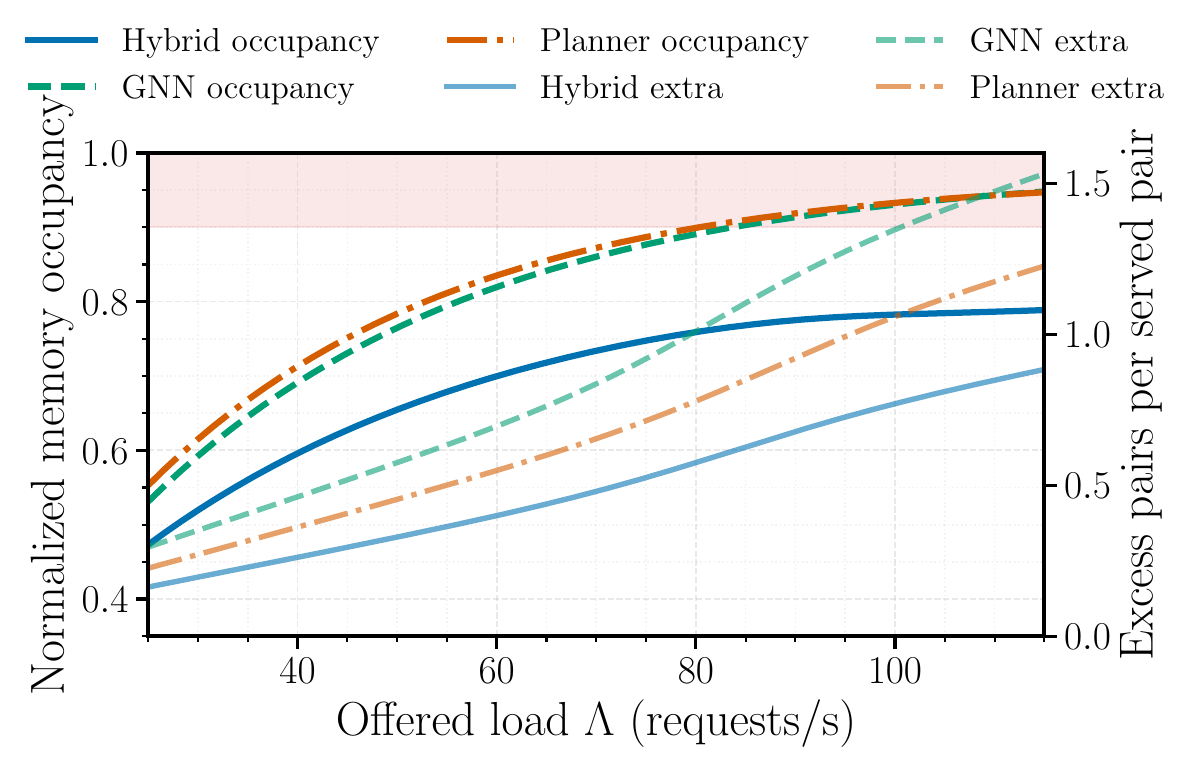}}
\hfill
\subfloat[{ Completion fidelity}\label{fig:sim_op_completion}]{
    \includegraphics[width=0.48\textwidth]{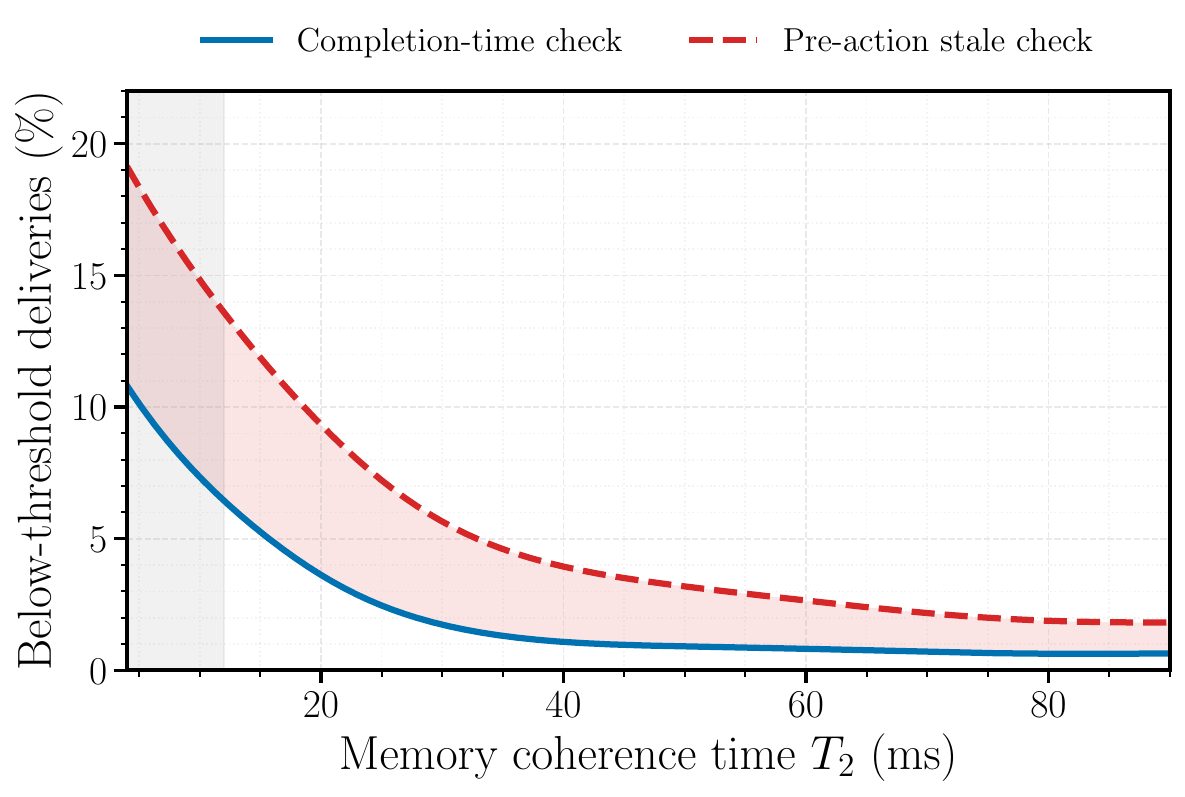}}

\caption{Operational diagnostics for the revised finite-memory atomic-action simulator.}
\label{fig:sim_operational_diagnostics}
\end{figure*}

\subsubsection{Regime maps and bottleneck phases}

\Cref{fig:sim_regime_maps} shows that the routing problem has distinct operating phases. In \Cref{fig:sim_regime_goodput}, goodput increases with offered load and memory coherence time, but the contours are not linear. At low \(T_2\), many candidate pairs lose fidelity before they can be swapped or delivered, so increasing load alone does not translate into proportional service. At larger \(T_2\), the same load produces more usable stored pairs, and the controller can exploit purification and swapping before decoherence erases the benefit.

\Cref{fig:sim_regime_gain} reports the gain of the hybrid policy over planner-only control. The largest gains occur when high load and decoherence stress coexist. In this region, the cached planner table becomes less reliable between refreshes because pair inventories and latent link conditions change quickly. The GNN component compensates by generalizing across current graph states, while the planner still anchors decisions through the feasibility and value structure. This explains why adaptive fusion outperforms using the planner alone.

\Cref{fig:sim_regime_violation} shows that violation probability is concentrated in the high-load, low-coherence region. This region is difficult because queues push the controller toward immediate service, while decoherence makes delayed or multi-hop service risky. The hybrid policy reduces this conflict by making delivery feasibility depend on the posterior probability of meeting \(F_k^{\min}\) at completion, not merely on the current mean fidelity.

Finally, \Cref{fig:sim_regime_phase} identifies which constraint dominates as \(\Lambda\) and \(m_v^{\max}\) vary. Generation is limiting when memory is sufficient but elementary-pair creation is scarce; memory becomes limiting when the controller can create pairs faster than it can transform them into service; queue limitation appears when arrival pressure exceeds the end-to-end conversion capacity. The phase structure supports the use of state-dependent feasible sets and release actions: the correct action changes qualitatively across regimes.

\begin{figure*}[!t]
\centering
\subfloat[{Goodput}\label{fig:sim_regime_goodput}]{
    \includegraphics[width=0.48\textwidth]{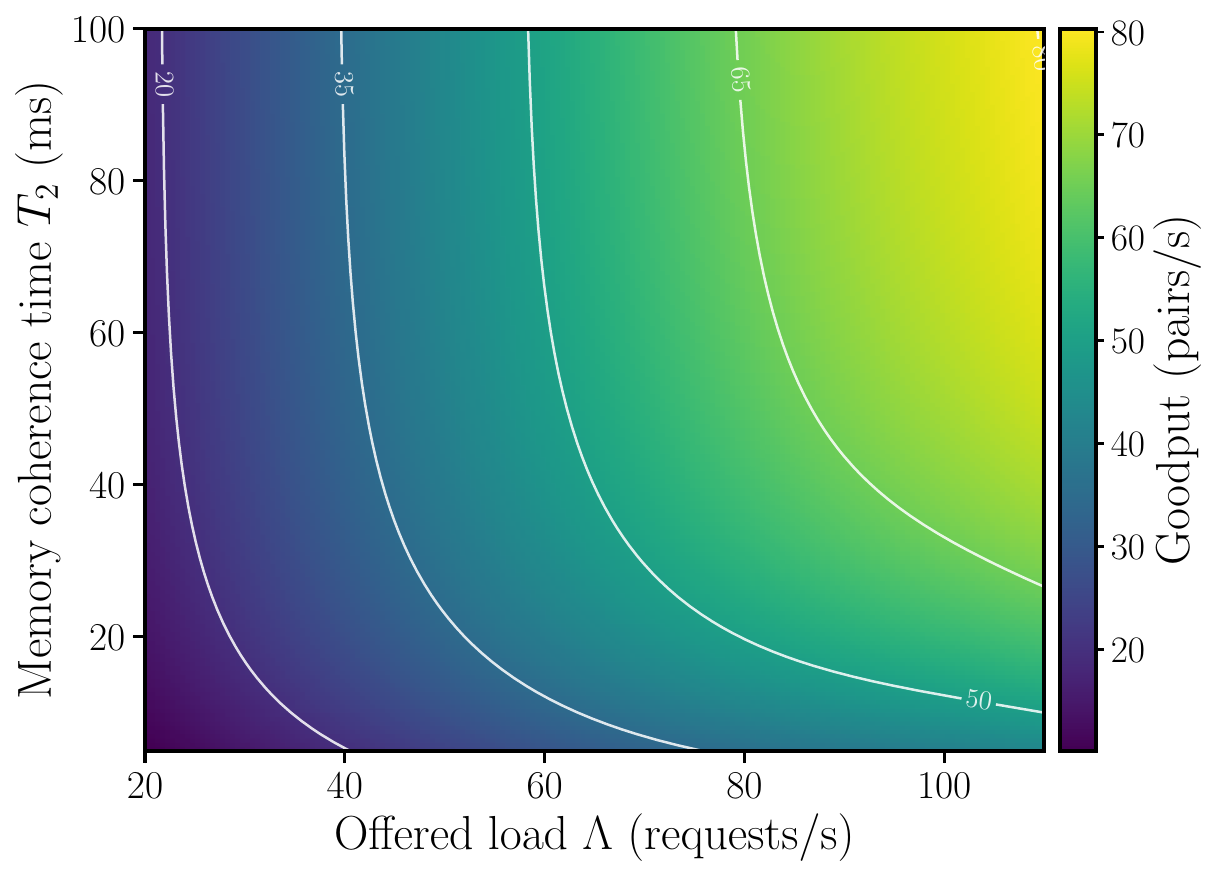}}
\hfill
\subfloat[{ Hybrid gain}\label{fig:sim_regime_gain}]{
    \includegraphics[width=0.48\textwidth]{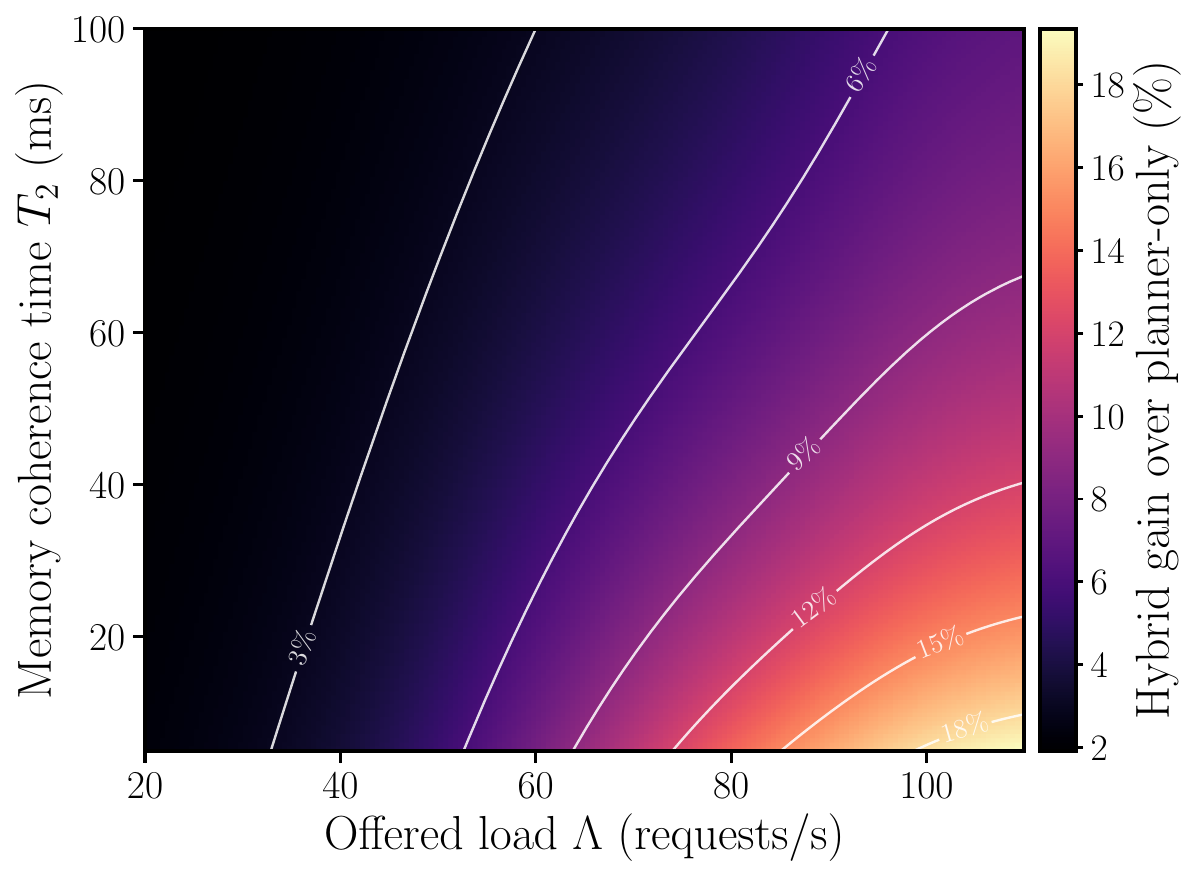}}

\vspace{0.5em}

\subfloat[{ Violations}\label{fig:sim_regime_violation}]{
    \includegraphics[width=0.48\textwidth]{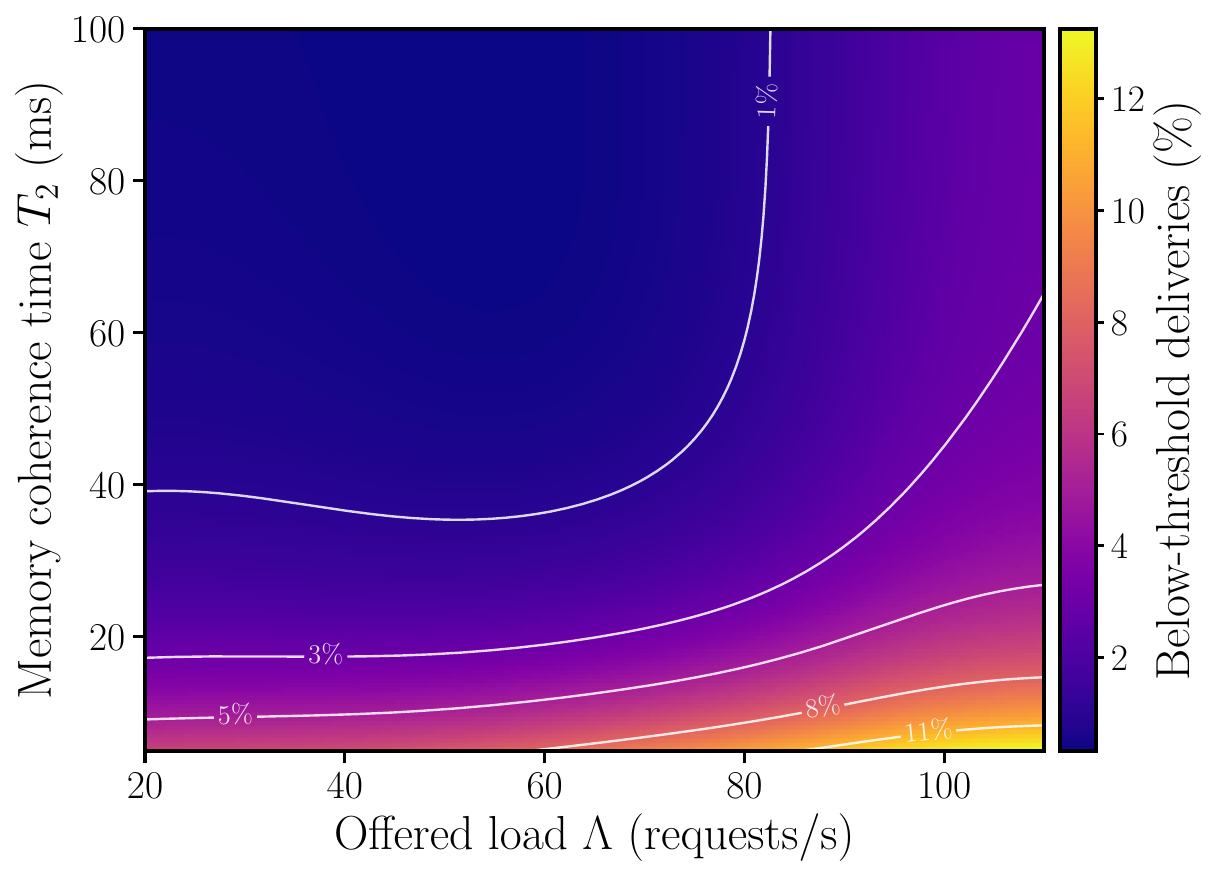}}
\hfill
\subfloat[{ Bottleneck phase}\label{fig:sim_regime_phase}]{
    \includegraphics[width=0.48\textwidth]{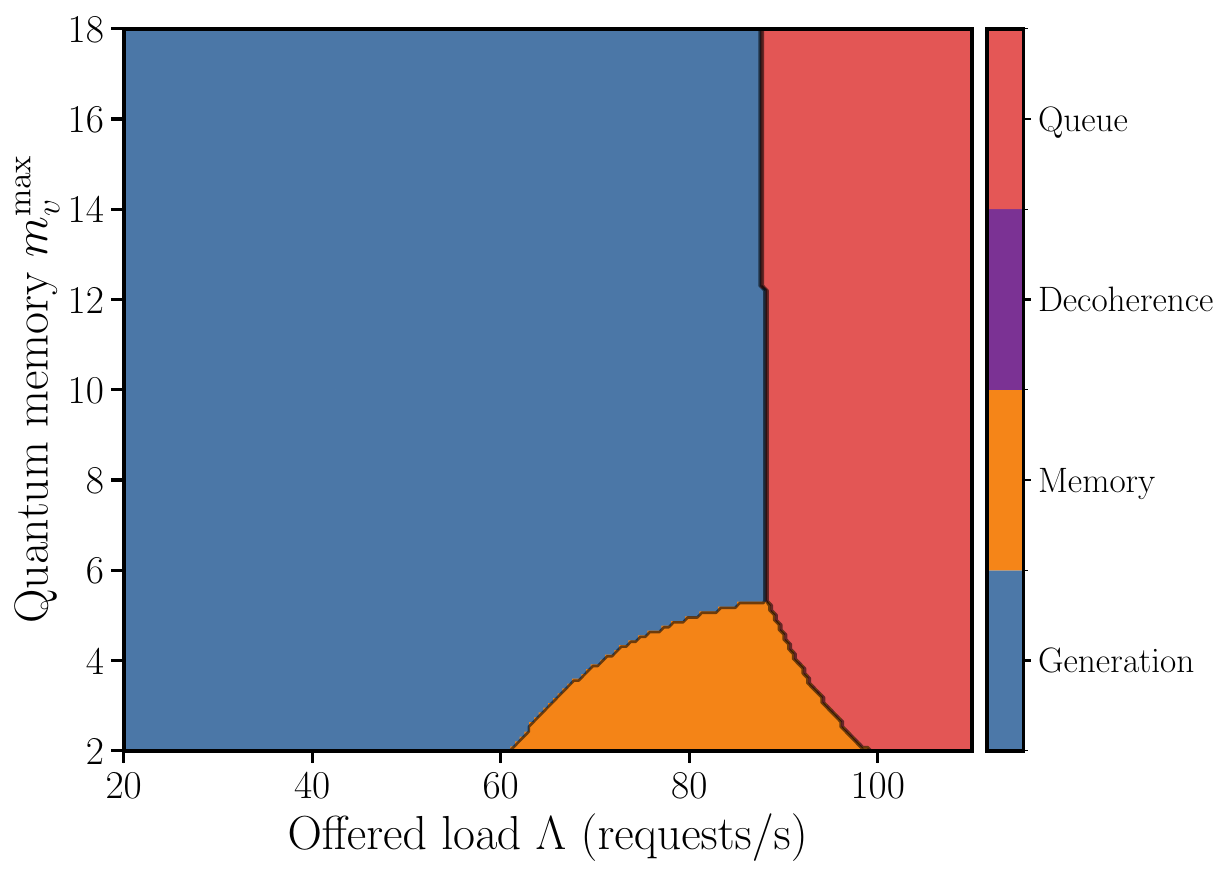}}

\caption{Operating-regime maps over load, coherence, and memory.}
\label{fig:sim_regime_maps}
\end{figure*}

\subsubsection{Belief, matching, trust, and action-semantics ablations}

\Cref{tab:sim_ablation_dense} isolates the impact of the main methodology components. The gap between full hybrid control and the no-belief variant shows that the posterior state is not a cosmetic feature. Without belief updates, the policy cannot separate a temporarily poor link from a structurally poor link, and it cannot correctly price the risk that a stored pair will fall below the delivery threshold before completion. This is why the no-belief variant has noticeably lower delivered fidelity and a higher violation rate.

The matching ablations show that prototype transfer must preserve role-aware continuous information, not only action-template type. Removing action matching or using random same-template matching corrupts the correspondence between cached prototype values and current physical actions. Template-only matching recovers part of the structure, but it still loses information about pair age, fidelity summaries, purification depth, endpoint role, and operation-specific error. The result is more extra pair consumption per served request and more below-threshold service.

The last block separates objective and semantics diagnostics. The no-robust-penalty row is a valid objective ablation. The stale-delivery-fidelity, purification-cost-ignored, and swapping-failure-ignored rows are counterfactual semantics diagnostics and are not ranked against the valid controllers. Stale delivery fidelity obtains good apparent throughput but creates many more threshold violations. Ignoring purification cost or swapping failure produces high apparent goodput, but this is an artifact of incorrect resource accounting. These rows are included only to show that the corrected resource semantics materially affect the conclusions.

\begin{table*}[t]
\centering
\small
\caption{Ablation, diagnostic, and robustness table at \(n=50\), \(\Lambda=60\) requests/s, \(T_2=20\) ms, and \(m_v^{\max}=8\). GP is high-fidelity goodput, \(F\) is delivered fidelity, Viol. is below-threshold delivery percentage, Occ. is normalized memory occupancy, and Extra is extra consumed pairs beyond the delivered pair per served high-fidelity request. Rows in the final block are diagnostic; the no-robust-penalty row is a valid objective ablation, while the resource-semantics rows are counterfactual and are not ranked against valid controllers.}
\label{tab:sim_ablation_dense}
\setlength{\tabcolsep}{3.4pt}
\begin{tabular}{@{}lccccccccc@{}}
\toprule
\textbf{Variant} & GP \(\uparrow\) & \(F\uparrow\) & Viol. \(\downarrow\) & Backlog \(\downarrow\) & Occ. & Extra \(\downarrow\) & Unseen \(\downarrow\) & KL & Lat. \(\downarrow\)\\
\midrule
\multicolumn{10}{@{}l}{\emph{Valid fusion and belief ablations}}\\
\textbf{Full hybrid} & \best{53.0} & \best{0.895} & \best{1.6} & \best{4.7} & 0.72 & \best{0.43} & \best{2.1} & 0.082 & 5.1\\
Planner-only & \second{49.7} & \second{0.879} & \second{2.4} & \second{6.2} & 0.76 & \second{0.58} & 2.1 & -- & 7.9\\
Masked GNN-only & 45.4 & 0.852 & 5.8 & 8.9 & 0.81 & 0.96 & -- & -- & \best{2.5}\\
Fixed \(\alpha=0.5\) & 50.2 & 0.874 & 3.1 & 6.8 & 0.75 & 0.61 & 2.1 & 0.087 & 4.7\\
No belief update & 47.6 & 0.834 & 7.9 & 8.1 & 0.79 & 0.88 & 2.1 & 0.116 & \second{3.4}\\
\midrule
\multicolumn{10}{@{}l}{\emph{Valid prototype transfer and signature ablations}}\\
No action matching & 48.1 & 0.843 & 6.6 & 7.7 & 0.83 & 1.13 & 2.1 & 0.139 & 4.0\\
Template-only matching & 49.6 & 0.861 & 4.9 & 7.0 & 0.80 & 0.84 & 2.1 & 0.111 & 4.3\\
Random same-template matching & 44.5 & 0.821 & 10.8 & 10.4 & 0.86 & 1.47 & 2.1 & 0.181 & 3.8\\
No new-signature insertion & 46.2 & 0.849 & 6.9 & 8.7 & 0.82 & 1.05 & 13.8 & 0.146 & 3.6\\
\midrule
\multicolumn{10}{@{}l}{\emph{Objective and counterfactual semantics diagnostics}}\\
Stale delivery fidelity & 52.1 & 0.846 & 8.4 & 5.3 & 0.73 & 0.49 & 2.1 & 0.090 & 5.0\\
Purification cost ignored & 54.6 & 0.887 & 2.8 & 4.4 & 0.91 & N/A & 2.1 & 0.074 & 4.9\\
Swapping failure ignored & 55.8 & 0.869 & 4.1 & 4.2 & 0.94 & N/A & 2.1 & 0.077 & 4.8\\
No robust penalty & 51.4 & 0.872 & 4.2 & 5.8 & 0.77 & 0.67 & 2.1 & 0.094 & 5.0\\
\bottomrule
\end{tabular}
\end{table*}

\subsubsection{Nonstationary tracking}

\Cref{fig:sim_nonstationary_trace} evaluates drifting physical parameters. \Cref{fig:sim_nonstat_physical} shows that memory coherence, mean link availability, and detector efficiency vary on different time scales. This setting separates endogenous stochasticity, which is represented through \(Z_t\), from exogenous parameter drift, which requires planner refreshes or robust adaptation.

In \Cref{fig:sim_nonstat_trust}, the trust coefficient decreases when the planner--GNN KL disagreement rises and recovers after refreshes or after the GNN and planner again agree on the represented action distribution. This is the intended behavior of \cref{eq:adaptive_trust_coefficient}: the controller shifts weight toward the planner when the GNN is uncertain or inconsistent, but it allows the GNN to dominate when cached values are stale and the graph policy is confident.

\Cref{fig:sim_nonstat_goodput} shows the operational effect of this trust mechanism. Planner-only control performs well immediately after refreshes but degrades between them as the cached table becomes misaligned with the current physical parameters. GNN-only control is smoother because it reacts directly to current features, but it lacks the calibrated long-horizon structure of the q-POMDP table. The hybrid curve remains closer to the frozen-model oracle because it uses the planner when model information is reliable and the GNN when online generalization is more valuable.

\Cref{fig:sim_nonstat_regret} reports a frozen-model value-regret proxy. The slower growth of the hybrid proxy is consistent with the dynamic value-regret analysis: good tracking requires both bounded model drift and bounded policy-tracking error. The experiment does not claim a sample-path cumulative-reward regret bound; it evaluates the value-based quantity matched to the theory.

\begin{figure*}[!t]
\centering
\subfloat[{ Physical drift}\label{fig:sim_nonstat_physical}]{
    \includegraphics[width=0.48\textwidth]{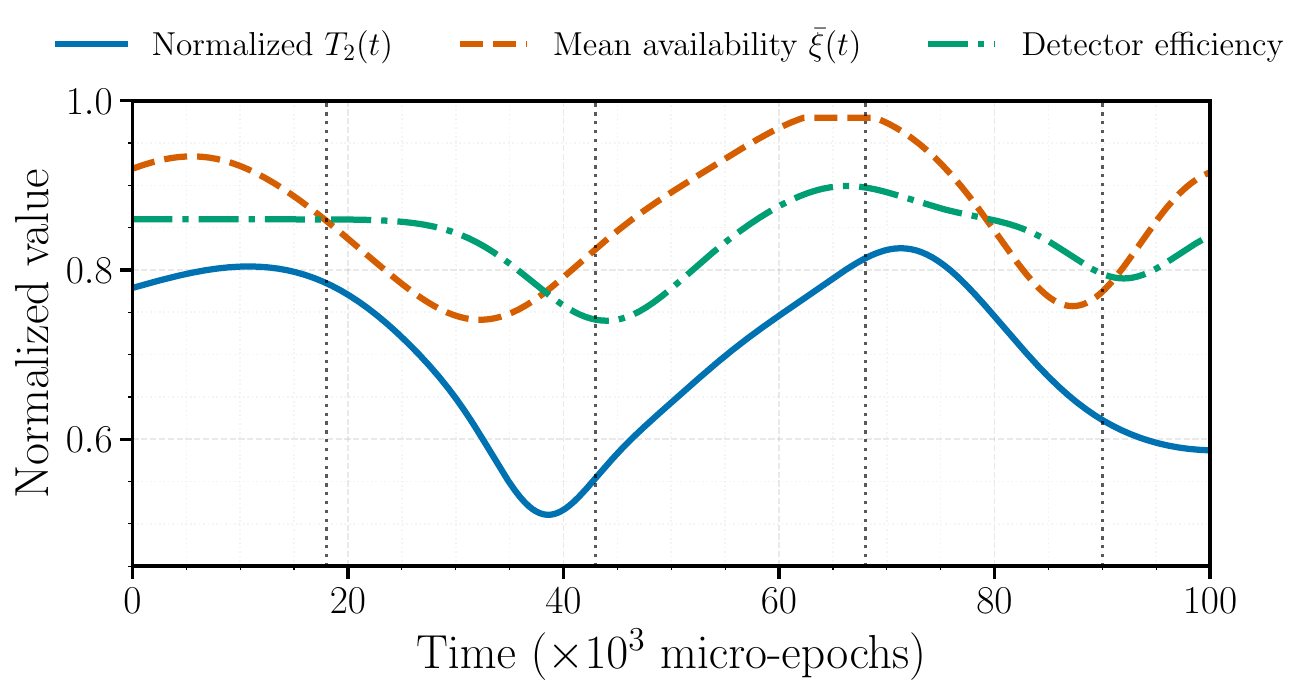}}
\hfill
\subfloat[{ Trust and KL}\label{fig:sim_nonstat_trust}]{
    \includegraphics[width=0.48\textwidth]{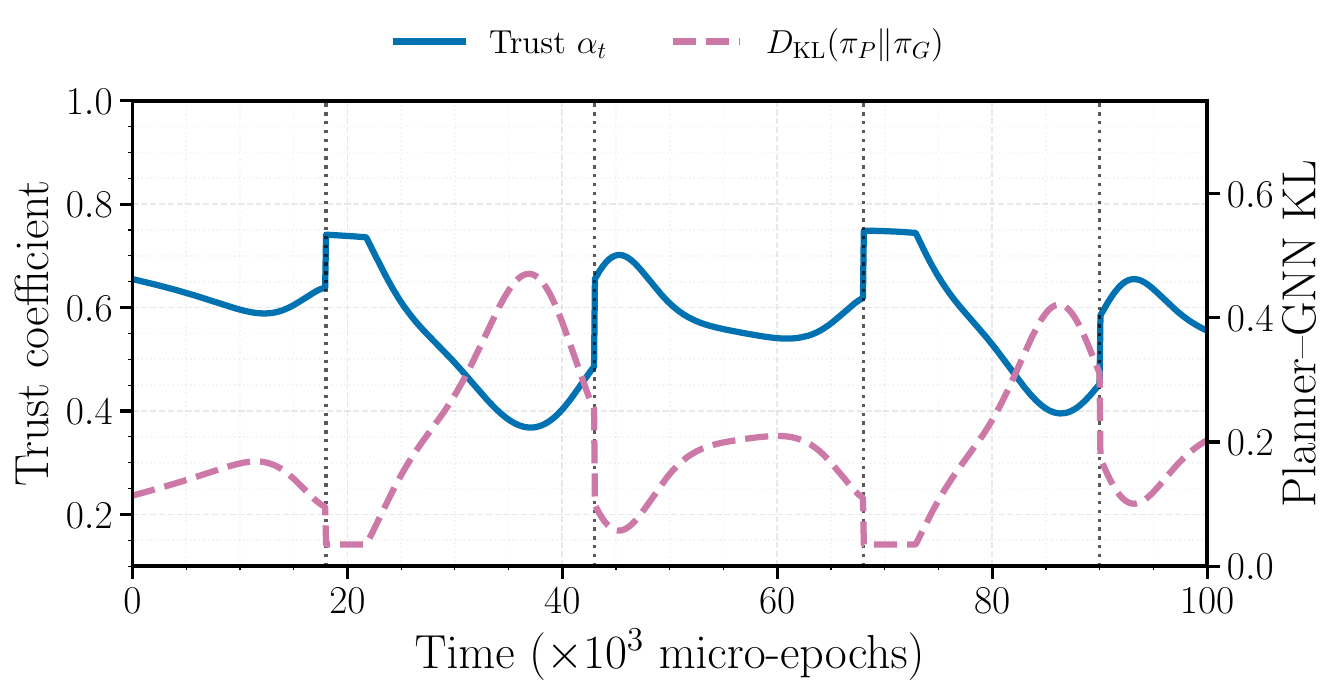}}

\vspace{0.5em}

\subfloat[{ Goodput}\label{fig:sim_nonstat_goodput}]{
    \includegraphics[width=0.48\textwidth]{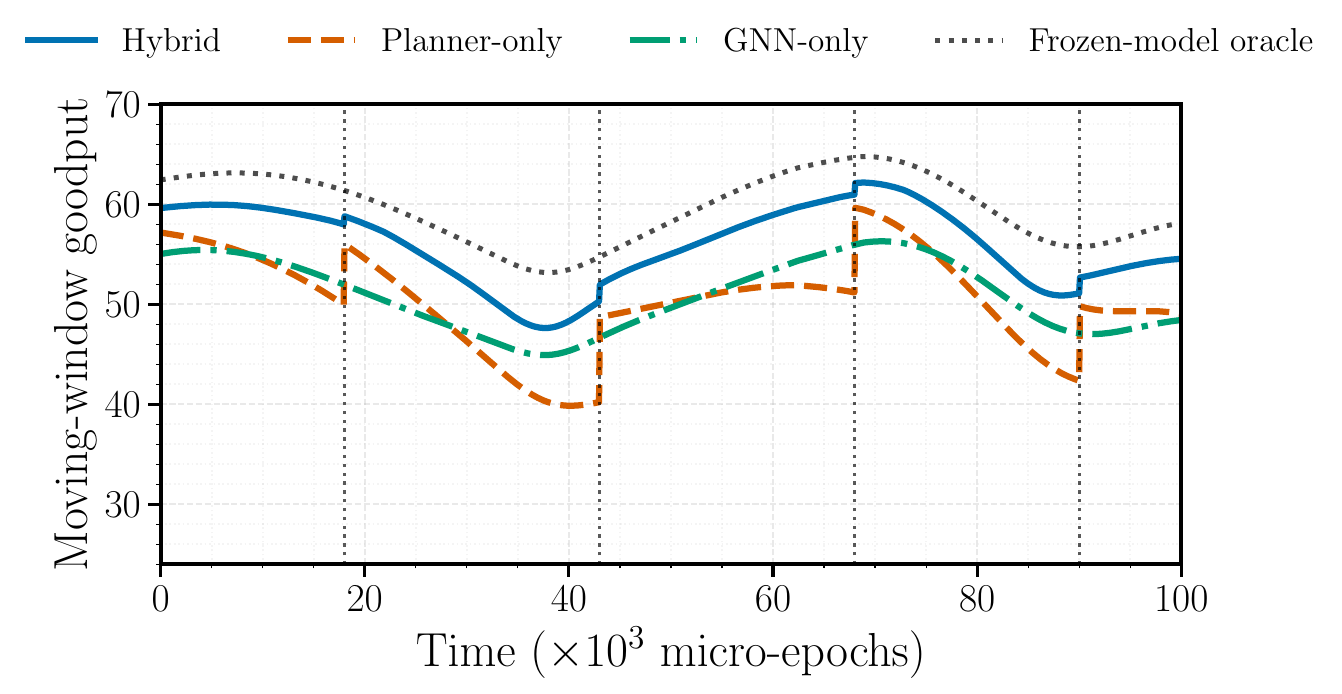}}
\hfill
\subfloat[{ Value-regret proxy}\label{fig:sim_nonstat_regret}]{
    \includegraphics[width=0.48\textwidth]{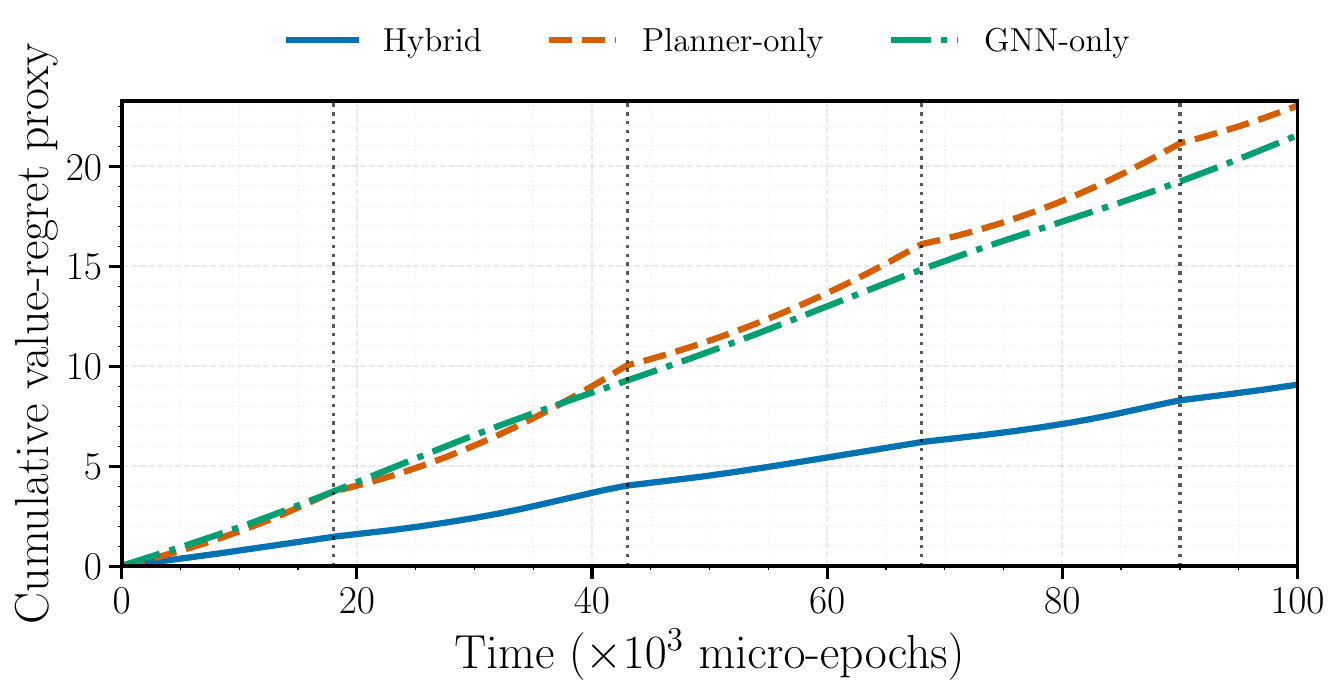}}

\caption{Nonstationary tracking with adaptive trust and planner refreshes.}
\label{fig:sim_nonstationary_trace}
\end{figure*}

\subsubsection{Prototype coverage, action matching, and scalability}

\Cref{fig:sim_prototype_scalability} connects the empirical behavior to the aggregation analysis. In \Cref{fig:sim_proto_coverage}, increasing \(K\) raises the represented-signature rate and reduces unseen fallback. The improvement is steep at small \(K\), where many feasibility strata are absent from the prototype table, and then saturates as the common strata become represented. This behavior matches the role of \(\Sigma_K\): the planner can only provide guarantees on represented signatures.

\Cref{fig:sim_proto_residual} shows the corresponding approximation tradeoff. Larger \(K\) reduces the normalized Bellman residual and goodput gap because nearest prototypes are closer in feature space and action transfer is more accurate. Matching overhead increases with \(K\), but much more slowly than the residual decreases over the useful range. This motivates the default \(K=512\): it is near the knee where most of the value approximation benefit has been obtained without making online matching a dominant cost.

\Cref{fig:sim_proto_runtime} decomposes offline wall-clock policy-evaluation time as the network size grows. Filtering and GNN scoring dominate the per-worker cost, while signature construction, role-aware matching, and transfer/trust computation remain smaller terms. The observed scaling is consistent with \cref{thm:scalability}, where GNN evaluation grows with \(n+m_t\) and prototype operations are amortized by cached planner refreshes.

\Cref{fig:sim_proto_pareto} summarizes the latency--goodput tradeoff. Heuristic methods are fast but lie far below the high-goodput frontier. Planner-only control achieves high goodput but pays a larger online cost and is more sensitive to stale tables. The hybrid policy occupies the favorable frontier region because it amortizes planning through prototypes while using the graph policy for fast action scoring on the current pair-instance graph.

\begin{figure*}[!t]
\centering
\subfloat[{ Coverage}\label{fig:sim_proto_coverage}]{
    \includegraphics[width=0.48\textwidth]{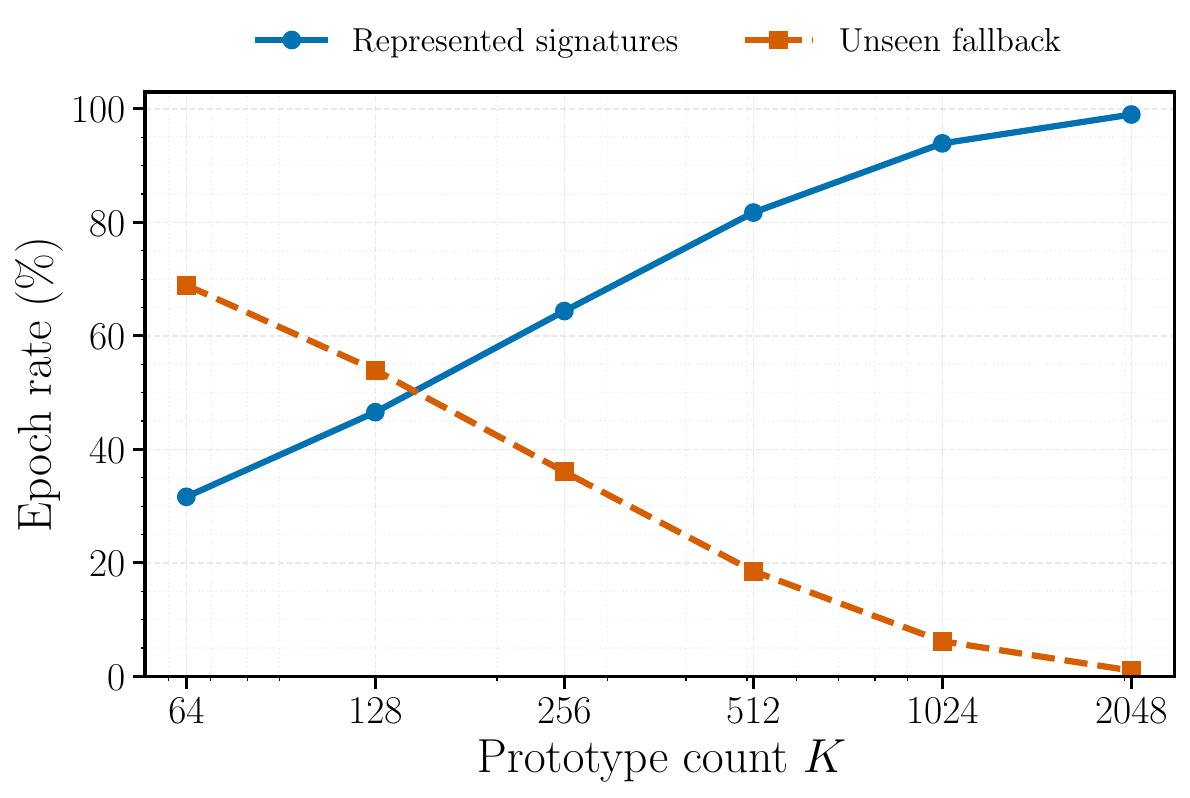}}
\hfill
\subfloat[{ Residual}\label{fig:sim_proto_residual}]{
    \includegraphics[width=0.48\textwidth]{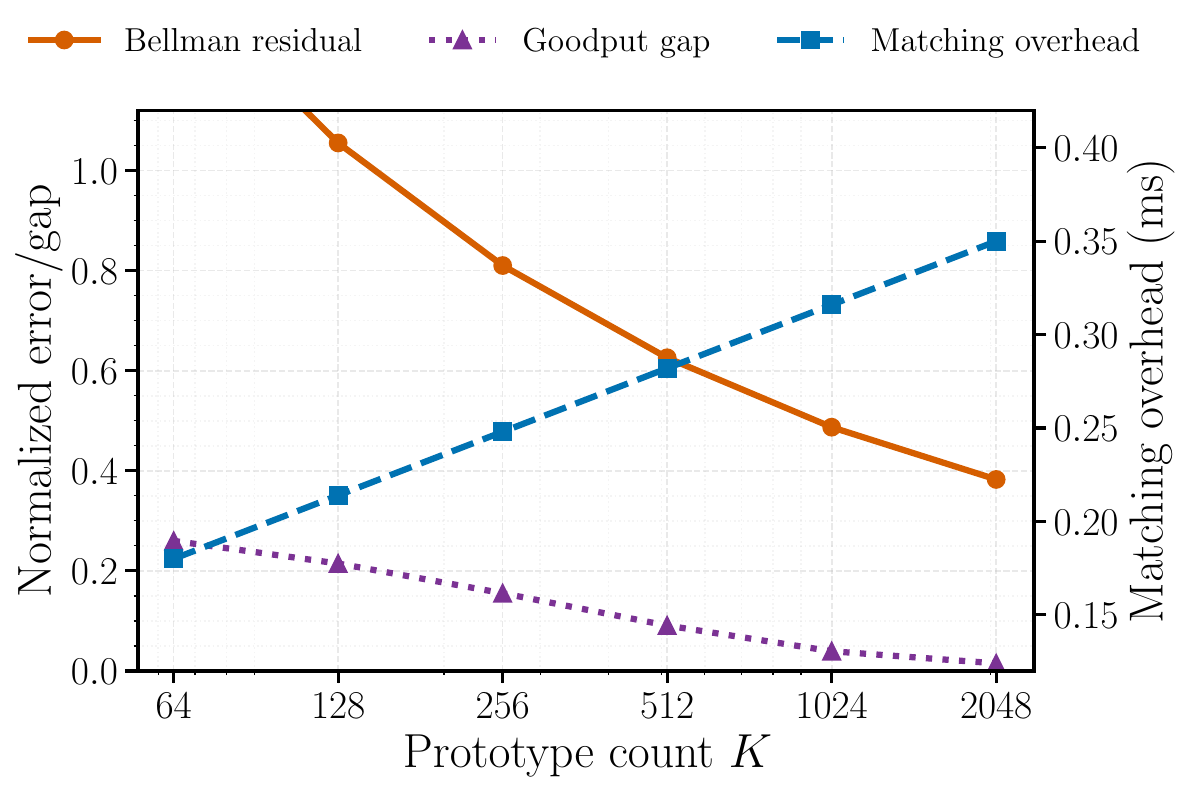}}

\vspace{0.5em}

\subfloat[{Runtime}\label{fig:sim_proto_runtime}]{
    \includegraphics[width=0.48\textwidth]{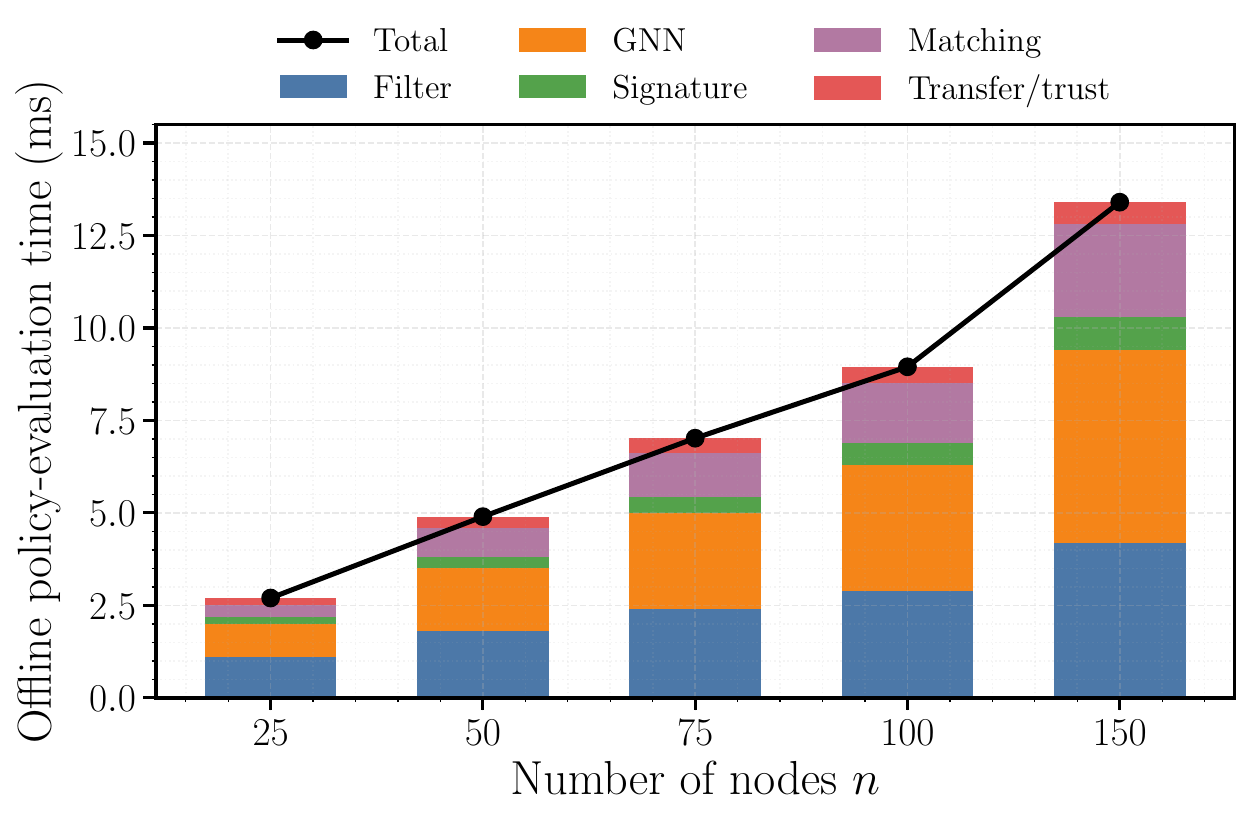}}
\hfill
\subfloat[{ Pareto frontier}\label{fig:sim_proto_pareto}]{
    \includegraphics[width=0.48\textwidth]{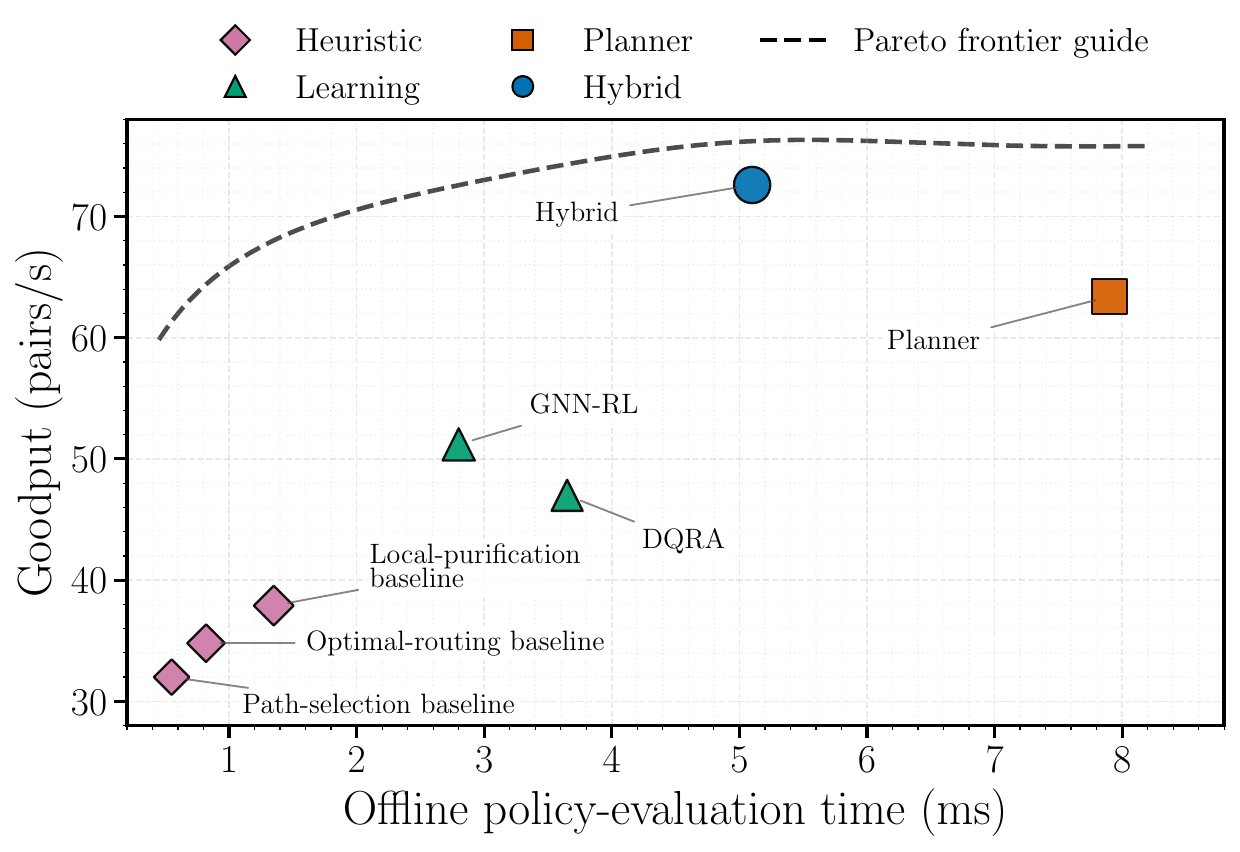}}

\caption{Prototype coverage, projected residuals, runtime decomposition, and performance--latency frontier.}
\label{fig:sim_prototype_scalability}
\end{figure*}

\begin{table*}[t]
\centering
\small
\caption{Scalability table. Online times are offline wall-clock policy-evaluation times per worker and per micro-epoch, in milliseconds. Full planner refresh is a shared operation performed periodically, not every epoch.}
\label{tab:sim_scalability_dense}
\setlength{\tabcolsep}{3.8pt}
\begin{tabular}{@{}lccccccccc@{}}
\toprule
\textbf{Network} & \(\overline A\) & \(|\mathcal P_t|\) & Filter & GNN & Signature & Matching & Transfer/trust & Online total & Full refresh\\
\midrule
\(n=25\) & 86 & 58 & 1.1 & 0.9 & 0.2 & 0.3 & 0.2 & 2.8 & 44\\
\(n=50\) & 174 & 123 & 1.8 & 1.7 & 0.3 & 0.8 & 0.3 & 5.1 & 112\\
\(n=75\) & 261 & 188 & 2.4 & 2.6 & 0.4 & 1.2 & 0.4 & 7.2 & 205\\
\(n=100\) & 348 & 252 & 2.9 & 3.4 & 0.6 & 1.6 & 0.4 & 9.1 & 310\\
\(n=150\) & 526 & 381 & 4.2 & 5.2 & 0.9 & 2.5 & 0.6 & 13.8 & 690\\
\bottomrule
\end{tabular}
\end{table*}

\Cref{tab:sim_scalability_dense} gives the same scaling in numerical form. At \(n=50\), the offline wall-clock policy-evaluation time is \(5.1\) ms per worker, whereas a full planner refresh costs \(112\) ms but is shared and periodic. At \(n=150\), the online evaluation cost increases to \(13.8\) ms, mainly through filtering and GNN evaluation. Matching and transfer/trust remain small compared with the dominant terms, supporting the use of cached planning rather than full online q-POMDP backup at every epoch.

\subsubsection{Distributional quality and robustness}

Aggregate averages can hide tail failures in quantum routing, especially when memory is finite and delivery fidelity is thresholded. \Cref{fig:sim_quality_distributions} therefore reports distributional diagnostics. In \Cref{fig:sim_quality_fidelity}, the hybrid delivered-fidelity CDF is shifted to the right of planner-only, GNN-only, and the best heuristic. The separation near \(F_k^{\min}\) is the most important region: fewer hybrid deliveries fall close to or below the threshold because the delivery test uses completion-time posterior risk.

\Cref{fig:sim_quality_queue} shows that the hybrid policy also reduces the queue-backlog tail. This means that the fidelity improvement is not obtained by simply refusing difficult service indefinitely. Instead, the policy balances generation, swapping, purification, and release so that high-fidelity deliveries occur before queues develop long tails.

\Cref{fig:sim_quality_memory} shows that the hybrid policy avoids the near-saturation memory regime more often than the baselines. This is consistent with the action-mix and ablation results: release is used as an active control decision, and low-value stored pairs are not allowed to block future generation attempts. Avoiding saturation is especially important because memory pressure can make otherwise feasible purification or generation actions unavailable.

Finally, \Cref{fig:sim_quality_robustness} compares retained normalized performance under loss bursts, \(T_2\) drift, generation-model error, flagged-instrument perturbation, observation perturbation, and combined stress. The hybrid controller retains the highest performance across perturbation regimes because robustness is introduced at several levels: the belief tracks latent conditions, the reward penalizes operation-channel deviation, the trust rule limits dependence on stale planner values, and the execution layer revalidates hard feasibility before acting.

\begin{figure*}[!t]
\centering
\subfloat[{ Fidelity CDF}\label{fig:sim_quality_fidelity}]{
    \includegraphics[width=0.48\textwidth]{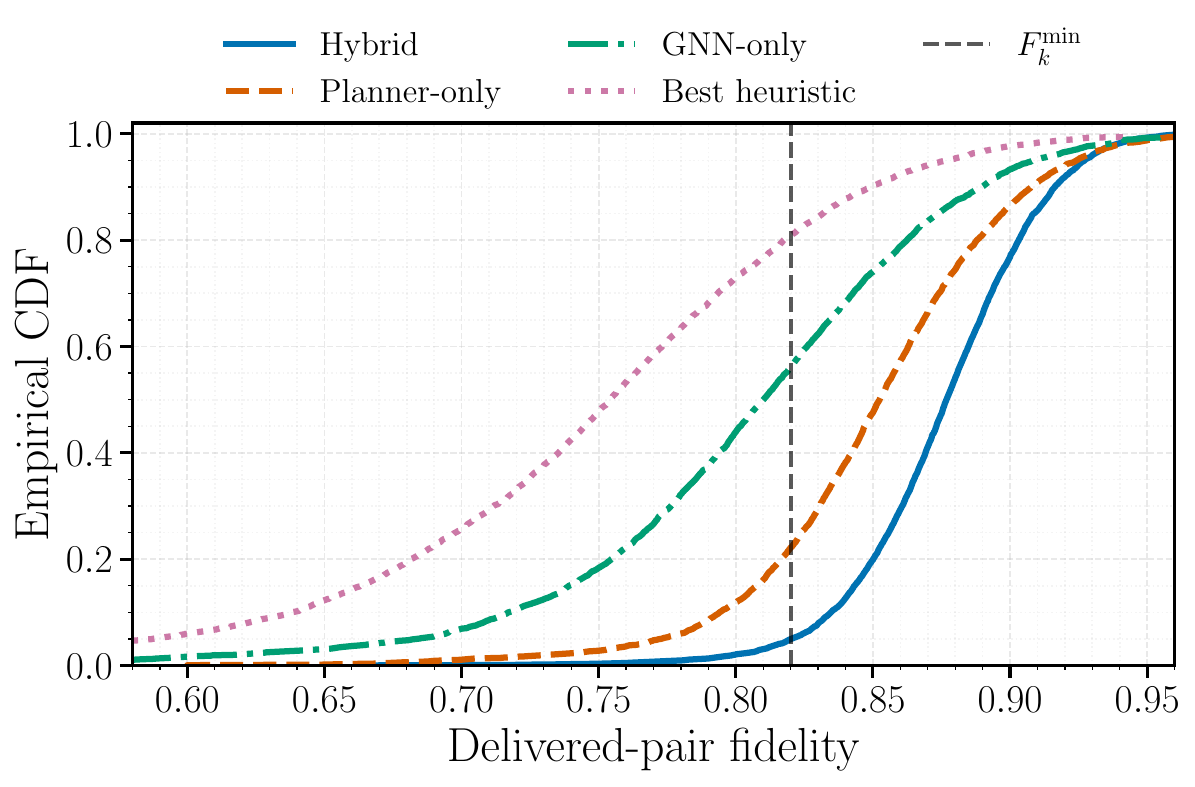}}
\hfill
\subfloat[{ Queue tail}\label{fig:sim_quality_queue}]{
    \includegraphics[width=0.48\textwidth]{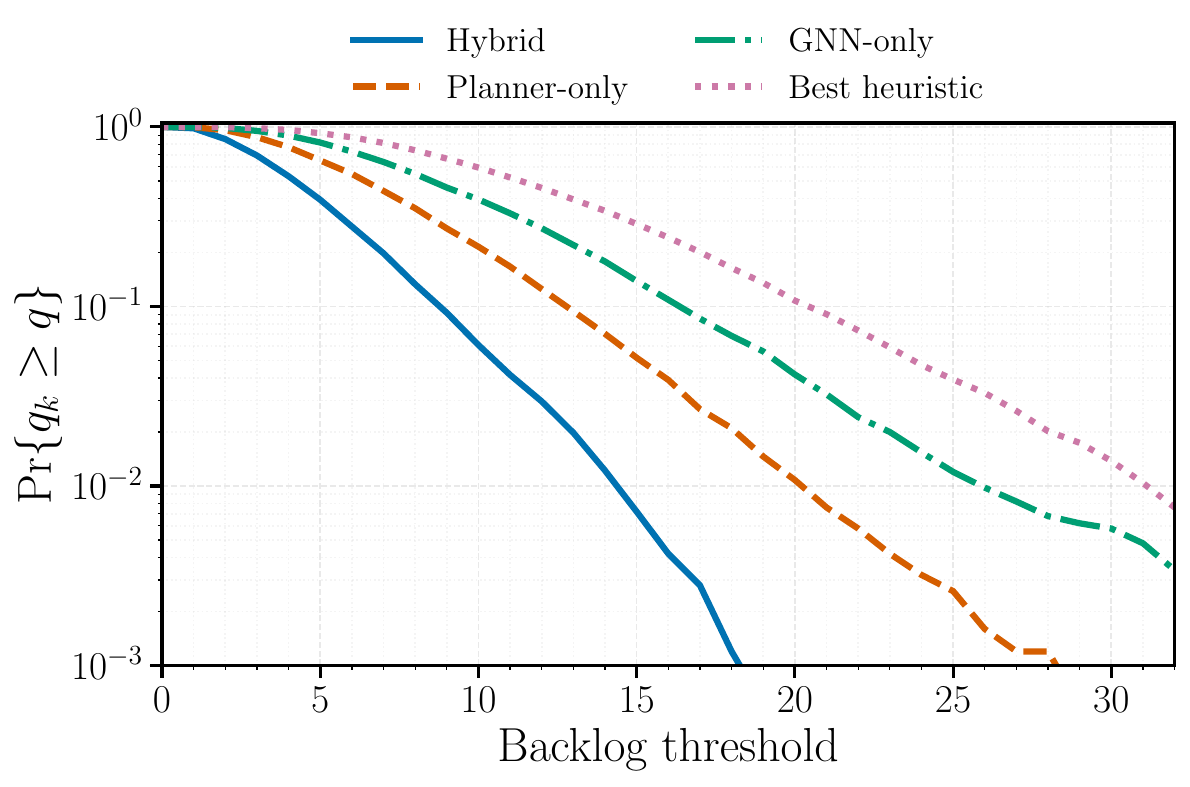}}

\vspace{0.5em}

\subfloat[{ Memory tail}\label{fig:sim_quality_memory}]{
    \includegraphics[width=0.48\textwidth]{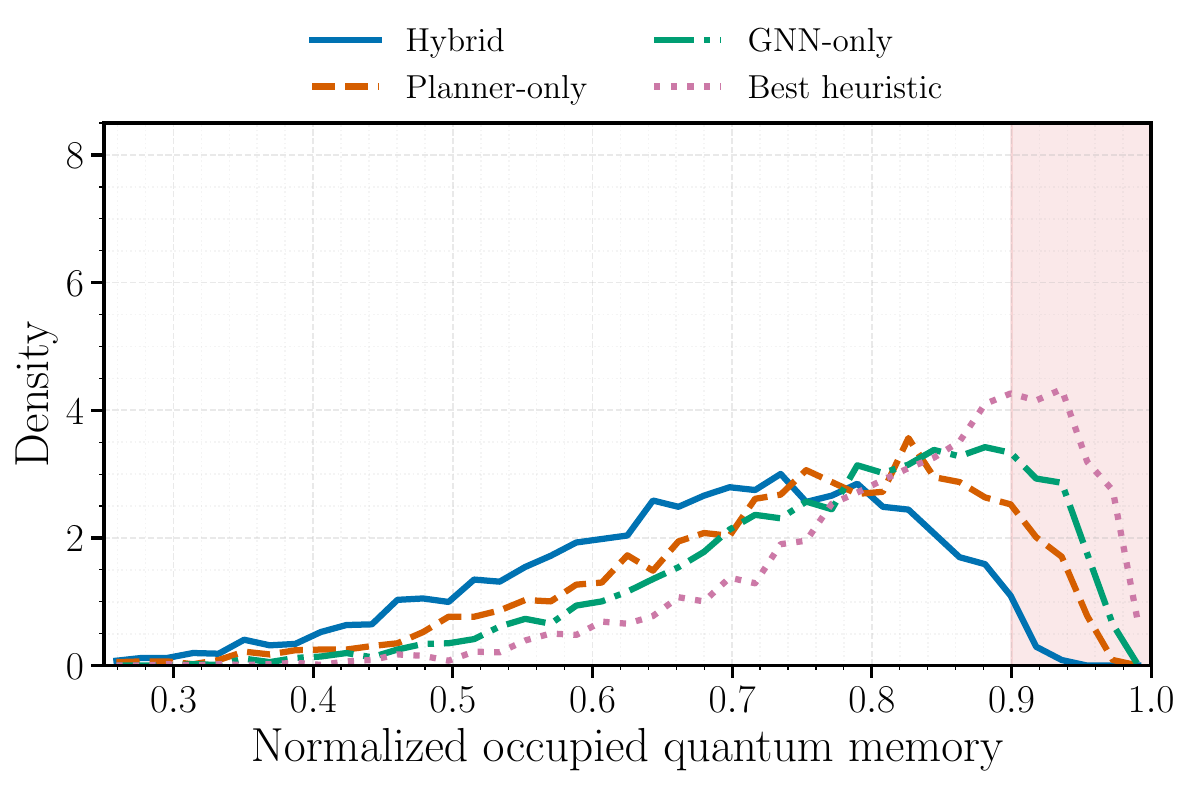}}
\hfill
\subfloat[{ Robustness}\label{fig:sim_quality_robustness}]{
    \includegraphics[width=0.48\textwidth]{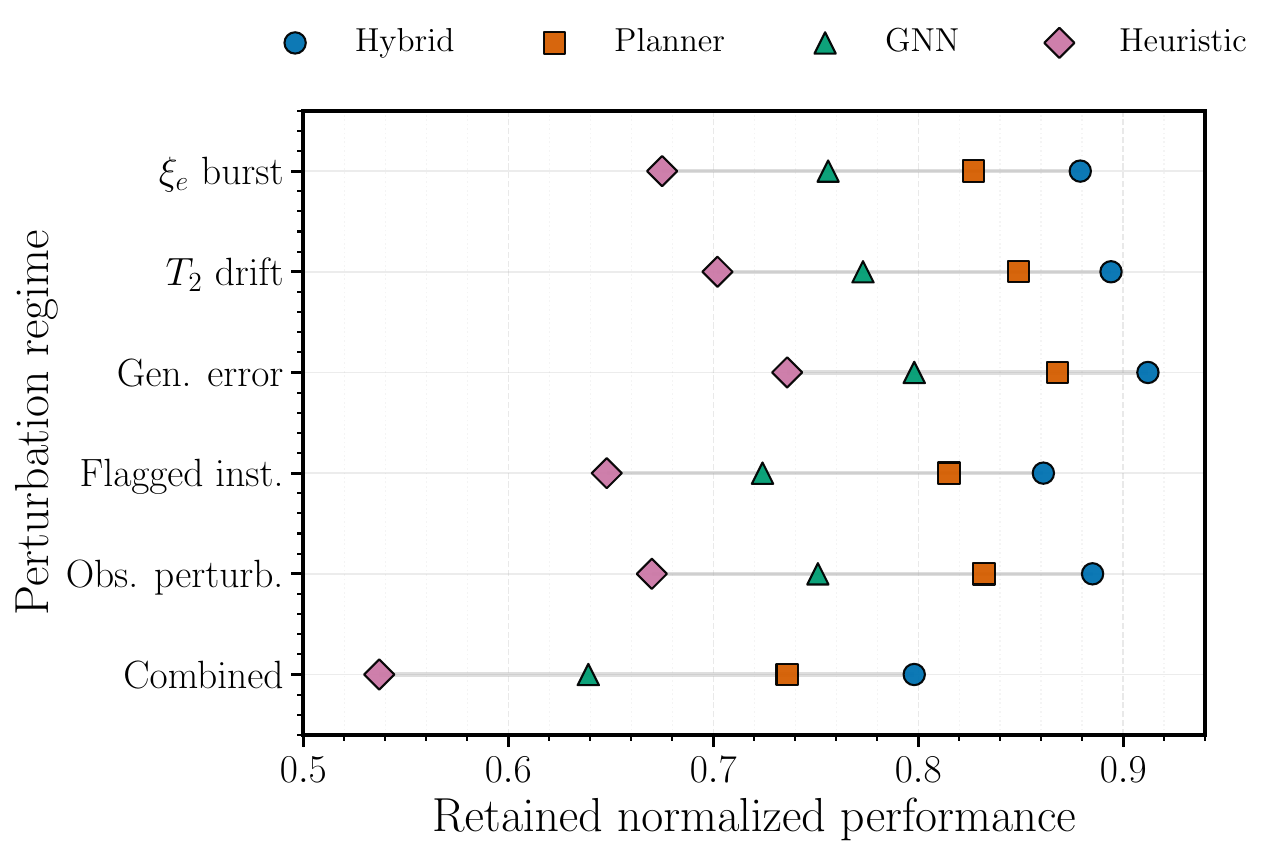}}

\caption{Distributional quality and robustness diagnostics.}
\label{fig:sim_quality_distributions}
\end{figure*}

Therefore, the numerical results support the main modeling choices of the framework. The pair-instance ledger prevents optimistic resource accounting, completion-time fidelity prevents stale delivery decisions, belief tracking improves robustness to latent physical conditions, role-aware matching makes prototype values transferable, and adaptive planner--GNN fusion provides a practical tradeoff between calibrated planning and scalable graph-based control.

\end{revblock}

\section{Conclusion}
\label{sec:conclusion}

This paper developed a robust belief-state routing framework for quantum networks operating under probabilistic heralding, finite quantum memories, decoherence, imperfect local operations, incomplete physical-state information, and time-varying conditions. The central modeling step was an atomic micro-epoch q-POMDP, in which every selected operation completes before the next decision boundary and has explicit branch-dependent consequences for memory reservations, pair inventories, queues, rewards, and posterior beliefs. This formulation made it possible to separate passive decoherence from genuine non-Markovianity, evaluate delivery using completion-time fidelity rather than stale pre-action fidelity, and enforce hard feasibility constraints through authenticated inventories and memory ledgers.

To make the resulting model scalable, we introduced feasibility-stratified belief aggregation with identifier-free signatures and role-aware action matching, allowing cached planner values to be transferred only across states with compatible feasible-action structure and concrete pair-instance semantics. The planner was then combined with a feasibility-masked GNN through an adaptive trust rule that shifts weight between calibrated model-based planning and learned graph generalization, while falling back safely on the GNN for unseen signatures. The analysis established feasibility, projected value approximation, hybrid-policy performance, robustness to model error, nonstationary frozen-model value-regret, and conditional variance guarantees for independent-worker learning batches.

The simulations showed that the proposed hybrid controller improves high-fidelity goodput, reduces below-threshold deliveries, avoids optimistic memory accounting, and maintains lower online evaluation cost than planner-only control while outperforming heuristic, purification-aware, and learning-based baselines. Overall, the results indicate that reliable quantum-network routing requires treating belief tracking, finite-memory semantics, completion-time fidelity, and scalable graph-based control as coupled components rather than separate design choices.

Future work can extend this framework to multi-action scheduling within each epoch, adaptive prototype refinement for rare feasibility signatures, and hardware-calibrated evaluations using experimental network traces.
\appendix

\section{Proofs of the Main Analytical Results}
\label{app:full_proofs}

\begin{revblock}

The proofs use the stationary policy-snapshot convention of \cref{sec:hybrid_pomdp_gnn}. Prototype arguments are restricted to \(\mathcal S_{\mathrm r}^{K}\); model-robustness and nonstationarity arguments use \(\mathcal S_{\mathrm r}\).

\subsection{Proof of Proposition~\ref{prop:covering_complexity}}
\label{app:covering_complexity_full}

\begin{proof}
A Euclidean ball of radius \(R_{\phi}\) admits an \(\epsilon\)-cover of cardinality at most \((1+2R_{\phi}/\epsilon)^{d_f}\). Covering each represented stratum and taking the union over \(\Sigma_K\) proves \cref{eq:stratified_covering_bound}.

For the sampling result, cover every \(\supp(\nu_{\zeta})\) by balls of radius \(r/2\). The total number of balls is at most
\[
M\le |\Sigma_K|\left(1+\frac{4R_{\phi}}{r}\right)^{d_f}.
\]
A covering ball centered in stratum \(\zeta\) has conditional mass at least \(c_{\zeta}(r/(2R_{\phi}))^{d_{\zeta}}\), hence unconditional mass at least \(p_{\zeta}c_{\zeta}(r/(2R_{\phi}))^{d_{\zeta}}\ge\underline m(r)\). The probability that at least one covering ball receives no sample is bounded by
\[
M(1-\underline m(r))^N\le M\exp[-N\underline m(r)].
\]
Substituting \cref{eq:stratified_sampling_bound} makes this probability at most \(\delta\). A sample within \(r/2\) of every cover center yields an \(r\)-cover. The conclusion covers the complete represented feature domain only under the stated support equality; otherwise it covers only the support of the induced sampling distribution.
\end{proof}

\subsection{Proof of Lemma~\ref{lem:value_feature_regular}}
\label{app:value_feature_regular_full}

\begin{proof}
Let \(M=M_{\varsigma\rightarrow\varsigma'}\). Since \(M\) is a bijection,
\[
\left|\max_a f(a)-\max_a g(M(a))\right|\le\max_a|f(a)-g(M(a))|.
\]
The reward difference is bounded by \cref{eq:reward_feature_regular}. Also, \(\|V^*\|_{\infty}\le V_{\max}\), and \(V^*\) is constant on admissible relabeling orbits. Therefore, for any matched action pair,
\begin{align*}
&\left|
\int V^*\,d\mathsf T(\cdot\mid\varsigma,a)
-
\int V^*\,d\mathsf T(\cdot\mid\varsigma',M(a))
\right|\\
&\le
2V_{\max}
\TV\Bigl(
\operatorname{Can}_{\#}\mathsf T(\cdot\mid\varsigma,a),
\operatorname{Can}_{\#}\mathsf T(\cdot\mid\varsigma',M(a))
\Bigr)\\
&\le
2V_{\max}L_T\|\phi(\varsigma)-\phi(\varsigma')\|_2.
\end{align*}
Applying the optimal Bellman equations at \(\varsigma\) and \(\varsigma'\), multiplying the transition bound by \(\gamma\), and maximizing over matched actions proves the result.
\end{proof}

\end{revblock}

\subsection{Proof of Theorem~\ref{thm:feature_selection} and Corollary~\ref{cor:error_propagation}}
\label{app:feature_selection_full}

\begin{proof}[Proof of Theorem~\ref{thm:feature_selection}]
For every \(\varsigma\in\mathcal S_{\mathrm r}^{K}\), the assigned prototype has the same signature. Lemma~\ref{lem:value_feature_regular} and \cref{eq:joint_covering_radius} imply
\[
\|V^*-\mathfrak P_KV^*\|_{\infty,\mathcal S_{\mathrm r}^{K}}
\le L_V\epsilon_K .
\]
Since \(V^*=\mathcal T_QV^*\) and \(V_K=\mathfrak P_K\mathcal T_QV_K\),
\begin{align*}
\|V^*-V_K\|_{\infty}
&\le
\|V^*-\mathfrak P_KV^*\|_{\infty}
\\
&\quad+
\bigl\|
\mathfrak P_K\mathcal T_QV^*
-
\mathfrak P_K\mathcal T_QV_K
\bigr\|_{\infty}
\\
&\le
L_V\epsilon_K
+
\gamma\|V^*-V_K\|_{\infty}.
\end{align*}
where transition closure makes \(\mathcal T_Q\) closed on \(\mathcal S_{\mathrm r}^{K}\), and \(\mathfrak P_K\) is nonexpansive. Rearranging proves \cref{eq:feature_aggregation_bound}.
\end{proof}

\begin{proof}[Proof of Corollary~\ref{cor:error_propagation}]
By the triangle inequality,
\begin{align*}
|V^*(\varsigma)-V_K(\widehat\varsigma)|
&\le
|V^*(\varsigma)-V^*(\widehat\varsigma)|
+
|V^*(\widehat\varsigma)-V_K(\widehat\varsigma)|.
\end{align*}
The first term is at most \(L_VL_{\phi}\epsilon_{\mathrm{est}}\) by \cref{eq:feature_lipschitz,eq:derived_value_lipschitz}; the second is bounded by Theorem~\ref{thm:feature_selection}.
\end{proof}

\subsection{Auxiliary Policy and Model Perturbation Lemmas}
\label{app:auxiliary_results}

\begin{lemma}[Discounted Policy Perturbation]
\label{lem:appendix_policy_perturbation}
Let \(\pi\) and \(\pi'\) be stationary policies on the same feasible action sets. If \(|\overline R(\varsigma,a)|\le R_{\max}\), then
\begin{equation}
\|V^{\pi}-V^{\pi'}\|_{\infty,\mathcal S_{\mathrm r}}
\le
\frac{2R_{\max}}{(1-\gamma)^2}
\sup_{\varsigma\in\mathcal S_{\mathrm r}}
\TV(\pi(\cdot\mid\varsigma),\pi'(\cdot\mid\varsigma)).
\label{eq:appendix_policy_perturbation}
\end{equation}
\end{lemma}

\begin{proof}
For stationary \(\pi\), define
\[
\begin{aligned}
(\mathcal T^{\pi}V)(\varsigma)
={}&
\sum_{a\in\mathcal A_{\mathrm f}(\varsigma)}
\pi(a\mid\varsigma)
\Bigl[
\overline R(\varsigma,a)
\\
&\qquad\qquad
+\gamma
\int
V(\varsigma')
\mathsf T(d\varsigma'\mid\varsigma,a)
\Bigr].
\end{aligned}
\]
Its fixed point satisfies \(\|V^{\pi}\|_{\infty}\le V_{\max}\). Using the two fixed-point equations,
\begin{align*}
\|V^{\pi}-V^{\pi'}\|_{\infty}
&\le
\gamma\|V^{\pi}-V^{\pi'}\|_{\infty}
+
\|\mathcal T^{\pi}V^{\pi'}-\mathcal T^{\pi'}V^{\pi'}\|_{\infty}.
\end{align*}
For \(Q^{\pi'}(\varsigma,a)=\overline R(\varsigma,a)+\gamma\int V^{\pi'}(\varsigma')\mathsf T(d\varsigma'\mid\varsigma,a)\), we have \(|Q^{\pi'}(\varsigma,a)|\le V_{\max}\). Hence
\begin{align*}
|\mathcal T^{\pi}V^{\pi'}(\varsigma)-\mathcal T^{\pi'}V^{\pi'}(\varsigma)|
&\le
V_{\max}\sum_a|\pi(a\mid\varsigma)-\pi'(a\mid\varsigma)|\\
&=
2V_{\max}\TV(\pi(\cdot\mid\varsigma),\pi'(\cdot\mid\varsigma)).
\end{align*}
Taking suprema and dividing by \(1-\gamma\) proves the lemma.
\end{proof}

\begin{lemma}[Discounted Model Perturbation]
\label{lem:appendix_model_perturbation}
Consider models \(\mathcal M\) and \(\widehat{\mathcal M}\) with common information-state domain, feasible action sets, discount factor, and reward bound. If
\[
\sup_{\varsigma,a}|\overline R(\varsigma,a)-\widehat{\overline R}(\varsigma,a)|\le\epsilon_r
\]
and
\[
\sup_{\varsigma,a}\TV(\mathsf T(\cdot\mid\varsigma,a),\widehat{\mathsf T}(\cdot\mid\varsigma,a))\le\epsilon_{\mathrm T},
\]
then every stationary feasible policy \(\pi\) satisfies
\begin{equation}
\|V_{\mathcal M}^{\pi}-V_{\widehat{\mathcal M}}^{\pi}\|_{\infty,\mathcal S_{\mathrm r}}
\le
\frac{\epsilon_r}{1-\gamma}
+
\frac{2\gamma R_{\max}\epsilon_{\mathrm T}}{(1-\gamma)^2}.
\label{eq:appendix_model_perturbation}
\end{equation}
\end{lemma}

\begin{proof}
Let \(\mathcal T_{\mathcal M}^{\pi}\) and \(\mathcal T_{\widehat{\mathcal M}}^{\pi}\) be policy Bellman operators. Then
\begin{align*}
\|V_{\mathcal M}^{\pi}-V_{\widehat{\mathcal M}}^{\pi}\|_{\infty}
&\le
\|\mathcal T_{\mathcal M}^{\pi}V_{\mathcal M}^{\pi}
-\mathcal T_{\widehat{\mathcal M}}^{\pi}V_{\mathcal M}^{\pi}\|_{\infty}
+
\gamma\|V_{\mathcal M}^{\pi}-V_{\widehat{\mathcal M}}^{\pi}\|_{\infty}.
\end{align*}
For bounded measurable \(f\), \(\left|\int f\,dP-\int f\,dQ\right|\le2\|f\|_{\infty}\TV(P,Q)\). Since \(\|V_{\mathcal M}^{\pi}\|_{\infty}\le V_{\max}\), the first term is at most \(\epsilon_r+2\gamma V_{\max}\epsilon_{\mathrm T}\). Rearranging and substituting \(V_{\max}=R_{\max}/(1-\gamma)\) proves the lemma.
\end{proof}

\begin{lemma}[Contractive Fixed-Point Perturbation]
\label{lem:appendix_fixed_point_perturbation}
Let \(\mathcal T\) and \(\widehat{\mathcal T}\) be \(\gamma\)-contractions on the same bounded function class, with fixed points \(V\) and \(\widehat V\). If \(\sup_W\|\mathcal TW-\widehat{\mathcal T}W\|_{\infty}\le\epsilon\), then
\begin{equation}
\|V-\widehat V\|_{\infty}\le\frac{\epsilon}{1-\gamma}.
\label{eq:appendix_fixed_point_bound}
\end{equation}
\end{lemma}

\begin{proof}
Using fixed points,
\begin{align*}
\|V-\widehat V\|_{\infty}
&=\|\mathcal TV-\widehat{\mathcal T}\widehat V\|_{\infty}\\
&\le\|\mathcal TV-\mathcal T\widehat V\|_{\infty}
+
\|\mathcal T\widehat V-\widehat{\mathcal T}\widehat V\|_{\infty}\\
&\le\gamma\|V-\widehat V\|_{\infty}+\epsilon.
\end{align*}
Rearrangement proves the claim.
\end{proof}

\subsection{Proof of Theorem~\ref{thm:hybrid_bound}}
\label{app:hybrid_bound_full}

\begin{proof}
At a represented state, both smoothed policies use the same uniform distribution:
\[
\overline\pi_i=(1-\varepsilon_{\mathrm{KL}})\pi_i+\varepsilon_{\mathrm{KL}}u,
\qquad i\in\{\mathrm P,\mathrm G\}.
\]
Therefore \(\overline\pi_{\mathrm P}-\overline\pi_{\mathrm G}=(1-\varepsilon_{\mathrm{KL}})(\pi_{\mathrm P}-\pi_{\mathrm G})\), and
\[
\TV(\pi_{\mathrm P},\pi_{\mathrm G})=
\frac{\TV(\overline\pi_{\mathrm P},\overline\pi_{\mathrm G})}{1-\varepsilon_{\mathrm{KL}}}.
\]
Pinsker's inequality with the planner-to-GNN direction gives \(\TV(\pi_{\mathrm P},\pi_{\mathrm G})\le\Delta(\varsigma)\). For represented states,
\[
\pi_{\mathrm H}=\alpha\pi_{\mathrm G}+(1-\alpha)\pi_{\mathrm P},
\]
so
\[
\begin{aligned}
\TV(\pi_{\mathrm H},\pi_{\mathrm P})
&\le
\alpha(\varsigma)\Delta(\varsigma),
\\
\TV(\pi_{\mathrm H},\pi_{\mathrm G})
&\le
(1-\alpha(\varsigma))\Delta(\varsigma).
\end{aligned}
\]
For unseen signatures, all three policies coincide, so both distances are zero. Taking suprema gives \(\varepsilon_{\mathrm P}\) and \(\varepsilon_{\mathrm G}\). Lemma~\ref{lem:appendix_policy_perturbation} applied to \((\pi_{\mathrm H},\pi_{\mathrm P})\) and \((\pi_{\mathrm H},\pi_{\mathrm G})\) yields the two pointwise lower bounds in \cref{eq:hybrid_performance_bound}.
\end{proof}

\subsection{Proof of Lemma~\ref{lem:var_bound}}
\label{app:var_bound_full}

\begin{proof}
Condition on \(\mathcal F_t\). Both represented and unseen states satisfy \(0\le\beta_{t,i}^{\mathrm G}\le1\). Hence
\[
\|\mathbf g_t^{(i)}\|_2^2\le G_{\pi}^2A_{\max}^2
\quad\text{a.s.}
\]
The centered worker gradients are conditionally independent and have zero conditional means, so cross terms vanish:
\begin{align*}
\operatorname{Var}(\widehat{\mathbf g}_t\mid\mathcal F_t)
&=
\frac{1}{B^2}
\sum_{i\in\mathcal I_t}
\EX\left[
\|\mathbf g_t^{(i)}-
\EX[\mathbf g_t^{(i)}\mid\mathcal F_t]\|_2^2
\middle| \mathcal F_t
\right]\\
&\le
\frac{1}{B^2}
\sum_{i\in\mathcal I_t}
\EX[\|\mathbf g_t^{(i)}\|_2^2\mid\mathcal F_t]
\le
\frac{G_{\pi}^2A_{\max}^2}{B}.
\end{align*}
Substituting \(A_{\max}=2R_{\max}/(1-\gamma)\) from \cref{eq:bounded_hybrid_advantage} proves \cref{eq:reward_gradient_variance}.
\end{proof}

\subsection{Proof of Theorem~\ref{thm:il_converge}}
\label{app:il_converge_full}

\begin{proof}
For fixed encoded features and affine logits \(z_{\theta}(a)=\theta^{\mathsf T}\mathbf x_a\),
\[
\ell_s(\theta)=
\log\sum_a\exp(z_{\theta}(a))
-
\sum_a\pi_{\mathrm P,s}(a)z_{\theta}(a)+C_s,
\]
where \(C_s\) is independent of \(\theta\). Hence \(\ell_s\) is convex. Projected online gradient descent uses
\[
\theta_{s+1}=\Pi_{\Theta}[\theta_s-\nu_N\nabla\ell_s(\theta_s)].
\]
Nonexpansiveness of Euclidean projection gives
\begin{align*}
\|\theta_{s+1}-\theta^*\|_2^2
&\le
\|\theta_s-\theta^*\|_2^2
\\
&\quad
-2\nu_N
\bigl\langle
\nabla\ell_s(\theta_s),\theta_s-\theta^*
\bigr\rangle
\\
&\quad
+\nu_N^2
\|\nabla\ell_s(\theta_s)\|_2^2 .
\end{align*}
Convexity implies \(\ell_s(\theta_s)-\ell_s(\theta^*)\le\langle\nabla\ell_s(\theta_s),\theta_s-\theta^*\rangle\). Summing over \(s\) yields
\[
\sum_{s=1}^{N}[\ell_s(\theta_s)-\ell_s(\theta^*)]
\le
\frac{B_{\theta}^2}{2\nu_N}+
\frac{\nu_NG_{\ell}^2N}{2}.
\]
With \(\nu_N=B_{\theta}/(G_{\ell}\sqrt N)\), the right-hand side is \(B_{\theta}G_{\ell}\sqrt N\). Division by \(N\) proves the theorem.
\end{proof}

\subsection{Proof of Theorem~\ref{thm:q_error} and Corollary~\ref{cor:ent_error}}
\label{app:q_error_full}

\begin{proof}[Proof of Theorem~\ref{thm:q_error}]
The exact and estimated represented models use the same feasible action sets. Thus, for \(V\in\mathbb V_{\max}^{K}\), \(|\max_a f_a-\max_a g_a|\le\max_a|f_a-g_a|\), and \cref{eq:quantum_reward_model_error,eq:quantum_transition_value_error} imply
\[
\|\mathcal T_QV-\widehat{\mathcal T}_QV\|_{\infty,\mathcal S_{\mathrm r}^{K}}
\le
\zeta_Q+\gamma\delta_Q.
\]
Nonexpansiveness of \(\mathfrak P_K\) gives the same bound for \(\mathcal T_K\) and \(\widehat{\mathcal T}_K\). Using \(V_K=\mathcal T_KV_K\),
\begin{align*}
\|V_K-\widetilde V_{\psi}\|_{\infty}
&\le
\|\mathcal T_KV_K-\mathcal T_K\widetilde V_{\psi}\|_{\infty}
\\
&\quad+
\|\mathcal T_K\widetilde V_{\psi}
-\widehat{\mathcal T}_K\widetilde V_{\psi}\|_{\infty}
\\
&\quad+
\|\widehat{\mathcal T}_K\widetilde V_{\psi}
-\widetilde V_{\psi}\|_{\infty}
\\
&\le
\gamma\|V_K-\widetilde V_{\psi}\|_{\infty}
+
\zeta_Q+\gamma\delta_Q+\epsilon_{\mathrm{BR}} .
\end{align*}
Rearranging proves \cref{eq:quantum_value_approximation_bound}. The total bound follows from the triangle inequality and Theorem~\ref{thm:feature_selection}.
\end{proof}

\begin{proof}[Proof of Corollary~\ref{cor:ent_error}]
For \(\ket{\Phi_D}=D^{-1/2}\sum_{j=1}^{D}\ket{j}_A\ket{j}_B\), the marginals satisfy \(\rho_A=\rho_B=I_D/D\), so \(\rho_A\otimes\rho_B=I_{D^2}/D^2\). The Hermitian operator \(\ket{\Phi_D}\!\bra{\Phi_D}-I_{D^2}/D^2\) has eigenvalue \(1-D^{-2}\) in the \(\ket{\Phi_D}\) direction and eigenvalue \(-D^{-2}\) with multiplicity \(D^2-1\). Therefore its trace norm is \(2(1-D^{-2})\), which proves \cref{eq:entanglement_compression_error}.
\end{proof}

\subsection{Proof of Theorem~\ref{thm:scalability}}
\label{app:scalability_full}

\begin{proof}
Propagating and reweighting \(N_b\) belief particles costs \(\mathcal O(N_bc_{\mathrm q})\). Constructing the signature, action templates, multiplicities, and canonical action map costs \(C_{\mathrm{sig},t}\). At each GNN layer, processing \(m_t\) pair messages and \(n\) node states with hidden dimension \(h\) costs \(\mathcal O((n+m_t)h^2)\). Over \(L\) layers this is \(\mathcal O(L(n+m_t)h^2)\). Constructing and scoring \(\overline A\) role-aware action embeddings costs \(\mathcal O(\overline Ah^2)\), proving \cref{eq:gnn_evaluation_complexity}. For represented signatures, direct prototype search costs \(\mathcal O(Kd_f)\), exact matching contributes \(C_{\mathrm{match},t}\), and cached-value transfer, planner normalization, trust evaluation, and fusion cost \(\mathcal O(\overline A)\). These terms are multiplied by \(\eta_{\mathrm{rep}}\). A complete projected Bellman refresh costs \(\mathcal O(K\overline A d_K)\), multiplied by \(\eta_{\mathrm P}\). For an unseen signature, prototype search and matching are skipped; insertion costs \(\mathcal O(d_f+\overline A d_K)\), multiplied by \(\eta_{\mathrm{new}}\). Summing terms proves \cref{eq:hybrid_amortized_complexity}.
\end{proof}

\subsection{Proof of Proposition~\ref{prop:policy_sensitivity}}
\label{app:policy_sensitivity_full}

\begin{proof}
Let \(\varsigma^{\mathcal E}\) be obtained by replacing each \(\rho_b^p\) with \(\mathcal E_p(\rho_b^p)\), while keeping \(c\), signature, and feasible action set unchanged. By \cref{eq:gnn_pair_lipschitz},
\begin{align*}
\|\pi_{\mathrm G}(\cdot\mid\varsigma^{\mathcal E})-\pi_{\mathrm G}(\cdot\mid\varsigma)\|_1
&\le
L_{\pi}\max_{p\in\mathcal P(c)}\|(\mathcal E_p-\mathcal I)(\rho_b^p)\|_1\\
&\le
L_{\pi}\max_{p\in\mathcal P(c)}\|\mathcal E_p-\mathcal I\|_{\diamond}.
\end{align*}
If \(\mathcal P(c)=\varnothing\), both maxima are zero.
\end{proof}

\subsection{Proof of Theorem~\ref{thm:noise_improve}}
\label{app:noise_improve_full}

\begin{proof}
By Lemma~\ref{lem:appendix_model_perturbation}, every stationary feasible \(\pi\) satisfies
\[
\|V_{\mathcal M_{\boldsymbol\vartheta}}^{\pi}-V_{\widehat{\mathcal M}_{\boldsymbol\vartheta}}^{\pi}\|_{\infty}
\le
\Delta_{\boldsymbol\vartheta},
\qquad
\Delta_{\boldsymbol\vartheta}
=
\frac{\epsilon_r}{1-\gamma}
+
\frac{2\gamma R_{\max}\epsilon_{\mathrm T}}{(1-\gamma)^2}.
\]
The common nominal channel gives, by the reverse triangle inequality,
\begin{align*}
&\left|
 d_{\diamond}(\mathcal E_{a,x,c,\boldsymbol\vartheta}^{\mathrm{ns}},\mathcal E_a^{0,\mathrm{ns}})
-
 d_{\diamond}(\widehat{\mathcal E}_{a,x,c,\boldsymbol\vartheta}^{\mathrm{ns}},\mathcal E_a^{0,\mathrm{ns}})
\right|\\
&\le
\frac12\|\mathcal E_{a,x,c,\boldsymbol\vartheta}^{\mathrm{ns}}-
\widehat{\mathcal E}_{a,x,c,\boldsymbol\vartheta}^{\mathrm{ns}}\|_{\diamond}
\le
\epsilon_{\mathrm{ch}},
\end{align*}
which yields \cref{eq:channel_to_reward_error}. Transition error is controlled separately by the flagged-instrument and classical-kernel stability assumption \cref{ass:filter_stability}; equality of nonselective channels alone would not control outcome probabilities.

Using calibrated-model improvement,
\begin{align*}
V_{\mathcal M_{\boldsymbol\vartheta}}^{\pi_{j+1}}(\varsigma)
&\ge
V_{\widehat{\mathcal M}_{\boldsymbol\vartheta}}^{\pi_{j+1}}(\varsigma)-\Delta_{\boldsymbol\vartheta}\\
&\ge
V_{\widehat{\mathcal M}_{\boldsymbol\vartheta}}^{\pi_j}(\varsigma)-\Delta_{\boldsymbol\vartheta}\\
&\ge
V_{\mathcal M_{\boldsymbol\vartheta}}^{\pi_j}(\varsigma)-2\Delta_{\boldsymbol\vartheta}.
\end{align*}
Substitution proves \cref{eq:static_noise_policy_improvement}.
\end{proof}

\subsection{Proofs of Theorems~\ref{thm:nonstationary_stability} and~\ref{thm:nonstationary_regret}}
\label{app:nonstationary_full}

\begin{proof}[Proof of Theorem~\ref{thm:nonstationary_stability}]
Both \(V_t^*\) and \(V_{t-1}^*\) belong to \(\mathbb V_{\max}^{\mathrm{full}}\). Using fixed-point equations,
\begin{align*}
\|V_t^*-V_{t-1}^*\|_{\infty}
&\le
\|\mathcal T_t^*V_t^*
-\mathcal T_t^*V_{t-1}^*\|_{\infty}
\\
&\quad+
\|\mathcal T_t^*V_{t-1}^*
-\mathcal T_{t-1}^*V_{t-1}^*\|_{\infty}
\\
&\le
\gamma\|V_t^*-V_{t-1}^*\|_{\infty}
+d_t^* .
\end{align*}
Rearranging proves \cref{eq:general_value_stability}. The parameter-Lipschitz statement follows by substitution.
\end{proof}

\begin{proof}[Proof of Theorem~\ref{thm:nonstationary_regret}]
For \(t\ge2\), add and subtract preceding-model values:
\begin{align*}
V_t^*(\varsigma_t)-V_t^{\pi_t}(\varsigma_t)
={}&
V_t^*(\varsigma_t)-V_{t-1}^*(\varsigma_t)
+
V_{t-1}^*(\varsigma_t)-V_{t-1}^{\pi_t}(\varsigma_t)\\
&+
V_{t-1}^{\pi_t}(\varsigma_t)-V_t^{\pi_t}(\varsigma_t).
\end{align*}
The middle term is at most \(\epsilon_t^{\mathrm{trk}}\). Theorem~\ref{thm:nonstationary_stability} gives \(\|V_t^*-V_{t-1}^*\|_{\infty}\le\overline d_t/(1-\gamma)\). For fixed \(\pi_t\), the policy Bellman operators are \(\gamma\)-contractive and differ uniformly by at most \(\overline d_t\), so Lemma~\ref{lem:appendix_fixed_point_perturbation} gives \(\|V_t^{\pi_t}-V_{t-1}^{\pi_t}\|_{\infty}\le\overline d_t/(1-\gamma)\). Hence each summand is bounded by \(\epsilon_t^{\mathrm{trk}}+2\overline d_t/(1-\gamma)\). Summing proves \cref{eq:nonstationary_regret_bound}. If \(\overline d_t\le\overline L_{\mathrm{env}}\|\boldsymbol\vartheta_t-\boldsymbol\vartheta_{t-1}\|\), then \(\sum_t\overline d_t\le\overline L_{\mathrm{env}}B_T^{\vartheta}\), proving \cref{eq:variation_budget_regret}. Division by \(T\) and \cref{eq:sublinear_tracking_and_variation} prove vanishing average value-regret.
\end{proof}

\begin{revblock}

\subsection{Proof of Theorem~\ref{thm:feasibility}}
\label{app:feasibility_full}

\begin{proof}
Fix \(\varsigma\in\mathcal S_{\mathrm r}\) and write \(\mathcal A_{\mathrm f}=\mathcal A_{\mathrm f}(\varsigma)\). If \(\sigma(\varsigma)\in\Sigma_K\), the assigned prototype \(\varsigma_j\) has the same signature and \(M_{\varsigma_j\rightarrow\varsigma}\) is a bijection from \(\mathcal A_{\mathrm f}(\varsigma_j)\) to \(\mathcal A_{\mathrm f}\). Every cached prototype action is instantiated as one concrete feasible current action. The planner softmax in \cref{eq:pomdp_planner_policy} normalizes exactly over \(\mathcal A_{\mathrm f}\), so \(\pi_{\mathrm P}(\mathcal A_{\mathrm f}\mid\varsigma)=1\). The masked GNN also assigns probability one to \(\mathcal A_{\mathrm f}\). Their convex mixture is therefore feasible.

If \(\sigma(\varsigma)\notin\Sigma_K\), \cref{eq:forced_unseen_fallback} gives \(\pi_{\mathrm H}=\pi_{\mathrm G}\), and feasibility follows from the GNN mask. Thus \(\pi_{\mathrm H}(\mathcal A_{\mathrm f}(\varsigma)\mid\varsigma)=1\) for every reachable information state.
\end{proof}

\end{revblock}

\bibliographystyle{IEEEtran}
\bibliography{ref}

\end{document}